\documentclass[longbibliography,aps,prl,twocolumn,superscriptaddress,showkeys,amsfonts,amssymb,amsmath,floatfix]{revtex4-2}
\usepackage{inputenc}
\usepackage{amsmath,amsbsy,amsfonts,revsymb}
\usepackage{graphicx}
\usepackage{bm}
\usepackage[dvipsnames]{xcolor}
\usepackage{chemformula}
\usepackage[separate-uncertainty=true,multi-part-units=single,range-phrase={ - }]{siunitx}
\usepackage{bm}
\usepackage{mathtools} 
\usepackage[colorlinks=true,linkcolor=blue,citecolor=blue,urlcolor=blue]{hyperref}
\usepackage{soul}

\usepackage[resetlabels]{multibib}

\newcites{Main}{References}
\newcites{SM}{}

\makeatletter
\def\maketitle{
\@author@finish
\title@column\titleblock@produce
\suppressfloats[t]}
\makeatother

\begin{document}

\preprint{APS/123-QED}

\title{Nature of metallic and insulating domains in the CDW system 1T-TaSe$_2$}

\author{M. Straub}
\affiliation{Department of Quantum Matter Physics, University of Geneva, 24 Quai Ernest-Ansermet, 1211 Geneva 4, Switzerland}

\author{F. Petocchi}
\affiliation{Department of Quantum Matter Physics, University of Geneva, 24 Quai Ernest-Ansermet, 1211 Geneva 4, Switzerland}

\author{C. Witteveen}
\affiliation{Department of Quantum Matter Physics, University of Geneva, 24 Quai Ernest-Ansermet, 1211 Geneva 4, Switzerland}
\affiliation{Department of Physics, University of Zürich, Winterthurerstr. 190, 8057 Zürich, Switzerland}

\author{F. B. Kugler}
\affiliation{Center for Computational Quantum Physics, Flatiron Institute, 162 5th Avenue, New York, New York 10010, USA}

\author{A. Hunter}
\affiliation{Department of Quantum Matter Physics, University of Geneva, 24 Quai Ernest-Ansermet, 1211 Geneva 4, Switzerland}

 \author{Y. Alexanian}
\affiliation{Department of Quantum Matter Physics, University of Geneva, 24 Quai Ernest-Ansermet, 1211 Geneva 4, Switzerland}

\author{G. Gatti}
\affiliation{Department of Quantum Matter Physics, University of Geneva, 24 Quai Ernest-Ansermet, 1211 Geneva 4, Switzerland}

\author{S. Mandloi}
\affiliation{Department of Quantum Matter Physics, University of Geneva, 24 Quai Ernest-Ansermet, 1211 Geneva 4, Switzerland}

\author{C. Polley}
\affiliation{MAX IV Laboratory, Lund University, Lund SE-221 00, Sweden}

\author{G. Carbone}
\affiliation{MAX IV Laboratory, Lund University, Lund SE-221 00, Sweden}

\author{J. Osiecki}
\affiliation{MAX IV Laboratory, Lund University, Lund SE-221 00, Sweden}

\author{F.O. von Rohr}
\affiliation{Department of Quantum Matter Physics, University of Geneva, 24 Quai Ernest-Ansermet, 1211 Geneva 4, Switzerland}

\author{A. Georges}
\affiliation{Department of Quantum Matter Physics, University of Geneva, 24 Quai Ernest-Ansermet, 1211 Geneva 4, Switzerland}
\affiliation{Center for Computational Quantum Physics, Flatiron Institute, 162 5th Avenue, New York, New York 10010, USA}
\affiliation{Collège de France, 11 place Marcelin Berthelot, 75005 Paris, France}
\affiliation{CPHT, CNRS, École Polytechnique, Institut Polytechnique de Paris, Route de Saclay, 91128 Palaiseau, France}

\author{F. Baumberger}
\affiliation{Department of Quantum Matter Physics, University of Geneva, 24 Quai Ernest-Ansermet, 1211 Geneva 4, Switzerland}
\affiliation{Swiss Light Source, Paul Scherrer Institut, CH-5232 Villigen PSI, Switzerland}
 
\author{A. Tamai}
\affiliation{Department of Quantum Matter Physics, University of Geneva, 24 Quai Ernest-Ansermet, 1211 Geneva 4, Switzerland}

\date{\today}

\begin{abstract}
We study the electronic structure of bulk 1T-TaSe$_2$ in the charge density wave phase at low temperature. Our spatially and angle resolved photoemission (ARPES) data show insulating areas coexisting with metallic regions characterized by a chiral Fermi surface and moderately correlated quasiparticle bands. Additionally, high-resolution laser ARPES reveals variations in the metallic regions, with series of low-energy states, whose energy, number and dispersion 
can be explained by
the formation of quantum well states of different thicknesses. Dynamical mean field theory calculations show that the observed rich behaviour 
can be rationalized
by assuming occasional stacking faults of the charge density wave. Our results indicate that the diverse electronic phenomena reported previously in 1T-TaSe$_2$ are dictated by the stacking arrangement and the resulting quantum size effects while correlation effects play a secondary role.

\end{abstract}

\maketitle

The tantalum-based transition-metal dichalcogenides TaX$_2$ (X = S or Se) in the 1T-polytype undergo a charge density wave (CDW) transition, characterized by the formation of a star-of-David (SOD) pattern.
The CDW rearranges the low-energy electronic structure into a narrow half-filled single band~\cite{Fazekas1979aa}. This renders 1T-TaX$_2$ model systems for correlated electron physics on a triangular lattice.
Depending on the strength of correlations, the half-filled band may undergo a Mott transition and display exotic correlated phases~\cite{Sahebsara2008,Laubach2015,ChenLei2019,wietek2021,Pickett2007,KangZhang2020}. 
Indeed, multiple phenomena and quantum phases have been reported.
In 1T-TaS$_2$ the CDW reconstruction induces a metal-insulator transition \cite{disalvo1977}.
A hidden metastable metallic phase can be accessed with light or voltage pulses~\cite{stojchevska2014, Hollander2015}. 
Superconductivity appears under hydrostatic pressure~\cite{Sipos2008}.
Moreover, the Kondo effect and signatures of a quantum spin liquid have been reported in monolayer TaS$_2$ and TaSe$_2$ at low temperature~\cite{Law2017aa, Murayama2020, ruan2021,chen2021, vano2021, Chen2022aa}.
However, a unifying picture of the different phases 
is lacking.

In the case of bulk 1T-TaSe$_2$ even elementary properties are not fully established.
Resistivity measurements show a metallic behaviour down to the lowest temperature \cite{disalvo1974a,wilson1974} consistent with a recent ARPES study \cite{tian2023}.
In contrast, earlier ARPES work found a significant suppression of the spectral weight towards the Fermi level, which was initially interpreted as evidence for a Mott insulating surface layer~\cite{perfetti2003}. A recent time-resolved ARPES study reported a gap as large as $\sim$ 0.7 eV and argued that bulk 1T-TaSe$_2$ is a charge-transfer insulator \cite{sayers2023}. 
Scanning tunneling microscopy (STM) showed that insulating and different, possibly correlated, metallic states can coexist at the surface on a micrometer scale \cite{chen2022,zhang2022,fei2022}. 
This behaviour has been interpreted in terms of a periodic Anderson model assuming coupling between a putative Mott insulating surface layer and the metallic bulk \cite{chen2022,chen2021}.

An important degree of freedom, often disregarded in 2D materials, is the stacking of CDW layers. X-ray diffraction finds that the buckled structure induced by the CDW in 1T-TaSe$_2$ orders periodically out-of-plane \cite{wiegers2001}, while in 1T-TaS$_2$ intertwined domains with different stacking configurations are observed down to low temperature \cite{scruby1975,wang2023a}.
STM studies and density functional theory (DFT) calculations indicate that some stacking configurations in TaS$_2$~\cite{ritschel2018,lee2019,butler2020,wu2022}, and possibly also stacking faults in TaSe$_2$~\cite{ge2010,wang2023}, may result in a band insulator independent of any Mott physics.

In this Letter, we present a microfocus ARPES study of bulk 1T-TaSe$_2$ in the CDW phase.  
We find that large parts of the surface are insulating. Small metallic regions display weakly correlated bands and a chiral Fermi surface in agreement with dynamical mean field theory (DMFT) calculations of the bulk electronic structure.
High-resolution laser ARPES shows series of sharp peaks a few meV from the Fermi level, consistent with STM data interpreted as evidence for strong correlation effects.
We demonstrate that our findings can be explained naturally assuming occasional stacking faults in the CDW and do not imply strong correlations.

\begin{figure*}[t]
\centering
\includegraphics[width=1.5\columnwidth]{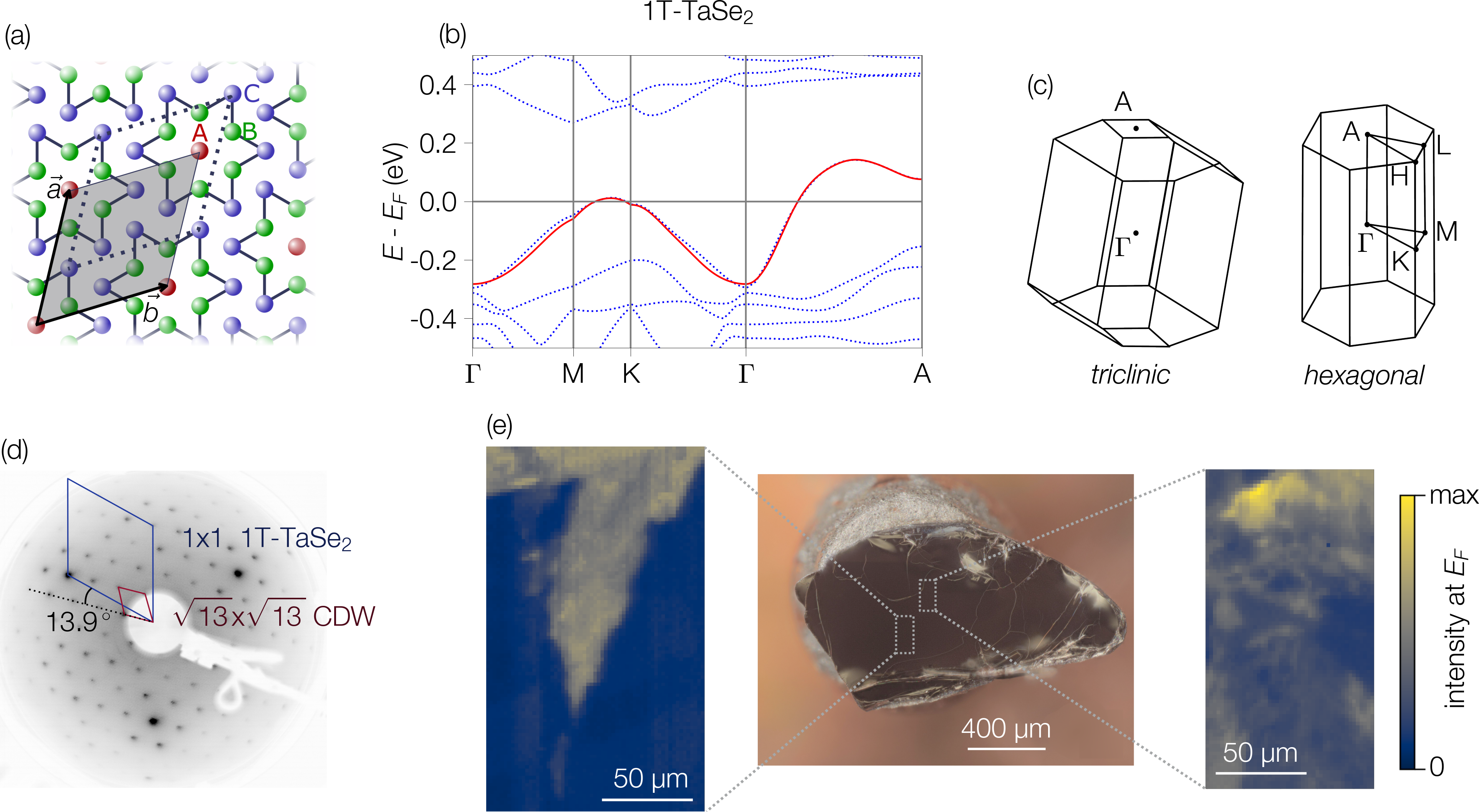}
\caption{Star-of-David CDW phase in 1T-TaSe$_2$. (a) Sketch of the real space reconstructed lattice. The selenium atoms are omitted for clarity. The CDW unit cell in the (a,b) plane is marked in grey. The dashed unit cell indicates the translation of the CDW between consecutive layers.
(b) DFT band structure of bulk 1T-TaSe$_2$ (blue dotted lines) and the single band effective tight-binding model derived from maximally localized Wannier functions (red line). 
(c) Triclinic BZ of the bulk stacking. Throughout this Letter we label $k$-space paths with the high symmetry points of the simple hexagonal BZ corresponding to AA stacking.
(d) LEED pattern of cleaved 1T-TaSe$_2$. The unit cells of the unreconstructed lattice and the CDW superstructure are marked in blue and red, respectively. (e) ARPES spatial scans showing the integrated photoemission intensity within an energy window of $\pm$ 5 meV around the Fermi level. The position and sizes of the spatial scans are shown in the central optical image.}
\label{fig:fig1}
\end{figure*}

We start by introducing the low-temperature CDW structure of 1T-TaSe$_2$ (Fig.~\ref{fig:fig1}(a)). The Ta atoms are arranged in a triangular lattice, sandwiched between two triangular layers of Se. The system undergoes a CDW transition at 473 K, but unlike 1T-TaS$_2$, it remains metallic, with the electronic specific heat indicating a small but finite density of states \cite{disalvo1974a}.
At the CDW transition the 12 Ta atoms forming a SOD move towards the central Ta atom while the neighboring Se planes buckle to help relieve the stress.
The CDW layers are then stacked such that the central Ta atom in one layer is aligned with an outermost Ta atom of a SOD in the following layer (AC stacking). In the resulting 3D unit cell the c-axis is inclined with respect to the (a,b) plane and the system becomes triclinic with $P\bar{1}$ space group. This stacking configuration breaks all spatial symmetries, other than translation and inversion symmetry.  
Fig.~\ref{fig:fig1}(b) displays the DFT band structure of the CDW structure. 
Even though one CDW unit cell contains 13 Ta and 26 Se atoms, the low-energy electronic structure is remarkably simple. The $d_{z^2}$ orbitals of the outer Ta atoms of the SOD hybridize and form six filled bonding bands and six empty antibonding bands. The remaining $d_{z^2}$ orbital of the central Ta atom forms a single half-filled nonbonding band.
Because of the triclinic stacking, the in-plane dispersion remains larger than in 1T-TaS$_2$, which is in line with the CDW transition in 1T-TaSe$_2$ not coinciding with a metal-insulator transition as observed in the sulfide \cite{ge2010}.

Fig.~\ref{fig:fig1}(d,e) illustrates key properties of our bulk single crystals cleaved at low temperature. 
Low-energy electron diffraction (LEED) shows a sharp pattern with the $\sqrt{13}\times\sqrt{13}$ periodicity of the CDW reconstruction. 
The unit cell 
is rotated by $13.9^\circ$ with respect to the undistorted lattice, which breaks the mirror symmetry and gives rise to chirality in plane \cite{yang2022}. We note that on most samples a single rotational domain (right handed or left handed) is found, indicating that our samples are of very good quality.
Details of sample growth and characterization are given in Supplemental Material
.
This picture is in sharp contrast with spatially-resolved photoemission intensity maps of the same samples measured with a microfocused laser (Fig.~\ref{fig:fig1}(e)). Monitoring the spectral intensity in a narrow window around the Fermi level ($E_F$), we find large regions with near zero intensity typical of an insulating state. These regions coexist with smaller areas of high intensity which display a metallic spectrum. Within these metallic regions, we find significant intensity variations on a few micrometer scale (Fig.~\ref{fig:fig1}(e), map on the right).
In the following we first present the electronic structure underlying the two different domains. Variations within the metallic areas will be addressed in Fig.~\ref{fig:fig4}.

\begin{figure*}[t]
\centering
\includegraphics[width=1\textwidth]{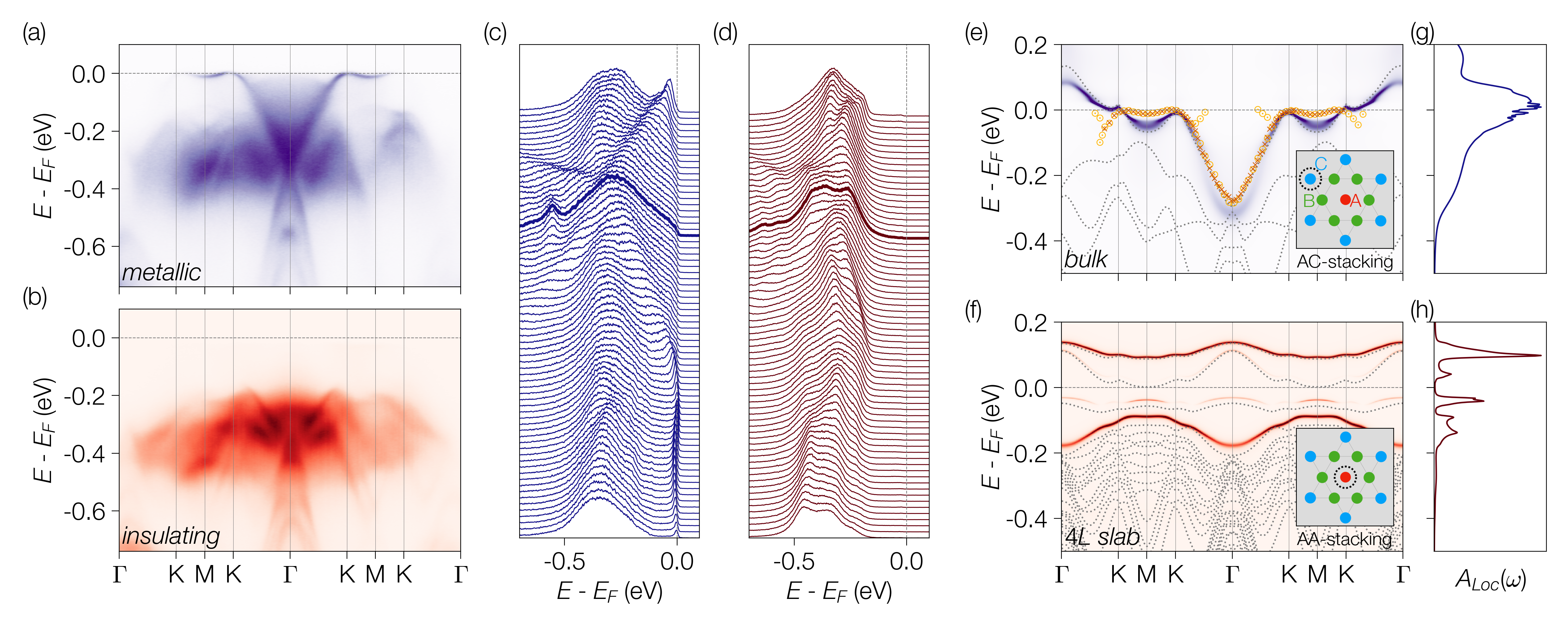}
\caption{Metallic and insulating band structures on the surface of 1T-TaSe$_2$. (a,c) ARPES spectra obtained on metallic, and (b,d) on insulating domains. Data were acquired with 50 eV s-polarized light on the same cleaved surface. (e,f) DMFT spectral functions of the bulk structure (AC stacking) and of a 4-layer slab where the topmost layer is aligned to the layer below (AA stacking). Dotted lines in (e,f) are the non-interacting bands calculated with DFT. Yellow markers in (e) indicate the ARPES dispersion of the metallic bands obtained with $s$ and $p$ polarization (round and cross markers, respectively). The local spectral functions of the two structures are displayed in (g) and (h).
}
\label{fig:fig2} 
\end{figure*}
Figure~\ref{fig:fig2}(a-d) shows ARPES data from metallic and insulating domains measured with a 10 $\mu$m spot at the BLOCH beamline of MAX IV.
At high energy, around -0.6 eV, both domains display rapidly dispersing hole-like bands centered at $\Gamma$, which derive from Se $p$ orbitals. Moving towards the Fermi level, in an energy window between about -0.4 and -0.2 eV, there are multiple sets of backfolded bands with dominant Ta $d_{z^2}$ character. In the insulating domains the photoemission intensity drops to near zero above -0.2 eV (Fig.~\ref{fig:fig2}(b,d)), which is in line with the gap observed in STS studies of bulk single crystals \cite{chen2022}. In the metallic domains this set of bands is $\sim$ 80 meV closer to $E_F$. Here the dominant feature is a fast dispersing band that resembles a Dirac cone centered at $\Gamma$ and reaches the Fermi level to form shallow electron-like pockets at the edges of the Brillouin zone (Fig.~\ref{fig:fig2}(a,c)).
In both domains, the in-plane superstructure of the CDW is manifested in the back folding of the Se 4$p$ bands (see Supplemental Material, Section VII). This rules out the possibility that the observation of two distinct band structures is due to the formation or absence of the CDW reconstruction in different regions of the crystal.
We also note that 
the metallic bands, 
and even the individual bands at higher energy, are not discernible in ARPES data taken with a spot size $> 100 \, \mu$m (see as example our He lamp data in Supplemental Material, Section V). This is fully consistent with previous ARPES studies not seeing the metallic states and interpreting the broad peak below $E_F$ as the lower Hubbard band of a Mott insulator \cite{perfetti2003,chen2020,nakata2021}. Clearly that interpretation was biased by the poor spatial resolution of those measurements.

We further investigate the different domains with DFT+DMFT calculations of the bulk unit cell and of two slab structures,
one where the AC stacking of the bulk is maintained for all layers and one where the top layer is translated to have two SODs on top of each other (AA stacking). 
Upon structural relaxation, the two slab structures are found to differ only by a few meV/atom in total energy. This is in line with STM studies that reported both AC and AA stacking, together with other possible configurations, at the surface of 1T-TaSe$_2$ \cite{zhang2022}.
For the DMFT calculations, DFT band structures were downfolded to effective low energy tight-binding models of the $d_{z^2}$ orbital localized on the central Ta atom of each cluster employing maximally-localized-Wannier-functions.
A local Hubbard interaction of $U=0.198$~eV was estimated using constrained random phase approximation (cRPA) and kept fixed throughout all our calculations. 
Further details are given in Supplemental Material, Sections III and IV.

The momentum-resolved DMFT spectral function of the bulk unit cell is shown in Fig.~\ref{fig:fig2}(e) with superimposed the ARPES dispersions obtained with different light polarizations on a metallic domain (square and round yellow markers).
The overall good agreement between theory and experiment suggests that those measurements are representative of the bulk structure.
The DMFT solution for the bulk unit cell lies in the Fermi liquid regime with moderate band renormalization (Z $\sim$ 0.66), in agreement with the metallic behavior seen in transport  \cite{disalvo1974a,wilson1974}.
Compared to DMFT, the ARPES data show higher renormalization of low-energy states (see for instance the dispersion of the shallow electron pocket centered at M). Whether this indicates stronger correlation effects in the bulk or derives from surface modifications of the band structure remains unclear. 
In Fig.~\ref{fig:fig2}(f) we show the surface spectral function of the AA-stacked slab. The strong hybridization between vertically aligned $d_{z^2}$ orbitals results already at the DFT level in a band insulating surface, with minimal changes introduced by DMFT. Albeit with smaller gap size, the resulting valence band dispersion is in qualitative agreement with our data from the insulating regions, suggesting that local variations in the CDW stacking are the main reason for the formation of insulating domains. These observations are in agreement with the existing literature on 1T-TaS$_2$ reporting experimental and theoretical evidence that AA stacking is responsible for an insulating surface \cite{wu2022,fei2022,lee2019,ritschel2015,ritschel2018,butler2020,Petocchi2022,wang2023a}. 

\begin{figure}[t]
\centering
\includegraphics[width=1\columnwidth]{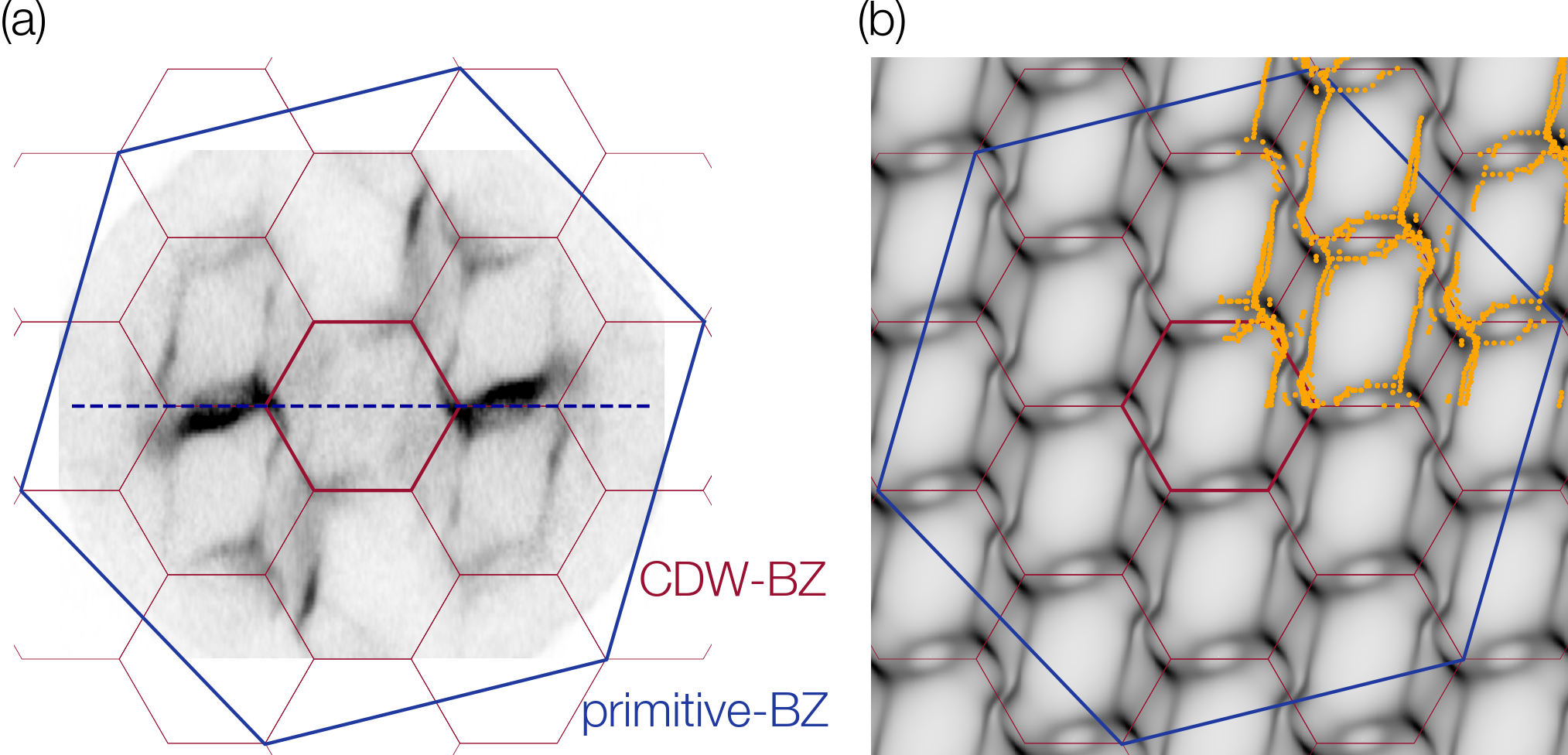}
\caption{(a) Fermi surface of bulk-like metallic 1T-TaSe$_{2}${} measured with 50 eV $p$-polarized light. The BZs of the unreconstructed lattice and of the CDW  are indicated by blue and red lines. The blue dashed line marks the momentum cut displayed in Fig.~\ref{fig:fig2}. (b) DMFT Fermi surface of a 100-layers slab mimicking a semi-infinite bulk structure. The yellow dots in the upper right corner are the extracted ARPES contours offset by the CDW reciprocal lattice vectors.}
\label{fig:fig3} 
\end{figure}

In Fig.~\ref{fig:fig3}(a) we present the Fermi surface of 1T-TaSe$_{2}$ as measured by ARPES with 50 eV photons. 
Using the free-electron final-state approximation and assuming an inner potential of 15 eV, this photon energy corresponds to a cut at $k_z \approx 4\pi/c$, near the center of the BZ. Our measurement shows well-defined sharp contours that form a 2-fold symmetric pattern lacking mirror symmetry. This is in agreement with the in-plane chiral structure of the CDW.
The Fermi surface consists of distorted hole-like parallelograms centered at the $\Gamma$ points of each CDW BZ. 
At two sides of the hexagonal BZ, where these contours overlap, small elliptical electron-like pockets form. The Luttinger volume is 0.84 $\pm$ 0.2 \;$e$/CDW unit cell corresponding to 0.012 holes per Ta atom, likely due to a similar number of Ta vacancies. 

In  Fig.~\ref{fig:fig3}(b) we compare the extracted ARPES contours to a DMFT spectral function of an AC-stacked slab artificially extended to 100 layers by means of an embedding Green's function constructed with continued fractions \cite{Petocchi2022}. 
Such a slab calculation is equivalent to a surface projection of the 3D bulk band structure. The good agreement between the two indicates that the ARPES data are closer to the $k_z$ integrated spectral weight rather than a specific cut through the 3D BZ. This is a natural consequence of the poor $k_z$ resolution of ARPES resulting from the strong damping of the final state \cite{Lindroos1996}. In the case of 1T-TaSe$_{2}$, the integration produces only limited broadening and clear contours remain visible because of the peculiar staircase-like out-of-plane dispersion of the bulk bands (see Section VIII in Supplemental Material).
Signatures of the $k_z$ integration are (i) the ARPES Fermi surface that follows the hexagonal periodicity of the projected Brillouin zone (Fig.~\ref{fig:fig3}(a)) and (ii) the high spectral intensity inside the ``Dirac cone" and the faint back-folded bands that are not seen in calculations of the bulk triclinic structure (compare Fig.~\ref{fig:fig2}(a) to Fig.~\ref{fig:fig2}(e)).
%


\begin{figure*}[t]
\centering
\includegraphics[width=0.95\textwidth]{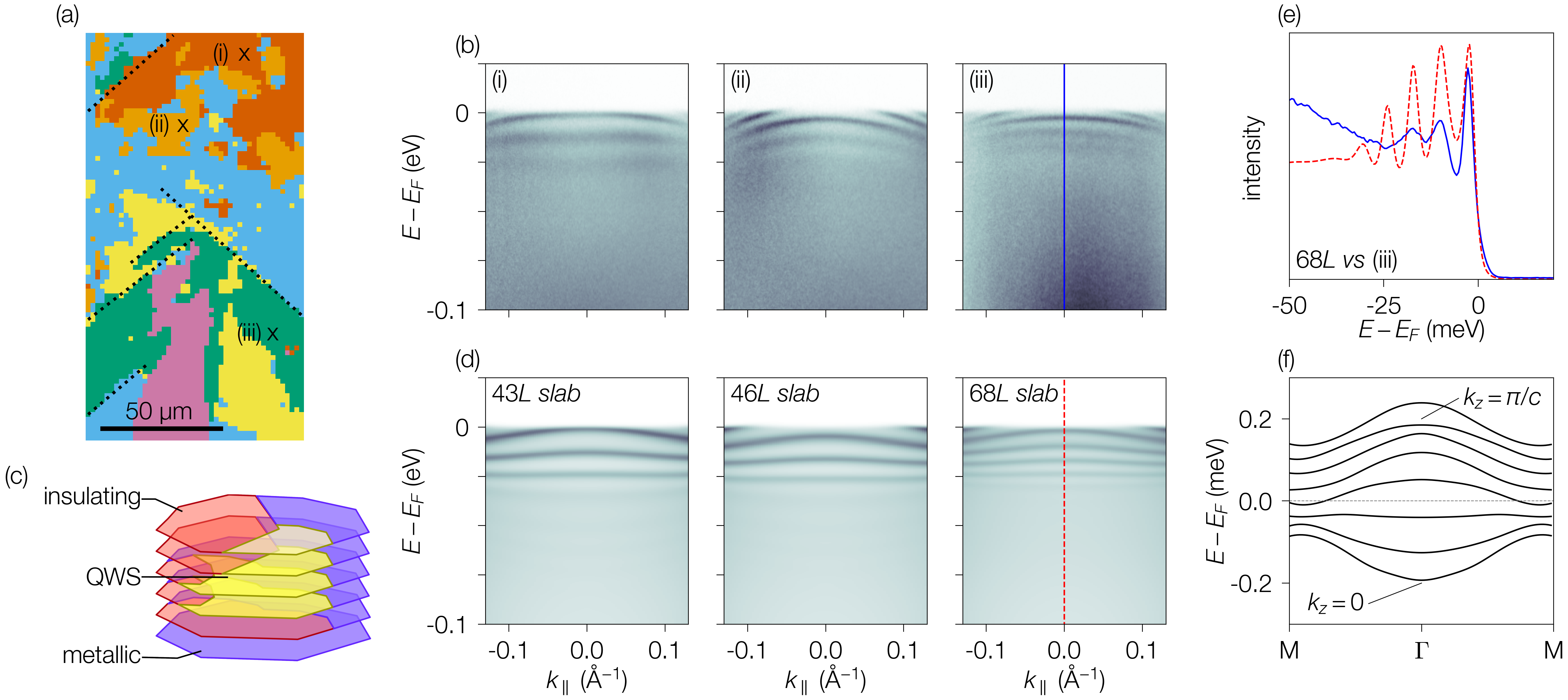}
\caption{QWSs in metallic 1T-TaSe$_2$. (a) K-means cluster map of the metallic area shown in Fig.~\ref{fig:fig1}(d) obtained using EDCs at $\Gamma$, integrated over $\pm$ 0.06 $\text{\AA}^{-1}$. (b) Dispersion plots of QWSs from three different positions indicated in panel (a). (c) Sketch of the origin of the QWS. Stacking faults within the material cause insulating layers and confine the metallic electrons in potential wells. (d) DMFT slab calculations for different numbers of layers selected to match the laser ARPES spectra in panel (b). (e) Comparison of ARPES and DMFT spectra at $\Gamma$ (solid blue and dashed red line, respectively). (f) Band structure calculation for 8-layer QWSs.}
\label{fig:fig4}
\end{figure*}
In the remainder of this Letter, we discuss variations of the electronic structure within a metallic region.
Fig.~\ref{fig:fig4}(a) shows a cluster map of the metallic area displayed in Fig.~\ref{fig:fig1}(d). By applying the K-means clustering algorithm to energy distribution curves (EDCs) at $\Gamma$ \cite{ekahana2023}, we identify 6 different domains which we highlight by different colors. The edges of the domains preferentially follow high symmetry directions of the unreconstructed 1T-TaSe$_2$ lattice. 
Beyond the bulk-like metallic and insulating domains described above, we systematically find regions displaying series of weakly dispersing states confined within 50 meV from the Fermi level. The number of states, their energy and dispersion vary as we scan a metallic region in micrometer steps. Representative dispersion plots from these regions are displayed in panel (b). 
The observation of these sharply defined features is remarkable given that according to our bulk band structure calculations one does not expect any band in this energy/momentum window. 

A natural explanation for this phenomenology is the formation of quantum well states (QWS). As we have shown before, stacking faults of the CDW can produce an insulating layer. The presence of such an insulating layer acts as a barrier within the material and confines the conduction electrons between that layer and the vacuum (Fig.~\ref{fig:fig4}(c)). 
Indeed a simple phase accumulation model for a particle in a box accurately reproduces the number and energy of the states observed in experiment (see Supplemental Materials, Section IX). 
We further studied the QWS theoretically by varying the number of continued fractions iterations used to extend the slab calculation beyond four layers (Supplemental Material, Section IV(a)). 
Figs.~\ref{fig:fig4}(d,e) show that this approach reproduces the key features of the data. Importantly, our analysis reveals that the near flat dispersion of the low-energy states is purely a band structure effect. 
The QWSs originate from the quantization of the out-of-plane dispersion, with the lowest energy band corresponding to $k_z = 0$ and the highest energy band close to $\pi / c$ 
(Fig.~\ref{fig:fig4}(f)). Along this path the bulk band crosses the Fermi level and the in-plane dispersion changes curvature. This naturally results in flat QWSs near the Fermi level. We note that the corresponding EDCs resemble the different STS spectra of bulk 1T-TaSe$_2$ showing a more or less pronounced zero-bias peak \cite{chen2022}. That peak was interpreted as a Kondo resonance resulting from strong correlation physics. The formation of QWSs documented here offers a simpler alternative explanation.

In conclusion, the observation of a metallic Fermi surface in  bulk 1T-TaSe$_2$ settles a longstanding apparent inconsistency between ARPES, STM and transport studies. Our work establishes that 1T-TaSe$_2$ in the CDW phase is a moderately correlated metal.
It remains an open question why the surface layer is predominantly insulating. Given that the energy difference between different stacking configurations is small \cite{wang2023}, it is possible that the absence of out-of-plane bonding in one direction favors an insulating over a metallic configuration.
We note that the crucial role played by stacking faults in this CDW system is not an exception. Recent studies of the kagome metal CsV$_3$Sb$_5$ \cite{jin2024}, as well as of NbS$_2$ and MoS$_2$ \cite{watson2024}, indicate that stacking faults are widespread in van der Waals materials and can affect numerous measurements in a non-trivial way.
Finally, our observation of QWSs with near flat dispersion suggests that direct control of the sample thickness, e.g. by mechanical exfoliation, is a promising way to engineer correlated states by controlling the conduction band width.

\begin{acknowledgements}
The experimental work was supported by the Swiss National Science Foundation (SNSF) grants 184998, 215548. We acknowledge MAX IV Laboratory for time on Beamline Bloch under Proposals 20220192, 20221276, 20230714. Research conducted at MAX IV, a Swedish national user facility, is supported by the Swedish Research council under contract 2018-07152, the Swedish Governmental Agency for Innovation Systems under contract 2018-04969, and Formas under contract 2019-02496. F.P. thanks Tommaso Gorni for fruitful discussions.
The Flatiron Institute is a division of the Simons Foundation. 
\end{acknowledgements}

%


\clearpage

\renewcommand{\thefigure}{S\arabic{figure}} 
\renewcommand{\theequation}{S\arabic{equation}} 
\renewcommand{\thetable}{S\arabic{table}}
\renewcommand*{\thesubsection}{\Alph{subsection}}

\setcounter{figure}{0} 
\setcounter{equation}{0} 
\setcounter{table}{0}
\setcounter{section}{0}

\title{Supplemental Material: Nature of metallic and insulating domains in the CDW system 1T-TaSe$_2$}

\date{\today}
\maketitle

\onecolumngrid

\section{Sample growth and characterization}

\begin{figure}[b]
\centering
\includegraphics[width=0.85\columnwidth]{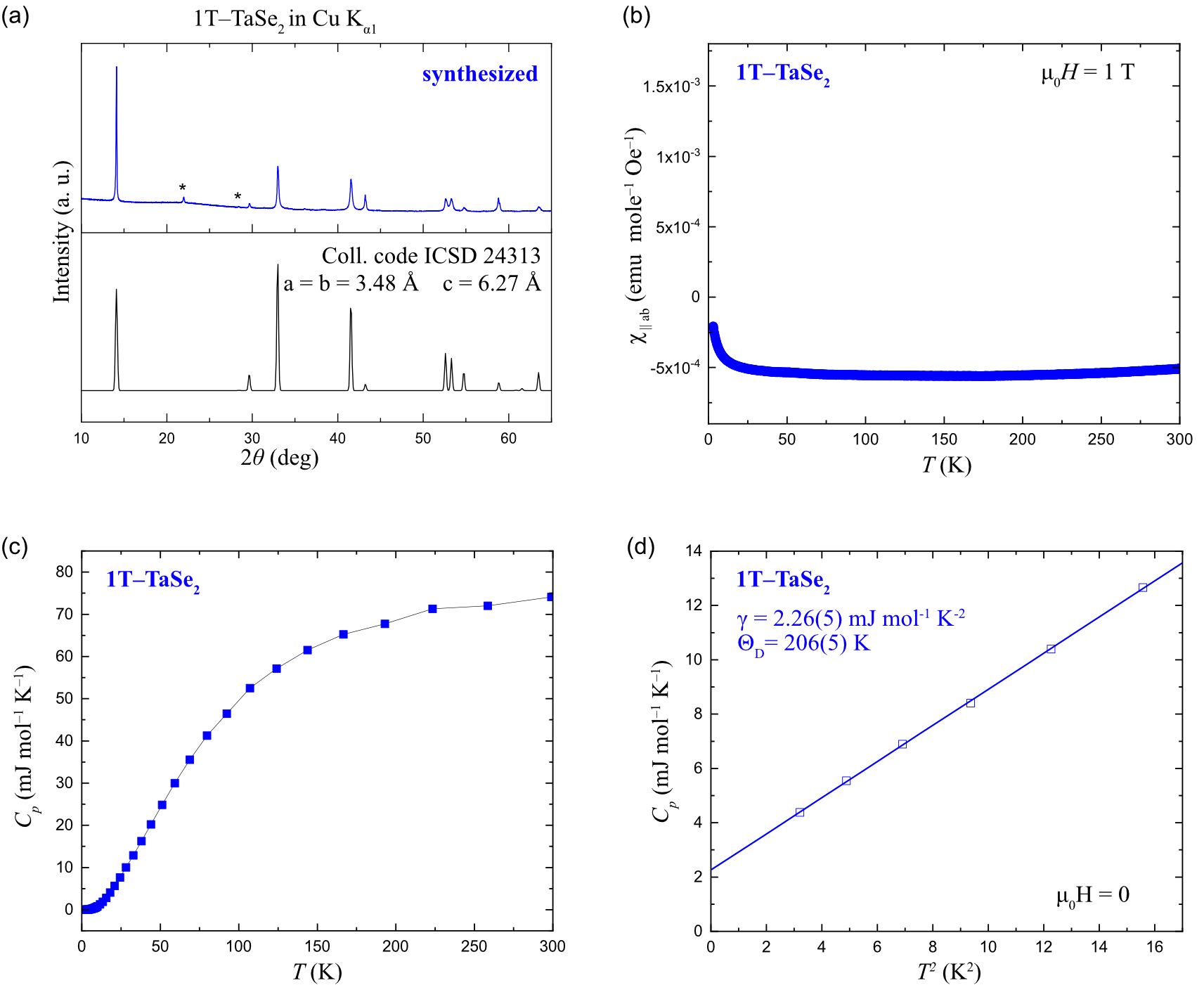}
\caption{Characterization of the synthesized 1T-TaSe$_2$ crystals. (a) PXRD pattern in blue. The asterisk mark the signal from SiO$_2$ used for grinding the crystals. (b) Susceptibility of a single crystal measured parallel to the ab-plane in an external field of 1 T. (c) Temperature dependence of the zero-field specific heat of a single crystals of 1T-TaSe$_2$. The discontinuity around 250 K originates from the Apiezon N grease. (d) Heat capacity data plotted as $C_p/T$ versus $T^2$ at zero field with the fit shown as blue line.}
\label{fig:sm_fig_Witteveen}
\end{figure}
Single crystals of 1T-TaSe$_2$ were grown by chemical vapor transport (CVT) using iodine as transport agent.
All elements were used as received. Stoichiometric amounts of tantalum (powder, Alfa Aesar, 99.99\%) and selenium (shots, Alfa Aesar, 99.999\%) were sealed together with 30 mg of iodine in an ampoule (l = 12 cm, $\varnothing_i$ = 9 mm) under vacuum. 
The ampoule was heated in a two zone furnace at a rate of 180 °C/h, in which the source zone and the growth zone were fixed at 1080 °C and 900 °C, kept at this temperature for 6 days and subsequently quenched in water to retain the 1T phase, resulting in large single crystals with a golden luster.
The crystals were confirmed to be phase pure by powder X-ray diffraction (PXRD) using a Panalytical Empyrean diffractometer with Cu K$\alpha$ radiation in transmission geometry equipped with a focusing mirror and a solid-state PIXcel detector. Thereby, a crystal was ground with SiO$_2$ powder and then loaded into a 0.5 mm capillary tube. The PXRD pattern is shown in Fig.~\ref{fig:sm_fig_Witteveen} (a) in the top panel, with a reference pattern on the bottom.

Fig.~\ref{fig:sm_fig_Witteveen}(b) shows the characteristic diamagnetic susceptibility of a 1T-TaSe$_2$ single crystal. The measurement was taken on a field cooled single crystal of 6.0 mg weight in a Physical Property Measurements in a cryogen-free system (PPMS DynaCool) equipped with a 9 T magnet from 1.8 K to 300 K and an external field $\mu\textit{H}_0$ of 1 T applied parallel to the ab-plane.
Specific-heat measurements were carried out in a Physical Property Measurements in a cryogen-free system (PPMS DynaCool) equipped with a 9 T magnet in zero field from 300 K to 1.8 K. A crystal was cut to a suitable size and mounted with Apiezon N grease onto the heat capacity puck to ensure good thermal contact. The specific heat as a function of temperature is shown in Fig.~\ref{fig:sm_fig_Witteveen} (c). In Fig.~\ref{fig:sm_fig_Witteveen} (d), is the fit of the specific heat using $C_p/T = \gamma + \beta T^2$ from 1.8 to 4 K, resulting in $\gamma$ = 2.26  mJ~mol$^{–1}$~K$^{–2}$ and the lattice term $\beta$ = 0.6654~mJ~mol$^{–1}$. The latter was used to calculate the Debye temperature according to  $\Theta_D = [12\pi^4NR/(5\beta)]^{1/3}$, where $N$ is the number of atoms per formula unit and $R$ is the gas constant, resulting in $\Theta_D$ = 206 K. These values fit very well with the existing literature \cite{disalvo1974SM}.

\section{Experimental Methods}
Spatially resolved ARPES measurements were performed at the Bloch beamline of the MaxIV synchrotron facility \cite{Polley2024aa}, using linearly polarized light between 30 and 100~eV and a probing spot size of 10 x 15~$\mu$m$^2$. The samples were cleaved in UHV with the aid of a ceramic top post and measured at the base temperature of the setup (18~K) with a ScientaOmicron DA30 analyzer. The total energy resolution was 10~meV.

Laser ARPES measurements were performed in a custom-built micro ARPES system in Geneva \cite{cucchi2018}. This setup is based on a continuous wave laser with a photon energy of 6.01~eV and an MB Scientific A-1 analyzer. The total energy resolution was set to 1.5 meV and the the spatial resolution was 4 $\mu$m. Complementary spatially integrated measurements were performed with a MB Scientific He Lamp with a spot size of $\sim$ 0.5~mm$^2$. Samples were cleaved and measured at the base temperature of 6 K.

\section{DFT calculations}

\begin{figure}[b]
\centering
\includegraphics[width=1.0\columnwidth]{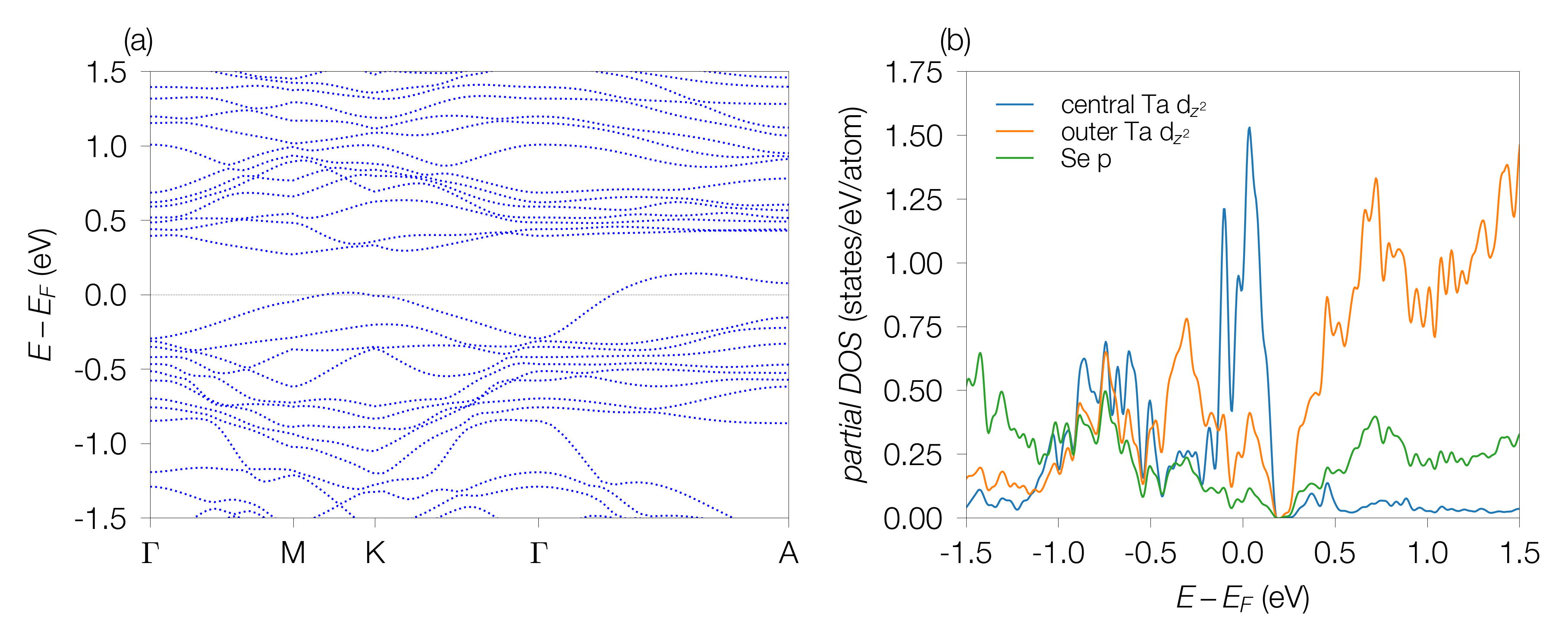}
\caption{(a) DFT band structure calculation. (b) Partial DOS for the $d_{z^2}$ orbital of the central Ta atom (blue), the $d_{z^2}$ orbitals of the other Ta atoms (orange), and the $p$-orbitals of the Se atoms (green).}
\label{fig:fig_DFT}
\end{figure}

Electronic structure calculations were carried out using the Quantum ESPRESSO package \cite{QEref}. The exchange-correlation functional was approximated by the GGA parametrisation \cite{GGAref}. The interactions between ions and valence electrons were described using projector-augmented-wave pseudopotentials, including the $s$ and $p$ semicore states of the Ta atoms explicitly. We performed calculations for monolayer, bulk and four-layer supercells of 1T-TaSe$_2$ in the CDW phase. The building block of our electronic structure calculations was the $\sqrt{13}\times\sqrt{13}$ super-cell containing 13 Ta and 26 Se sites in distorted positions resulting in the Star-of-David (SOD) buckling. DFT calculations were benchmarked so as to exactly reproduce the monolayer bandstructure of Ref.~[\onlinecite{KangZhang2020SM}]. The bulk unit cell was constructed by aligning the Ta site at the center of the SOD (position A in Fig.~1(a) of the main text) with the undistorted position of the outermost Ta site of adjacent unit cell (position C in Fig.~1(a) of the main text) in the vertical direction. We refer to this vertical arrangement as the AC-stacking. The overall positions were further adjusted via full cell relaxation with a force threshold of 0.02~eV/$\text{\AA}$ starting from an initial inter-layer distance $c=6.267$~$\text{\AA}$. For the bulk setup we used a wave-function cutoff of 60 Ry, charge-density cutoff of 500 Ry and a 6$\times$6$\times$6 $\mathbf{k}$-grid with Marzari-Vanderbilt smearing of 0.01 Ry. 

The resulting DFT band structure is shown in Fig.~\ref{fig:fig_DFT}. The density of states (DOS) close to the Fermi level is dominated by the $d_{z^2}$ orbital of the central Ta atom. Significant hybridization between the Ta $d$ and Se $p$ orbitals is evident across the entire energy range, with increasing Se $p$-orbital character at higher binding energies.

We explicitly investigated the effects of a stacking fault 
by constructing two four-layers AC-stacked supercells for a total of 156 atoms each. In one of the two structures we included a defected position in one of the external layers, placed so as to have vertically aligned SOD (AA-stacking). Inversion symmetry was further broken by allowing for structural relaxation only on two layers of the system with a force threshold of 0.05 eV/$\text{\AA}$. The vacuum was included via a spacing of 18 $\text{\AA}$ between super-cells in the $z$ direction. Calculations in layered setup were performed with wave-function cutoff of 40 Ry, charge-density cutoff of 400 Ry and a 6$\times$6$\times$1 $\mathbf{k}$-grid with Marzari-Vanderbilt smearing of 0.02 Ry. We checked that the cutoff reductions are not affecting appreciably the final band structures.

\section{DMFT calculations}
DFT band structures were downfolded to effective low energy tight-binding models employing maximally-localized-Wannier-functions optimized with the Wannier90 code \cite{W90ref}. We included only the band close to the Fermi level and localized on the Ta ion at the centre of each SOD contained in the DFT unit  cell. We used a 14$\times$14$\times$1 $\mathbf{k}$-grids for the monolayer, 10$\times$10$\times$10 for the bulk and 10$\times$10$\times$1 for the slabs. A single Wannier function with $d_{z^2}$ character was extracted for the monolayer and bulk setups, four for the slabs. The resulting Hamiltonians were used as input for subsequent DMFT calculations. We estimate the magnitude of the Hubbard local interaction with cRPA performed on the monolayer with the RESPACK code \cite{RespackRef}. Using the default parameters on top of the Wannier tight-binding model we obtained a value of $U=$0.198~eV. Slight modulations of $U$ are to be expected in the different structures since screening channels are modified by vacuum and layer-resolution. However the computational cost of cRPA calculations grows severely with the number of sites and assessing the optimal interaction is beyond the scope of this work. We therefore assumed the same interaction strength for both the bulk and slab structures.
\begin{figure}[h]
\centering
\includegraphics[width=1.0\columnwidth]{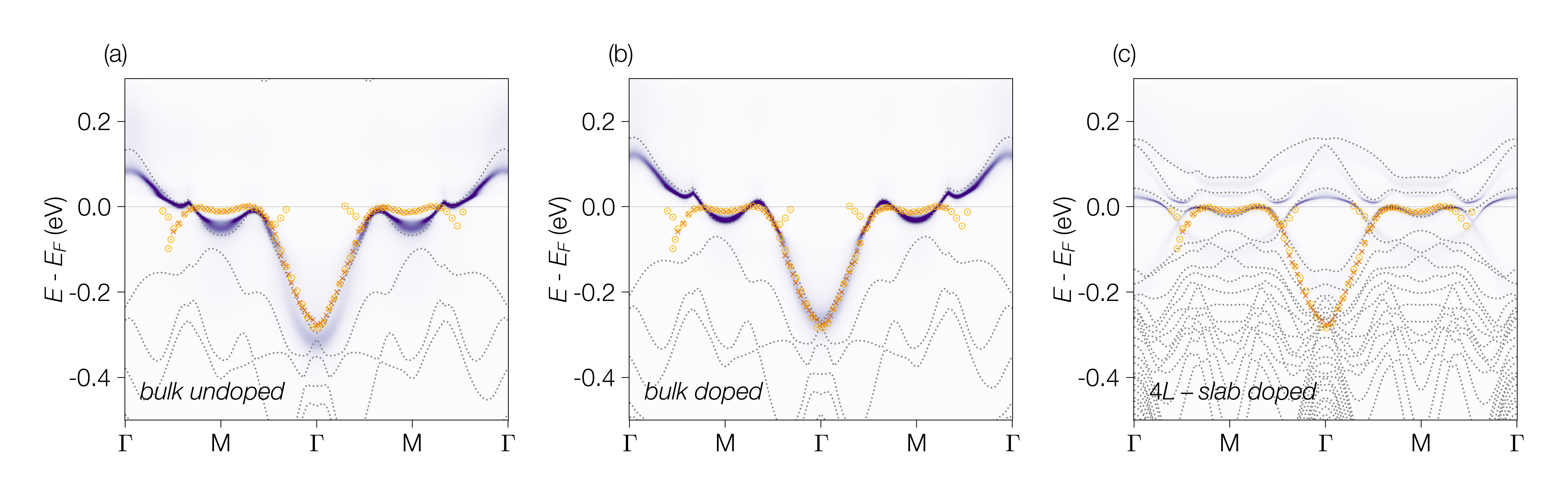}
\caption{DMFT spectral function of (a) bulk 1T-TaSe$_2$, (b) doped bulk with 0.8 electrons per SOD and (c) four-layers AC-stacked supercell with the same doping. The orange markers depict the quasiparticle dispersion extracted from the ARPES spectra in Fig.~2(a) of the main text and in Fig.~\ref{fig:fig_6}(b) (we use round and cross markers for $s$ and  $p$ polarization, respectively).}
\label{fig:fig_bulk_slabs}
\end{figure}

All the DMFT calculations were performed at T$\sim$15~K ($\beta$=800~eV$^{-1}$). In monolayer and bulk calculations the density has been kept fixed at one electron per site, each site is representative of the effective single-band model associated to a SOD. The impurity problem was solved with ct-QMC and the self-energy computed with improved estimators. For the monolayer and bulk setups the lattice Green's function has been corrected with a single local self-energy.
Remarkably the same interaction provided a Mott insulating monolayer and a Fermi liquid in the bulk with a moderate quasiparticle renormalization factor of $Z\sim$0.66. This result is in agreement with the existing knowledge of a thickness-dependent metal-insulator transition for this compound \cite{tian2023SM} and with the gap amplitude measured in monolayers \cite{STM_ml}. We studied the slab structures with a real-space extension of DMFT in which one impurity per layer has been solved yielding four independent self-energies coupled by the inversion of the lattice Green's function.

The optimal match with the experimental Fermi surface of Fig.~3 of the main text was obtained with a slight off-stoichiometric filling of 0.8 electrons per SOD. In Fig.~\ref{fig:fig_bulk_slabs} the DMFT spectra of bulk, undoped and doped, as well as of the doped four-layers slab are compared to the dispersion extracted from the ARPES data (orange markers). As  discussed in the main text, the experimental data show higher low energy renormalization compared to the bulk DMFT calculations. In the four-layers slab calculation, the electron pockets at the M-points are shallower and the agreement with the measurements improves.

\subsection{Semi-infinite bulk embedding}
The limitation to four-layers slabs was dictated by the significant computational cost of including a CDW reconstructed in-plane supercell. To break the inversion symmetry of our slabs beyond the selective relaxation constraints, we simulate the presence of a bulk on one side of the systems. We achieve that by mimicking the presence of an arbitrary large number of layers at a negligible computational cost with an approach equivalent to the one presented in Ref.~\onlinecite{Petocchi2022SM}. The method relies on the mathematical properties that tridiagonal matrices can be inverted by means of continued fractions. In practice we exploit this property when we construct the layer-resolved lattice Green's function $G$. We account for the presence of additional diagonal entries $\varepsilon$ in $G^{-1}$, without modifying the default rank of 4, via the embedding potential $G_E$:
\begin{align*}
G^{-1}&=\left(i\omega_{n}+\mu\right)\mathbf{I}_{N}-\left(\begin{array}{cccc|cc} \mathcal{H}_{1,\mathbf{k}}+\Sigma_{1} & \ldots & \ldots & \ldots & 0 & 0\\ \ldots & \mathcal{H}_{2,\mathbf{k}}+\Sigma_{2} & \ldots & \ldots & 0 & 0\\ \ldots & \ldots & \mathcal{H}_{3,\mathbf{k}}+\Sigma_{3} & t_{\mathbf{k}} & 0 & 0\\ \ldots & \ldots & t_{\mathbf{k}}^{*} & \mathcal{H}_{4,\mathbf{k}}+\Sigma_{4} & t_{\mathbf{k}} & 0\\ \hline 0 & 0 & 0 & t_{\mathbf{k}}^{*} & \mathcal{H}_{4,\mathbf{k}}+\Sigma_{4} & t_{\mathbf{k}}\\ 0 & 0 & 0 & 0 & t_{\mathbf{k}}^{*} & \ldots \end{array}\right)\\&=\left(i\omega_{n}+\mu\right)\mathbf{I}_{4}-\left(\begin{array}{cccc} \mathcal{H}_{1,\mathbf{k}}+\Sigma_{1} & \ldots & \ldots & \ldots\\ \ldots & \mathcal{H}_{2,\mathbf{k}}+\Sigma_{2} & \ldots & \ldots\\ \ldots & \ldots & \mathcal{H}_{3,\mathbf{k}}+\Sigma_{3} & \ldots\\ \ldots & \ldots & \ldots & \mathcal{H}_{4,\mathbf{k}}+\Sigma_{4}+G_{E} \end{array}\right)
\end{align*}
where $\mathbf{I}_{N}$ is the identity matrix of rank $N$. $G_E$ can be computed recursively if multiple entries in $G^{-1}$ have to be included simultaneously. In our implementation, at every DMFT self-consistency loop, we corrected the element of the inverse Green's function corresponding to the layer at the interface with the bulk with:
$$G_{E}\left(\mathbf{k},i\omega_{n}\right)=\frac{t_{\mathbf{k}}t_{\mathbf{k}}^{*}}{i\omega_{n}+\mu-\mathcal{H}_{4,\mathbf{k}}-\Sigma_{4}-\frac{t_{\mathbf{k}}t_{\mathbf{k}}^{*}}{i\omega_{n}+\mu-\mathcal{H}_{4,\mathbf{k}}-\Sigma_{4}-\frac{t_{\mathbf{k}}t_{\mathbf{k}}^{*}}{i\omega_{n}+\mu-\mathcal{H}_{4,\mathbf{k}}-\Sigma_{4}-\ldots}}}$$
where $t_{\mathbf{k}}$ is the hopping between the third and fourth layer, $\mathcal{H}_{4,\mathbf{k}}$ is the dispersion of the fourth layer and $\Sigma_{4}$ its local self-energy. This is a numerically cheap way to increase the extension of the slabs beyond their original size. The procedure comes with several approximations, namely the fact that the inter- and intra-layer dispersions are considered homogeneous after the third plane and that every hopping beyond the nearest neighbor are ignored, but its accurate enough to capture the physics of the quantum well states as shown in the main text.
\begin{figure}[b]
\centering
\includegraphics[width=1.0\columnwidth]{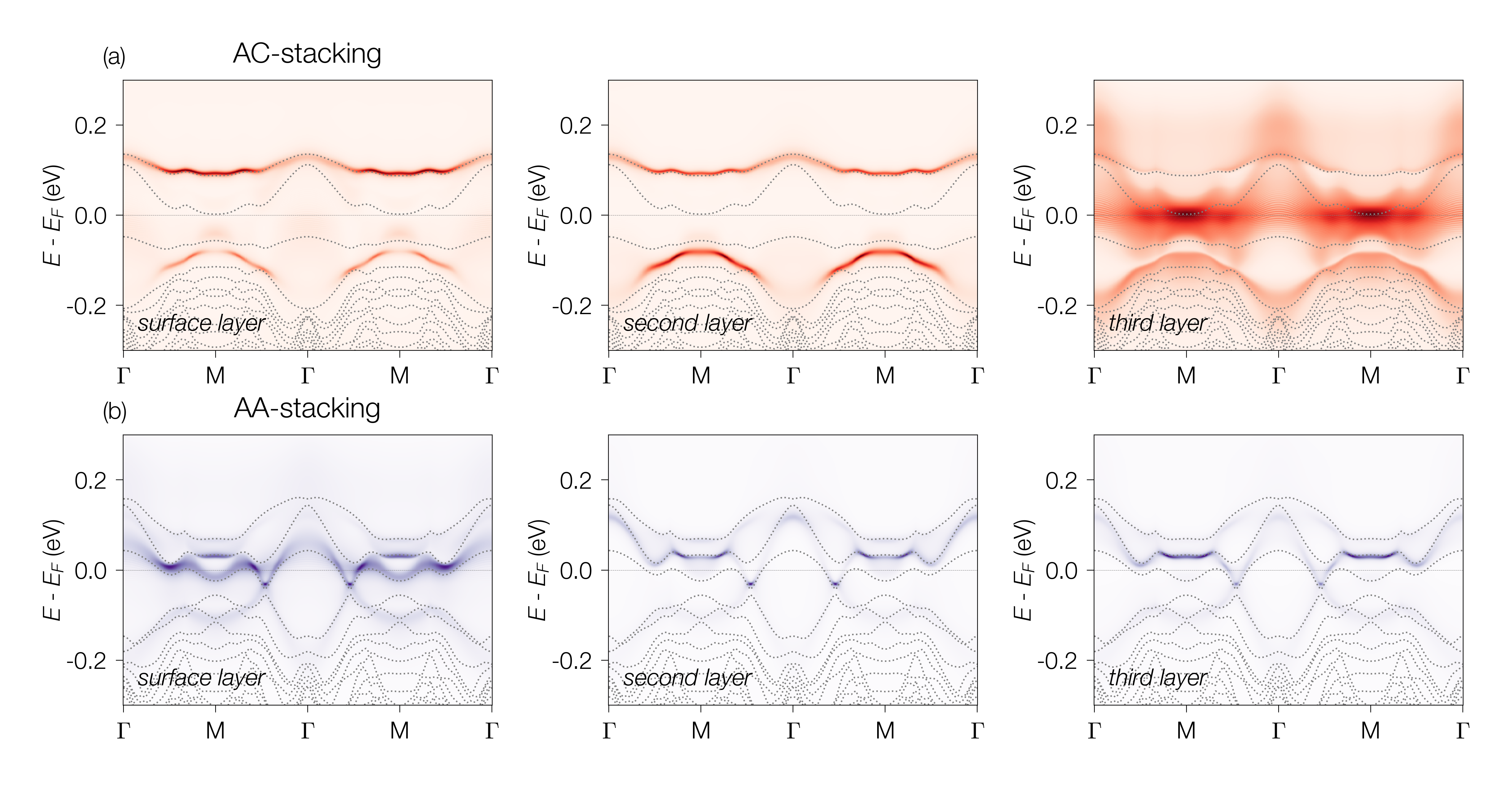}
\caption{DMFT layer-resolved spectral functions of a semi-infinite slab using 100 layers and different arrangements of the surface layer: (a) AA stacking fault between surface and second layer; (b) normal AC-stacking for all layers.}
\label{fig:fig_slabs}
\end{figure} 

The extension of the insulating four-layers slab in this manner demonstrates that the insulating behavior of the surface remains intact with this construction. As illustrated in Fig.~\ref{fig:fig_slabs}(a), the spectra of the surface and second layer exhibit a gap comparable to the one found in the four-layers structure presented in the main text. From the third layer onwards, the structure becomes metallic, matching the expected behavior of the bulk. Without stacking fault, all layers are metallic (Fig.~\ref{fig:fig_slabs}(b)).

\newpage
\section{Comparison of helium lamp, synchrotron and laser ARPES data}

\begin{figure}[h]
\centering
\includegraphics[width=1.0\columnwidth]{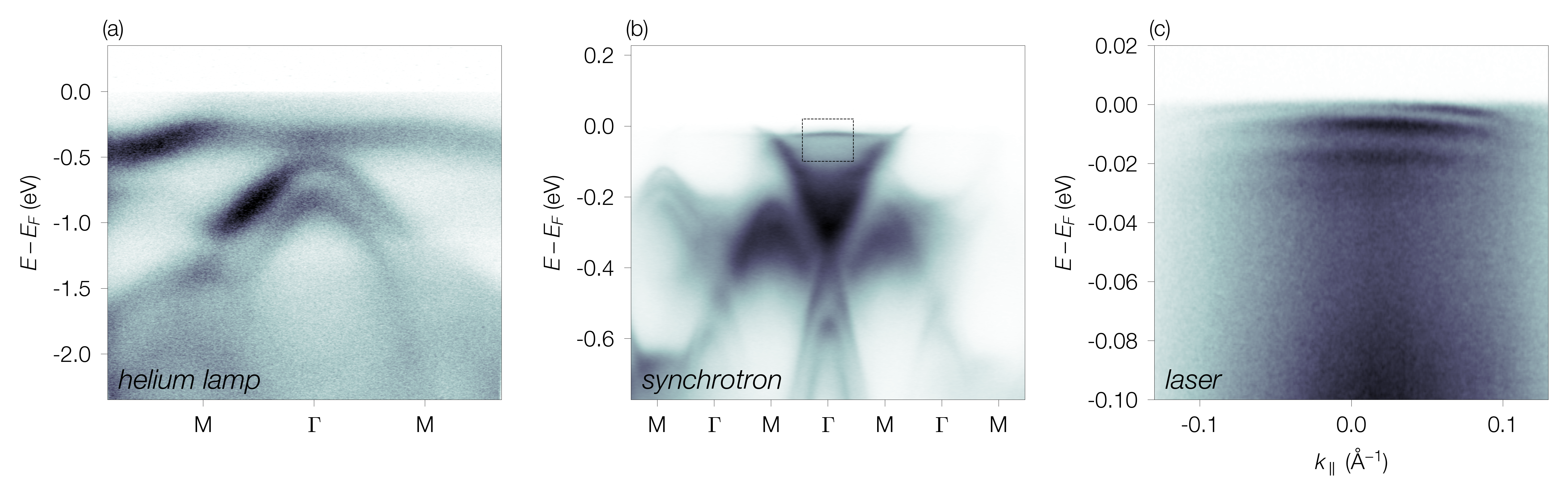}
\caption{(a) He-lamp data (He-I$\alpha$ radiation, 0.5~mm$^2$ spot). (b) Synchrotron data measured at the Bloch beamline (50~eV photon energy, 10 x 15 $\mu$m$^2$ spot). (c) 6~eV laser data obtained with a 4 $\mu$m$^2$ spot. All measurements use $p$-polarized light and are aligned along the $\Gamma$-M direction of the CDW unit cell. The dashed squared box in panel (b) indicates the energy/momentum window probed with the laser in panel (c).}
\label{fig:fig_1}
\end{figure}

Fig.~\ref{fig:fig_1} shows a comparison of ARPES data from bulk 1T-TaSe$_2$ obtained with different light sources and spot sizes.
The He-lamp data, measured with a spot size of approximately 0.5~mm$^2$, result from a superposition of insulating and metallic domains (panel (a)).
Towards the Fermi level, the dominating feature is a weakly dispersing broad peak centered at -400~meV and spanning the entire measured momentum space. This broad feature was interpreted in previous ARPES studies as the lower Hubbard band of a Mott-insulator \cite{perfetti2003SM,chen2020SM,nakata2021SM}. In contrast, our spatially-resolved measurements identify this feature as a mere convolution of different dispersing bands. No coherent band crossing the Fermi level can be observed in the He-lamp data. However, the presence of small metallic regions manifest itself in a ubiquitous and well-defined step at the Fermi level.
Measurements from the same samples using the micro-spot of the Bloch beamline (panel (b)) or our focused laser (panel (c)) show regions with dispersing bands up to the Fermi level, in agreement with the metallic behaviour of bulk 1T-TaSe$_2$. 

\section{Synchrotron ARPES with $p$ and $s$-polarized light}

\begin{figure}[h]
\centering
\includegraphics[width=1.0\columnwidth]{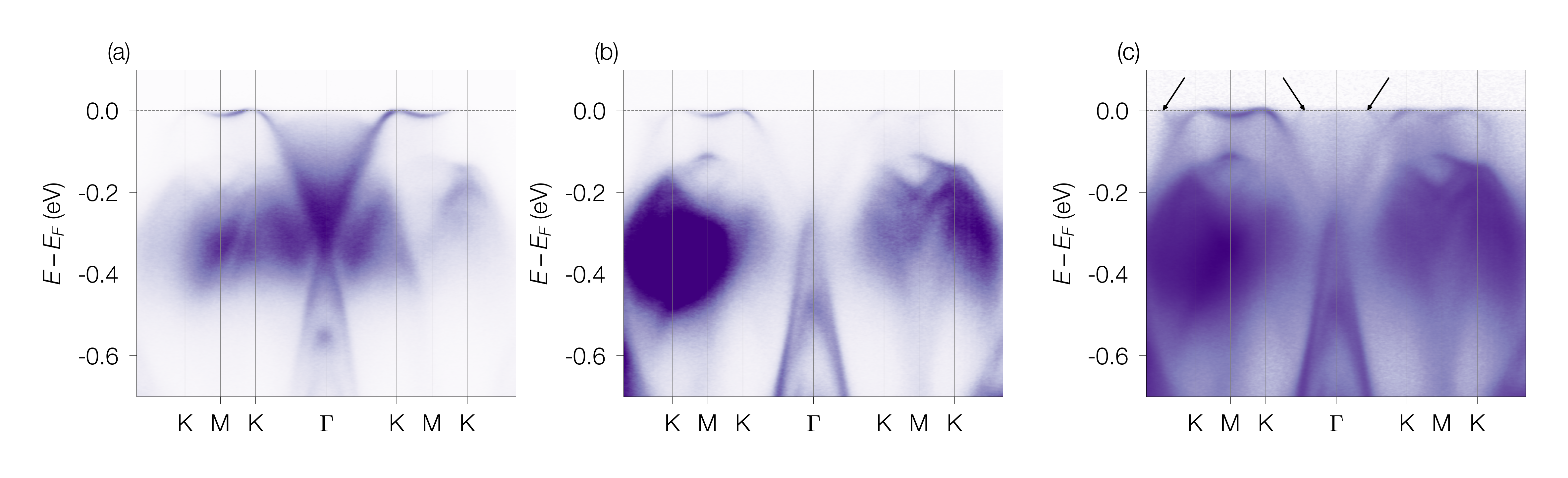}
\caption{(a,b) ARPES dispersion from a metallic domain measured with 50~eV  and $p$ and $s$-polarized, respectively. (c) The same data of panel (b) displayed with a logarithmic image contrast. The arrows point at weak back-folded bands that become visible with this procedure.} 
\label{fig:fig_6}
\end{figure}

Fig.~\ref{fig:fig_6}(a,b) shows the same dispersion plot measured  with two different polarizations of the light on a metallic domain. Using $s$- instead of $p$-polarized light, the relative intensity of the bands in the region near the Fermi level changes drastically, which is likely due to the different orbital character of these states. In particular, the intensity of the ``Dirac cone" centered at $\Gamma$ is significantly reduced with $s$-polarized light, which is consistent with the even symmetry of the underlying $d_{z^2}$ orbital.  

In Fig.~\ref{fig:fig_6}(c) we display the same data of panel (b) with a logarithmic contrast. This procedure enhances the contribution of weak features and allows us to identify additional back-folded bands, which are not present in calculations of the bulk band structure (see Fig.~2(e) in the main text). However, these bands are well reproduced in calculations where the spectral weight is integrated along $k_z$ over the whole BZ (see Section VIII).

\section{CDW periodicity in ARPES constant energy maps}
\begin{figure}[h]
\centering
\includegraphics[width=0.8\columnwidth]{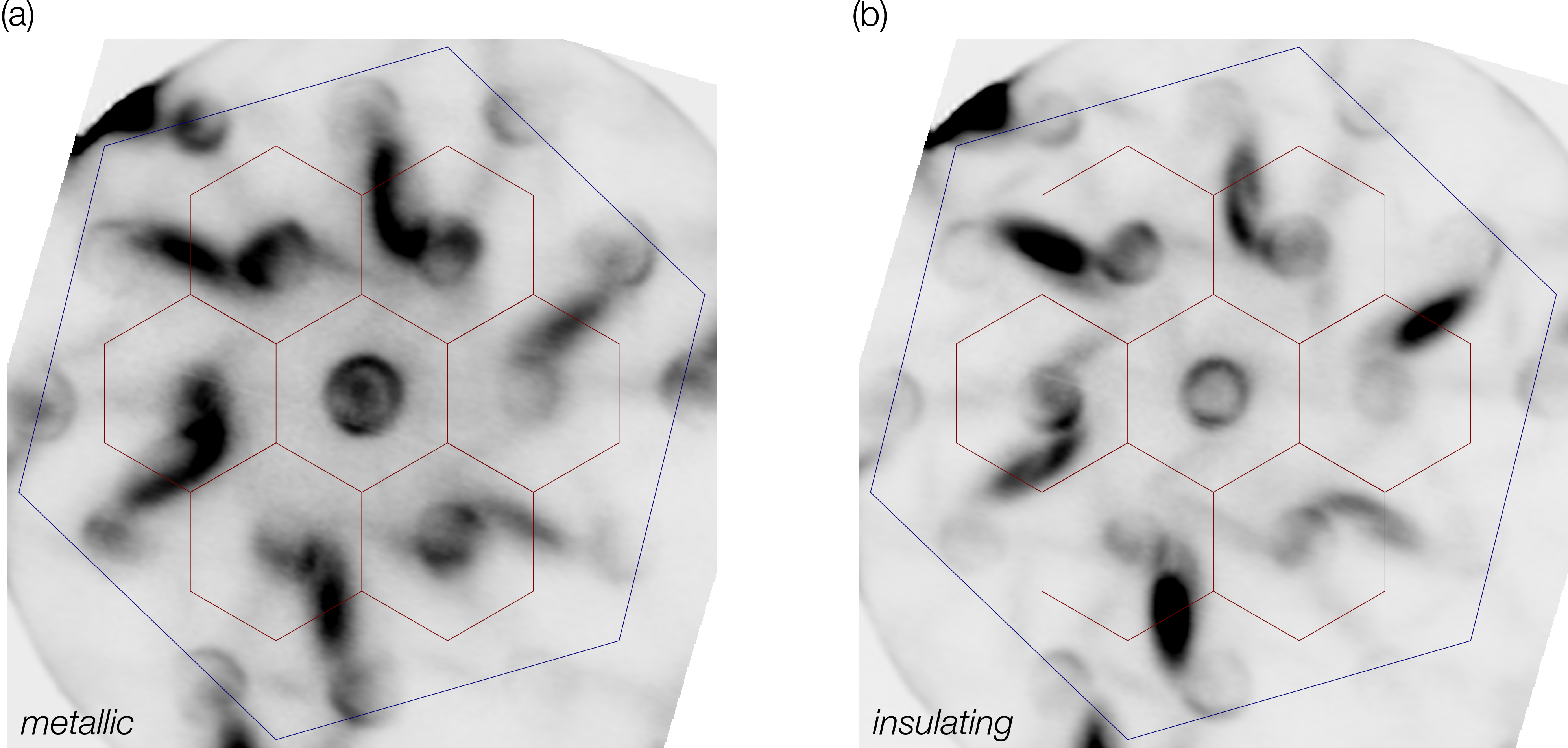}
\caption{Constant energy 
maps measured on (a) a metallic domain 0.57~eV below the Fermi level and (b) an insulating region at 0.59~eV. The BZs of the CDW superstructure and of the undistorted lattice are marked in red and blue, respectively. Both measurements were performed with 84~eV photon energy and s-polarized light.}
\label{fig:figS2}
\end{figure}

Fig.~\ref{fig:figS2} shows constant energy maps measured on metallic and insulating domains at energies with a strong contribution from the Se-$p$ orbitals. 
In both domains we find that the bands deriving from these states follow the periodicity of the CDW reconstruction and form small circular contours at the center of each CDW BZs. This is clear evidence that insulating and metallic domains have the same in-plane superstructure. 

On the edge of the second CDW BZ, intense oval shapes are observed. The intensity of these features lack the periodicity of the CDW but appear to have the same symmetry of the undistorted lattice. Noticeably, the oval structures at the M-point, in conjunction with the circular features at $\Gamma$, give rise to a windmill pattern characteristic of a chiral structure. 

\section{3D Electronic Structure of 1T-T\lowercase{a}S\lowercase{e}$_2$}\label{sec:3D}
\begin{figure}[h]
\centering
\includegraphics[width=1.0\columnwidth]{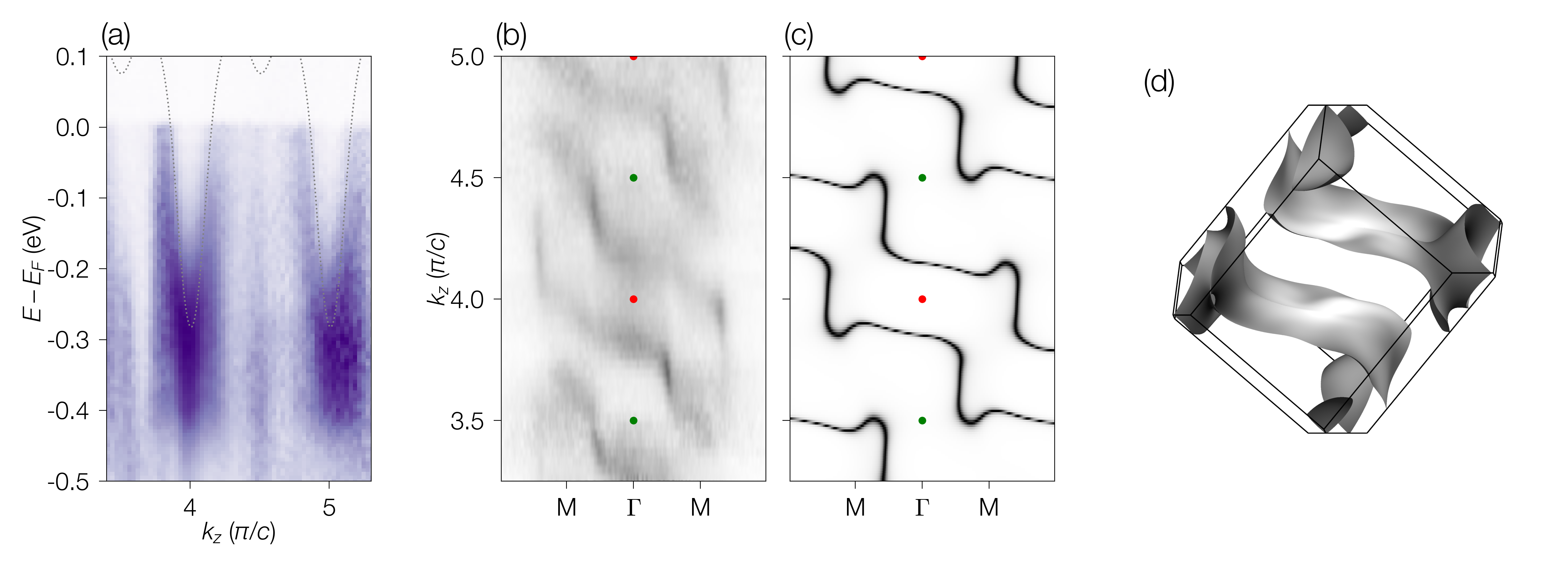}
\caption{(a) Out-of-plane dispersion measured at zero in-plane momentum, compared to a DFT band structure calculation for the bulk unit cell along the same direction (grey dotted line).  (b) Fermi surface along the out-of-plane momentum direction and the CDW high symmetry direction M. (c) Bulk DMFT calculation of the same momentum space. Red and green circles mark the positions of the $\Gamma$ and A points, respectively.  (d) DFT Fermi surface in the 3D triclinic Brillouin zone.}
\label{fig:figS3}
\end{figure}

Fig.~\ref{fig:figS3}(a) shows the out-of-plane dispersion of a metallic region measured at zero in-plane momentum with photon energies between 30 and 100 eV. Using the free-electron final-state approximation and assuming an inner potential of 15 eV, we find a dispersing band with minima at $\sim$ -0.3 eV at the center of each BZ. This is agreement with DFT calculations predicting considerable dispersion along this direction (grey dotted line).

The out-of-plane Fermi surface measured along the $\Gamma$M direction of the CDW is well-reproduced by Fermi surface calculations  of the bulk structure (Fig.~\ref{fig:figS3}(b-d)). This provides additional evidence that the metallic band structure reported in this study is characteristic of bulk 1T-TaSe$_2$ with AC stacked CDW layers.
The ARPES Fermi surface displays multiple sharp vertical contours with seemingly minimal dispersion in $k_z$ covering approximately half of a BZ. These contours are connected by diagonal stripes parallel to each other, which appear broad due to the poor momentum resolution along $k_z$. Upon examination of the 3D Fermi surface calculated with DFT (Fig. \ref{fig:figS3}(d)), it becomes evident that these features originate from the tilted staircase-like Fermi surface.

We note that the Fermi surface has a peculiar symmetry. In the distorted crystal structure, the only symmetry operation, apart from translation, is inversion symmetry. This can be directly observed in Fig.~\ref{fig:figS3}(b) where each $\Gamma$ point is an inversion center. The ARPES data presented in the main text were measured with 50~eV photons, which corresponds to a cut at $k_z \approx 4 \frac{\pi}{c}$, near the center of the BZ. 
In that particular case the three-dimensional inversion symmetry is translated to a two-fold rotational symmetry in the $k_x$-$k_y$ plane. Ultimately, the surface of the crystal breaks the inversion symmetry. The resulting surface band structure is chiral, however the bulk crystal is not regarded as chiral since inversion symmetry is not a proper operation of the bulk unit cell \cite{fecher2022, Louat2024aa}.

\begin{figure}[h]
\centering
\includegraphics[width=1.0\columnwidth]{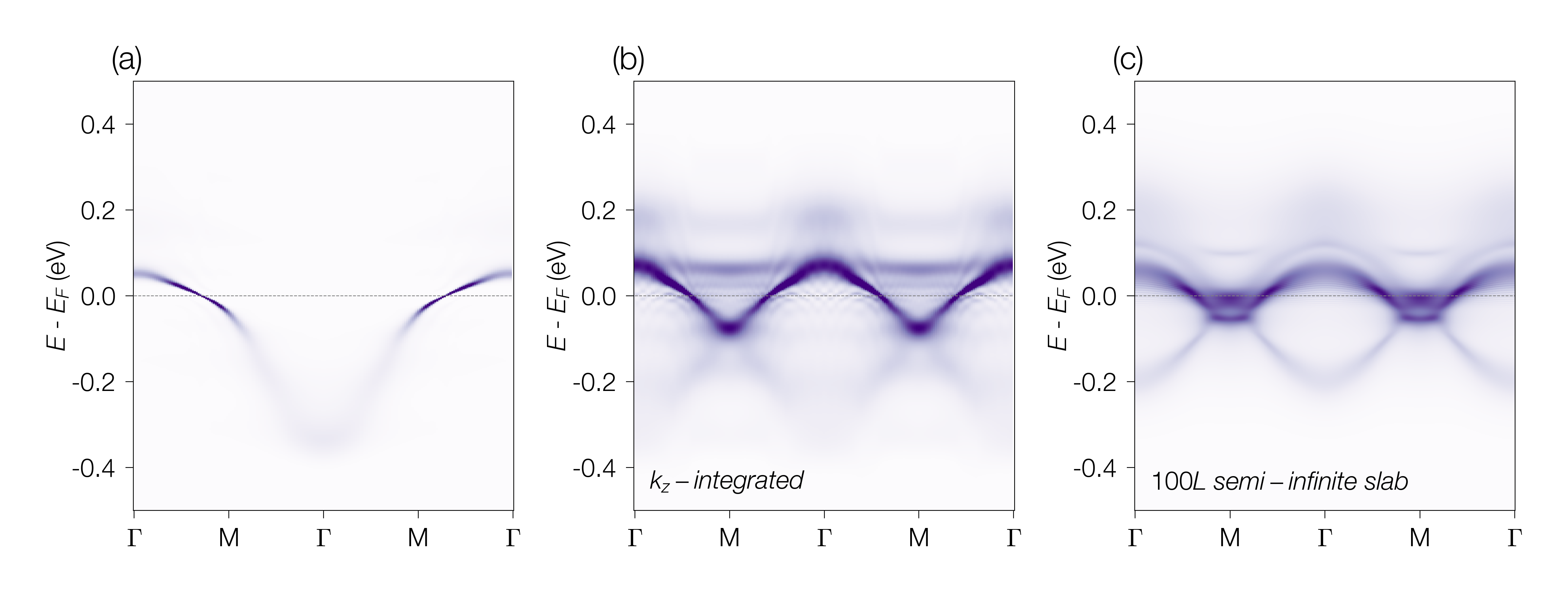}
\caption{DMFT spectra function of (a) the bulk spectra at $k_z = 0$ and (b) integrated over the whole BZ in the $k_z$-direction. (c) DMFT spectra of the surface layer in a semi-infinite slab using 100 layer in the bulk stacking configuration.}
\label{fig:fig_kz_integration}
\end{figure}

Finally, we discuss the origin of the back-folded bands described in Section VI. These features are a natural consequence of the poor $k_z$ resolution of ARPES, which result in an integration of the band structure along the same direction. 
In Fig.~\ref{fig:fig_kz_integration}, we explicitly demonstrate this effect in DMFT spectral function calculations.
Panel (a) shows the bulk band at $k_z$ = 0 and panel (b) the same cut integrated along $k_z$. In general for a band dispersing in 3D one would expect the integration to result in a complete broadening of the spectral features. However, in 1T-TaSe$_2$ the staircase-like dispersion of the metallic band results in well-defined bands even after full $k_z$ integration. In addition to the V-shaped band in panel (a), the $k_z$-integrated dispersion shows back-folded bands forming a hole-like pocket centered at $\Gamma$. In Fig.~\ref{fig:fig_kz_integration}(c), the DMFT spectra of a surface of an 100-layers slab is shown, featuring a similar hole-like band at $\Gamma$. Note, that in a slab calculations, $k_z$ is also effectively integrated. The presence of these hole-like bands in both bulk and slab calculations indicates these state should not be consider as surface states derived by the relaxation of the top layers but rather a result of the $k_z$-integration.
The same argument applies to the apparent hexagonal periodicity of the in-plane Fermi surface, which is well reproduced by the 100-layers slab calculation (see Fig.~3 of the main text).

\section{Quantum well states}

A simple approach for quantifying QWSs is provided by the 
model of a particle in a potential well \cite{margot2023,milun2002,kawakami1999}. This problem can be approximated in the so-called phase accumulation model \cite{milun2002}. In this approach, the particle is treated semi-classically as moving back and forth between the barriers, acquiring an additional phase on each reflection. In order to form a standing wave solution, the total accumulated phase of the particle after a full cycle must be a multiple of $2 \pi$. This leads to the following quantization condition:
\begin{align}
    2 k_z(E) L + \Phi = 2 \pi n,
\end{align}
where $k_z$ is the momentum along the well dimension, L is the length of the potential well and $\Phi$ is the total phase shift attained at each reflections.
In our case, the length of the potential of well, $L$, is determined by the number of AC stacked CDW layers, $N$, and the thickness per layer, $c = 6.27$~$\text{\AA}$.
For simplicity, we assumed that the phase shift $\Phi$ remains constant throughout the energy range of interest. A value of $\Phi=0.9$ was selected such that a hypothetical monolayer structure ($N = 1$) results in a single QWS at an energy position close to the Fermi level. 
This choice replicates the DFT calculations of monolayer 1T-TaSe$_2$. We note, however, that the precise value of $\Phi$ has a relatively minor impact on the result when the number of layers is large. 
\begin{figure}[h]
\centering
\includegraphics[width=1.0\columnwidth]{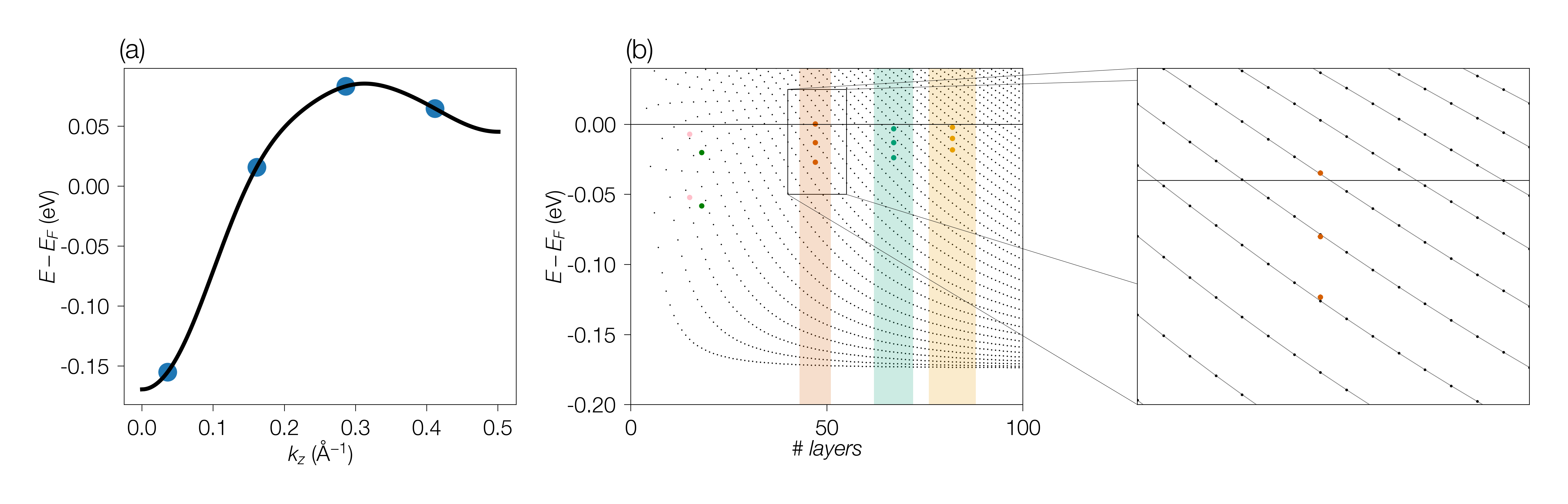}
\caption{(a) DFT band dispersion from $\Gamma$ to A scaled by the quasiparticle residue Z. The blue dots show an exemplary quantization for four layers. (b) Structure plot illustrating the calculation of the energy position of the QWS at the Gamma point as a function of the thickness of the potential well. The colored dots mark the energy positions and estimated thickness for the three QWSs presented in the main text. The uncertainty, depicted as colored bars, results from the experimental energy resolution.}
\label{fig:figS5}
\end{figure}

The discretized values of $k_z$ obtained in Eq.~(1) were used to determine the energy positions of the QWSs using the bulk DFT $E(k_z)$ dispersion.
However, at low energy the quasiparticle dispersion is renormalized due to electron-electron interactions. In order to account for this effect the DFT dispersion was renormalized by the quasiparticle residue Z = 0.66 obtained by DMFT. 
The result of this analysis is shown in Fig.~\ref{fig:figS5}. In panel(b) the energy positions of the QWSs are plotted against the number of layers. The energies of the three exemplary QWSs presented in the main text are marked with colored dots. The corresponding number of layers is obtained by comparing the energy difference between the first and second QWS with respect to the Fermi level. Upon renormalization of the DFT dispersion, the thickness of the quantum wells obtained with this toy model is comparable to the values in the main text, which we derived from comparisons to the DMFT slab calculations.

\begin{figure}[]
\centering
\includegraphics[width=1.0\columnwidth]{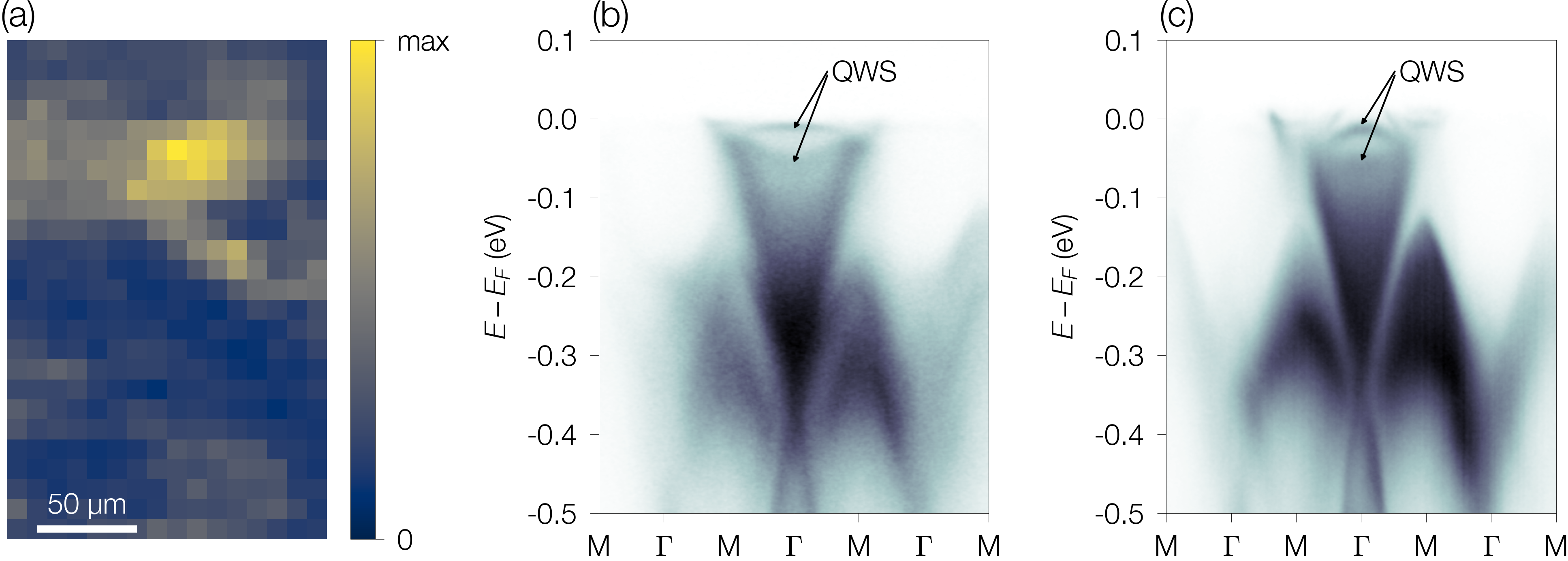}
\caption{(a) ARPES spatial scan on the surface of 1T-TaSe$_2$ measured at the Bloch beamline. The color scale corresponds to the photoemission intensity at the Fermi level. (b,c) ARPES dispersion plots of two  QWSs measured with 50~eV photon energy.}
\label{fig:figS4}
\end{figure}

The number of QWSs observed in our ARPES experiments varies significantly from sample to sample, consistent with the random arrangement and density of stacking faults in a single crystal. 
In addition to the laser data presented in the main text, we show in Fig.~\ref{fig:figS4} QWSs measured at the Bloch beamline. 
Similar to the laser data, a varying spectral intensity at the Fermi level is observed within the metallic domains (Fig.~\ref{fig:figS4}(a)). Figs.~\ref{fig:figS4}(b,c) show dispersion plot of two distinct QWSs. 
According to the toy model presented above, the energy positions of these QWSs correspond to a thickness of 15 (panel (b)) and 19 layers (panel (c)), respectively.
The QWSs are visible only at low energies and close to $\Gamma$. These measurements were performed on two different crystals and show that the observation of QWSs states is a general occurrence independent of sample and photon energy used. 


\begin{thebibliography}{56}%
\makeatletter
\providecommand \@ifxundefined [1]{%
 \@ifx{#1\undefined}
}%
\providecommand \@ifnum [1]{%
 \ifnum #1\expandafter \@firstoftwo
 \else \expandafter \@secondoftwo
 \fi
}%
\providecommand \@ifx [1]{%
 \ifx #1\expandafter \@firstoftwo
 \else \expandafter \@secondoftwo
 \fi
}%
\providecommand \natexlab [1]{#1}%
\providecommand \enquote  [1]{``#1''}%
\providecommand \bibnamefont  [1]{#1}%
\providecommand \bibfnamefont [1]{#1}%
\providecommand \citenamefont [1]{#1}%
\providecommand \href@noop [0]{\@secondoftwo}%
\providecommand \href [0]{\begingroup \@sanitize@url \@href}%
\providecommand \@href[1]{\@@startlink{#1}\@@href}%
\providecommand \@@href[1]{\endgroup#1\@@endlink}%
\providecommand \@sanitize@url [0]{\catcode `\\12\catcode `\$12\catcode `\&12\catcode `\#12\catcode `\^12\catcode `\_12\catcode `\%12\relax}%
\providecommand \@@startlink[1]{}%
\providecommand \@@endlink[0]{}%
\providecommand \url  [0]{\begingroup\@sanitize@url \@url }%
\providecommand \@url [1]{\endgroup\@href {#1}{\urlprefix }}%
\providecommand \urlprefix  [0]{URL }%
\providecommand \Eprint [0]{\href }%
\providecommand \doibase [0]{https://doi.org/}%
\providecommand \selectlanguage [0]{\@gobble}%
\providecommand \bibinfo  [0]{\@secondoftwo}%
\providecommand \bibfield  [0]{\@secondoftwo}%
\providecommand \translation [1]{[#1]}%
\providecommand \BibitemOpen [0]{}%
\providecommand \bibitemStop [0]{}%
\providecommand \bibitemNoStop [0]{.\EOS\space}%
\providecommand \EOS [0]{\spacefactor3000\relax}%
\providecommand \BibitemShut  [1]{\csname bibitem#1\endcsname}%
\let\auto@bib@innerbib\@empty
\bibitem [{\citenamefont {Fazekas}\ and\ \citenamefont {Tosatti}(1979)}]{Fazekas1979aa}%
  \BibitemOpen
  \bibfield  {author} {\bibinfo {author} {\bibfnamefont {P.}~\bibnamefont {Fazekas}}\ and\ \bibinfo {author} {\bibfnamefont {E.}~\bibnamefont {Tosatti}},\ }\bibfield  {title} {\bibinfo {title} {{Electrical, structural and magnetic properties of pure and doped 1T-TaS$_2$}},\ }\href {https://doi.org/10.1080/13642817908245359} {\bibfield  {journal} {\bibinfo  {journal} {Philosophical Magazine B}\ }\textbf {\bibinfo {volume} {39}},\ \bibinfo {pages} {229} (\bibinfo {year} {1979})}\BibitemShut {NoStop}%
\bibitem [{\citenamefont {Sahebsara}\ and\ \citenamefont {S\'en\'echal}(2008)}]{Sahebsara2008}%
  \BibitemOpen
  \bibfield  {author} {\bibinfo {author} {\bibfnamefont {P.}~\bibnamefont {Sahebsara}}\ and\ \bibinfo {author} {\bibfnamefont {D.}~\bibnamefont {S\'en\'echal}},\ }\bibfield  {title} {\bibinfo {title} {{Hubbard Model on the Triangular Lattice: Spiral Order and Spin Liquid}},\ }\href {https://doi.org/10.1103/PhysRevLett.100.136402} {\bibfield  {journal} {\bibinfo  {journal} {Physical Review Letters}\ }\textbf {\bibinfo {volume} {100}},\ \bibinfo {pages} {136402} (\bibinfo {year} {2008})}\BibitemShut {NoStop}%
\bibitem [{\citenamefont {Laubach}\ \emph {et~al.}(2015)\citenamefont {Laubach}, \citenamefont {Thomale}, \citenamefont {Platt}, \citenamefont {Hanke},\ and\ \citenamefont {Li}}]{Laubach2015}%
  \BibitemOpen
  \bibfield  {author} {\bibinfo {author} {\bibfnamefont {M.}~\bibnamefont {Laubach}}, \bibinfo {author} {\bibfnamefont {R.}~\bibnamefont {Thomale}}, \bibinfo {author} {\bibfnamefont {C.}~\bibnamefont {Platt}}, \bibinfo {author} {\bibfnamefont {W.}~\bibnamefont {Hanke}},\ and\ \bibinfo {author} {\bibfnamefont {G.}~\bibnamefont {Li}},\ }\bibfield  {title} {\bibinfo {title} {Phase diagram of the {{Hubbard}} model on the anisotropic triangular lattice},\ }\href {https://doi.org/10.1103/PhysRevB.91.245125} {\bibfield  {journal} {\bibinfo  {journal} {Physical Review B}\ }\textbf {\bibinfo {volume} {91}},\ \bibinfo {pages} {245125} (\bibinfo {year} {2015})}\BibitemShut {NoStop}%
\bibitem [{\citenamefont {Chen}\ \emph {et~al.}(2019)\citenamefont {Chen}, \citenamefont {Qu}, \citenamefont {Li}, \citenamefont {Chen}, \citenamefont {Gong}, \citenamefont {von Delft}, \citenamefont {Weichselbaum},\ and\ \citenamefont {Li}}]{ChenLei2019}%
  \BibitemOpen
  \bibfield  {author} {\bibinfo {author} {\bibfnamefont {L.}~\bibnamefont {Chen}}, \bibinfo {author} {\bibfnamefont {D.-W.}\ \bibnamefont {Qu}}, \bibinfo {author} {\bibfnamefont {H.}~\bibnamefont {Li}}, \bibinfo {author} {\bibfnamefont {B.-B.}\ \bibnamefont {Chen}}, \bibinfo {author} {\bibfnamefont {S.-S.}\ \bibnamefont {Gong}}, \bibinfo {author} {\bibfnamefont {J.}~\bibnamefont {von Delft}}, \bibinfo {author} {\bibfnamefont {A.}~\bibnamefont {Weichselbaum}},\ and\ \bibinfo {author} {\bibfnamefont {W.}~\bibnamefont {Li}},\ }\bibfield  {title} {\bibinfo {title} {{Two-temperature scales in the triangular-lattice Heisenberg antiferromagnet}},\ }\href {https://doi.org/10.1103/PhysRevB.99.140404} {\bibfield  {journal} {\bibinfo  {journal} {Physical Review B}\ }\textbf {\bibinfo {volume} {99}},\ \bibinfo {pages} {140404} (\bibinfo {year} {2019})}\BibitemShut {NoStop}%
\bibitem [{\citenamefont {Wietek}\ \emph {et~al.}(2021)\citenamefont {Wietek}, \citenamefont {Rossi}, \citenamefont {\ifmmode~\check{S}\else \v{S}\fi{}imkovic}, \citenamefont {Klett}, \citenamefont {Hansmann}, \citenamefont {Ferrero}, \citenamefont {Stoudenmire}, \citenamefont {Sch\"afer},\ and\ \citenamefont {Georges}}]{wietek2021}%
  \BibitemOpen
  \bibfield  {author} {\bibinfo {author} {\bibfnamefont {A.}~\bibnamefont {Wietek}}, \bibinfo {author} {\bibfnamefont {R.}~\bibnamefont {Rossi}}, \bibinfo {author} {\bibfnamefont {F.}~\bibnamefont {\ifmmode~\check{S}\else \v{S}\fi{}imkovic}}, \bibinfo {author} {\bibfnamefont {M.}~\bibnamefont {Klett}}, \bibinfo {author} {\bibfnamefont {P.}~\bibnamefont {Hansmann}}, \bibinfo {author} {\bibfnamefont {M.}~\bibnamefont {Ferrero}}, \bibinfo {author} {\bibfnamefont {E.~M.}\ \bibnamefont {Stoudenmire}}, \bibinfo {author} {\bibfnamefont {T.}~\bibnamefont {Sch\"afer}},\ and\ \bibinfo {author} {\bibfnamefont {A.}~\bibnamefont {Georges}},\ }\bibfield  {title} {\bibinfo {title} {{{Mott}} insulating states with competing orders in the triangular lattice {{Hubbard}} model},\ }\href {https://doi.org/10.1103/PhysRevX.11.041013} {\bibfield  {journal} {\bibinfo  {journal} {Physical Review X}\ }\textbf {\bibinfo {volume} {11}},\ \bibinfo {pages} {041013} (\bibinfo {year} {2021})}\BibitemShut {NoStop}%
\bibitem [{\citenamefont {Lee}\ \emph {et~al.}(2007)\citenamefont {Lee}, \citenamefont {Kune\ifmmode~\check{s}\else \v{s}\fi{}}, \citenamefont {Scalettar},\ and\ \citenamefont {Pickett}}]{Pickett2007}%
  \BibitemOpen
  \bibfield  {author} {\bibinfo {author} {\bibfnamefont {K.-W.}\ \bibnamefont {Lee}}, \bibinfo {author} {\bibfnamefont {J.}~\bibnamefont {Kune\ifmmode~\check{s}\else \v{s}\fi{}}}, \bibinfo {author} {\bibfnamefont {R.~T.}\ \bibnamefont {Scalettar}},\ and\ \bibinfo {author} {\bibfnamefont {W.~E.}\ \bibnamefont {Pickett}},\ }\bibfield  {title} {\bibinfo {title} {{Correlation effects in the triangular lattice single-band system {L}i$_{x}${N}b{O}$_{2}$}},\ }\href {https://doi.org/10.1103/PhysRevB.76.144513} {\bibfield  {journal} {\bibinfo  {journal} {Physical Review B}\ }\textbf {\bibinfo {volume} {76}},\ \bibinfo {pages} {144513} (\bibinfo {year} {2007})}\BibitemShut {NoStop}%
\bibitem [{\citenamefont {Zhang}\ \emph {et~al.}(2020)\citenamefont {Zhang}, \citenamefont {Si}, \citenamefont {Lian}, \citenamefont {Zhou},\ and\ \citenamefont {Sun}}]{KangZhang2020}%
  \BibitemOpen
  \bibfield  {author} {\bibinfo {author} {\bibfnamefont {K.}~\bibnamefont {Zhang}}, \bibinfo {author} {\bibfnamefont {C.}~\bibnamefont {Si}}, \bibinfo {author} {\bibfnamefont {C.-S.}\ \bibnamefont {Lian}}, \bibinfo {author} {\bibfnamefont {J.}~\bibnamefont {Zhou}},\ and\ \bibinfo {author} {\bibfnamefont {Z.}~\bibnamefont {Sun}},\ }\bibfield  {title} {\bibinfo {title} {{Mottness collapse in monolayer 1T-TaSe$_2$ with persisting charge density wave order}},\ }\href {https://doi.org/10.1039/d0tc01719a} {\bibfield  {journal} {\bibinfo  {journal} {Journal of Materials Chemistry C}\ }\textbf {\bibinfo {volume} {8}},\ \bibinfo {pages} {9742} (\bibinfo {year} {2020})}\BibitemShut {NoStop}%
\bibitem [{\citenamefont {Di~Salvo}\ and\ \citenamefont {Graebner}(1977)}]{disalvo1977}%
  \BibitemOpen
  \bibfield  {author} {\bibinfo {author} {\bibfnamefont {F.~J.}\ \bibnamefont {Di~Salvo}}\ and\ \bibinfo {author} {\bibfnamefont {J.~E.}\ \bibnamefont {Graebner}},\ }\bibfield  {title} {\bibinfo {title} {The low temperature electrical properties of {{1T-TaS}}$_2$},\ }\href {https://doi.org/10.1016/0038-1098(77)90961-9} {\bibfield  {journal} {\bibinfo  {journal} {Solid State Communications}\ }\textbf {\bibinfo {volume} {23}},\ \bibinfo {pages} {825} (\bibinfo {year} {1977})}\BibitemShut {NoStop}%
\bibitem [{\citenamefont {Stojchevska}\ \emph {et~al.}(2014)\citenamefont {Stojchevska}, \citenamefont {Vaskivskyi}, \citenamefont {Mertelj}, \citenamefont {Kusar}, \citenamefont {Svetin}, \citenamefont {Brazovskii},\ and\ \citenamefont {Mihailovic}}]{stojchevska2014}%
  \BibitemOpen
  \bibfield  {author} {\bibinfo {author} {\bibfnamefont {L.}~\bibnamefont {Stojchevska}}, \bibinfo {author} {\bibfnamefont {I.}~\bibnamefont {Vaskivskyi}}, \bibinfo {author} {\bibfnamefont {T.}~\bibnamefont {Mertelj}}, \bibinfo {author} {\bibfnamefont {P.}~\bibnamefont {Kusar}}, \bibinfo {author} {\bibfnamefont {D.}~\bibnamefont {Svetin}}, \bibinfo {author} {\bibfnamefont {S.}~\bibnamefont {Brazovskii}},\ and\ \bibinfo {author} {\bibfnamefont {D.}~\bibnamefont {Mihailovic}},\ }\bibfield  {title} {\bibinfo {title} {Ultrafast {{Switching}} to a {{Stable Hidden Quantum State}} in an {{Electronic Crystal}}},\ }\href {https://doi.org/10.1126/science.1241591} {\bibfield  {journal} {\bibinfo  {journal} {Science}\ }\textbf {\bibinfo {volume} {344}},\ \bibinfo {pages} {177} (\bibinfo {year} {2014})}\BibitemShut {NoStop}%
\bibitem [{\citenamefont {Hollander}\ \emph {et~al.}(2015)\citenamefont {Hollander}, \citenamefont {Liu}, \citenamefont {Lu}, \citenamefont {Li}, \citenamefont {Sun}, \citenamefont {Robinson},\ and\ \citenamefont {Datta}}]{Hollander2015}%
  \BibitemOpen
  \bibfield  {author} {\bibinfo {author} {\bibfnamefont {M.~J.}\ \bibnamefont {Hollander}}, \bibinfo {author} {\bibfnamefont {Y.}~\bibnamefont {Liu}}, \bibinfo {author} {\bibfnamefont {W.-J.}\ \bibnamefont {Lu}}, \bibinfo {author} {\bibfnamefont {L.-J.}\ \bibnamefont {Li}}, \bibinfo {author} {\bibfnamefont {Y.-P.}\ \bibnamefont {Sun}}, \bibinfo {author} {\bibfnamefont {J.~A.}\ \bibnamefont {Robinson}},\ and\ \bibinfo {author} {\bibfnamefont {S.}~\bibnamefont {Datta}},\ }\bibfield  {title} {\bibinfo {title} {{Electrically Driven Reversible Insulator-Metal Phase Transition in 1T-TaS$_2$}},\ }\href {https://doi.org/10.1021/nl504662b} {\bibfield  {journal} {\bibinfo  {journal} {Nano Letters}\ }\textbf {\bibinfo {volume} {15}},\ \bibinfo {pages} {1861} (\bibinfo {year} {2015})}\BibitemShut {NoStop}%
\bibitem [{\citenamefont {Sipos}\ \emph {et~al.}(2008)\citenamefont {Sipos}, \citenamefont {Kusmartseva}, \citenamefont {Akrap}, \citenamefont {Berger}, \citenamefont {Forr{\'o}},\ and\ \citenamefont {Tuti{\v s}}}]{Sipos2008}%
  \BibitemOpen
  \bibfield  {author} {\bibinfo {author} {\bibfnamefont {B.}~\bibnamefont {Sipos}}, \bibinfo {author} {\bibfnamefont {A.~F.}\ \bibnamefont {Kusmartseva}}, \bibinfo {author} {\bibfnamefont {A.}~\bibnamefont {Akrap}}, \bibinfo {author} {\bibfnamefont {H.}~\bibnamefont {Berger}}, \bibinfo {author} {\bibfnamefont {L.}~\bibnamefont {Forr{\'o}}},\ and\ \bibinfo {author} {\bibfnamefont {E.}~\bibnamefont {Tuti{\v s}}},\ }\bibfield  {title} {\bibinfo {title} {{From Mott state to superconductivity in 1T-TaS2}},\ }\href {https://doi.org/10.1038/nmat2318} {\bibfield  {journal} {\bibinfo  {journal} {Nature Materials}\ }\textbf {\bibinfo {volume} {7}},\ \bibinfo {pages} {960} (\bibinfo {year} {2008})}\BibitemShut {NoStop}%
\bibitem [{\citenamefont {Law}\ and\ \citenamefont {Lee}(2017)}]{Law2017aa}%
  \BibitemOpen
  \bibfield  {author} {\bibinfo {author} {\bibfnamefont {K.~T.}\ \bibnamefont {Law}}\ and\ \bibinfo {author} {\bibfnamefont {P.~A.}\ \bibnamefont {Lee}},\ }\bibfield  {title} {\bibinfo {title} {{1T-TaS$_2$ as a quantum spin liquid}},\ }\href {https://doi.org/10.1073/pnas.1706769114} {\bibfield  {journal} {\bibinfo  {journal} {Proceedings of the National Academy of Sciences}\ }\textbf {\bibinfo {volume} {114}},\ \bibinfo {pages} {6996} (\bibinfo {year} {2017})}\BibitemShut {NoStop}%
\bibitem [{\citenamefont {Murayama}\ \emph {et~al.}(2020)\citenamefont {Murayama}, \citenamefont {Sato}, \citenamefont {Taniguchi}, \citenamefont {Kurihara}, \citenamefont {Xing}, \citenamefont {Huang}, \citenamefont {Kasahara}, \citenamefont {Kasahara}, \citenamefont {Kimchi}, \citenamefont {Yoshida}, \citenamefont {Iwasa}, \citenamefont {Mizukami}, \citenamefont {Shibauchi}, \citenamefont {Konczykowski},\ and\ \citenamefont {Matsuda}}]{Murayama2020}%
  \BibitemOpen
  \bibfield  {author} {\bibinfo {author} {\bibfnamefont {H.}~\bibnamefont {Murayama}}, \bibinfo {author} {\bibfnamefont {Y.}~\bibnamefont {Sato}}, \bibinfo {author} {\bibfnamefont {T.}~\bibnamefont {Taniguchi}}, \bibinfo {author} {\bibfnamefont {R.}~\bibnamefont {Kurihara}}, \bibinfo {author} {\bibfnamefont {X.~Z.}\ \bibnamefont {Xing}}, \bibinfo {author} {\bibfnamefont {W.}~\bibnamefont {Huang}}, \bibinfo {author} {\bibfnamefont {S.}~\bibnamefont {Kasahara}}, \bibinfo {author} {\bibfnamefont {Y.}~\bibnamefont {Kasahara}}, \bibinfo {author} {\bibfnamefont {I.}~\bibnamefont {Kimchi}}, \bibinfo {author} {\bibfnamefont {M.}~\bibnamefont {Yoshida}}, \bibinfo {author} {\bibfnamefont {Y.}~\bibnamefont {Iwasa}}, \bibinfo {author} {\bibfnamefont {Y.}~\bibnamefont {Mizukami}}, \bibinfo {author} {\bibfnamefont {T.}~\bibnamefont {Shibauchi}}, \bibinfo {author} {\bibfnamefont {M.}~\bibnamefont {Konczykowski}},\ and\ \bibinfo {author} {\bibfnamefont {Y.}~\bibnamefont {Matsuda}},\ }\bibfield  {title} {\bibinfo {title}
  {{Effect of quenched disorder on the quantum spin liquid state of the triangular-lattice antiferromagnet 1T-TaS$_2$}},\ }\href {https://doi.org/10.1103/PhysRevResearch.2.013099} {\bibfield  {journal} {\bibinfo  {journal} {Phys. Rev. Res.}\ }\textbf {\bibinfo {volume} {2}},\ \bibinfo {pages} {013099} (\bibinfo {year} {2020})}\BibitemShut {NoStop}%
\bibitem [{\citenamefont {Ruan}\ \emph {et~al.}(2021)\citenamefont {Ruan}, \citenamefont {Chen}, \citenamefont {Tang}, \citenamefont {Hwang}, \citenamefont {Tsai}, \citenamefont {Lee}, \citenamefont {Wu}, \citenamefont {Ryu}, \citenamefont {Kahn}, \citenamefont {Liou}, \citenamefont {Jia}, \citenamefont {Aikawa}, \citenamefont {Hwang}, \citenamefont {Wang}, \citenamefont {Choi}, \citenamefont {Louie}, \citenamefont {Lee}, \citenamefont {Shen}, \citenamefont {Mo},\ and\ \citenamefont {Crommie}}]{ruan2021}%
  \BibitemOpen
  \bibfield  {author} {\bibinfo {author} {\bibfnamefont {W.}~\bibnamefont {Ruan}}, \bibinfo {author} {\bibfnamefont {Y.}~\bibnamefont {Chen}}, \bibinfo {author} {\bibfnamefont {S.}~\bibnamefont {Tang}}, \bibinfo {author} {\bibfnamefont {J.}~\bibnamefont {Hwang}}, \bibinfo {author} {\bibfnamefont {H.-Z.}\ \bibnamefont {Tsai}}, \bibinfo {author} {\bibfnamefont {R.~L.}\ \bibnamefont {Lee}}, \bibinfo {author} {\bibfnamefont {M.}~\bibnamefont {Wu}}, \bibinfo {author} {\bibfnamefont {H.}~\bibnamefont {Ryu}}, \bibinfo {author} {\bibfnamefont {S.}~\bibnamefont {Kahn}}, \bibinfo {author} {\bibfnamefont {F.}~\bibnamefont {Liou}}, \bibinfo {author} {\bibfnamefont {C.}~\bibnamefont {Jia}}, \bibinfo {author} {\bibfnamefont {A.}~\bibnamefont {Aikawa}}, \bibinfo {author} {\bibfnamefont {C.}~\bibnamefont {Hwang}}, \bibinfo {author} {\bibfnamefont {F.}~\bibnamefont {Wang}}, \bibinfo {author} {\bibfnamefont {Y.}~\bibnamefont {Choi}}, \bibinfo {author} {\bibfnamefont {S.~G.}\ \bibnamefont {Louie}}, \bibinfo {author}
  {\bibfnamefont {P.~A.}\ \bibnamefont {Lee}}, \bibinfo {author} {\bibfnamefont {Z.-X.}\ \bibnamefont {Shen}}, \bibinfo {author} {\bibfnamefont {S.-K.}\ \bibnamefont {Mo}},\ and\ \bibinfo {author} {\bibfnamefont {M.~F.}\ \bibnamefont {Crommie}},\ }\bibfield  {title} {\bibinfo {title} {Evidence for quantum spin liquid behaviour in single-layer {{1T-TaSe}}$_2$ from scanning tunnelling microscopy},\ }\href {https://doi.org/10.1038/s41567-021-01321-0} {\bibfield  {journal} {\bibinfo  {journal} {Nature Physics}\ }\textbf {\bibinfo {volume} {17}},\ \bibinfo {pages} {1154} (\bibinfo {year} {2021})}\BibitemShut {NoStop}%
\bibitem [{\citenamefont {Chen}\ \emph {et~al.}(2021)\citenamefont {Chen}, \citenamefont {Sodemann},\ and\ \citenamefont {Lee}}]{chen2021}%
  \BibitemOpen
  \bibfield  {author} {\bibinfo {author} {\bibfnamefont {C.}~\bibnamefont {Chen}}, \bibinfo {author} {\bibfnamefont {I.}~\bibnamefont {Sodemann}},\ and\ \bibinfo {author} {\bibfnamefont {P.~A.}\ \bibnamefont {Lee}},\ }\bibfield  {title} {\bibinfo {title} {Competition of spinon {{Fermi}} surface and heavy {{Fermi}} liquid states from the periodic {{Anderson}} to the {{Hubbard}} model},\ }\href {https://doi.org/10.1103/PhysRevB.103.085128} {\bibfield  {journal} {\bibinfo  {journal} {Physical Review B}\ }\textbf {\bibinfo {volume} {103}},\ \bibinfo {pages} {085128} (\bibinfo {year} {2021})}\BibitemShut {NoStop}%
\bibitem [{\citenamefont {Va{\v n}o}\ \emph {et~al.}(2021)\citenamefont {Va{\v n}o}, \citenamefont {Amini}, \citenamefont {Ganguli}, \citenamefont {Chen}, \citenamefont {Lado}, \citenamefont {Kezilebieke},\ and\ \citenamefont {Liljeroth}}]{vano2021}%
  \BibitemOpen
  \bibfield  {author} {\bibinfo {author} {\bibfnamefont {V.}~\bibnamefont {Va{\v n}o}}, \bibinfo {author} {\bibfnamefont {M.}~\bibnamefont {Amini}}, \bibinfo {author} {\bibfnamefont {S.~C.}\ \bibnamefont {Ganguli}}, \bibinfo {author} {\bibfnamefont {G.}~\bibnamefont {Chen}}, \bibinfo {author} {\bibfnamefont {J.~L.}\ \bibnamefont {Lado}}, \bibinfo {author} {\bibfnamefont {S.}~\bibnamefont {Kezilebieke}},\ and\ \bibinfo {author} {\bibfnamefont {P.}~\bibnamefont {Liljeroth}},\ }\bibfield  {title} {\bibinfo {title} {Artificial heavy fermions in a van der {{Waals}} heterostructure},\ }\href {https://doi.org/10.1038/s41586-021-04021-0} {\bibfield  {journal} {\bibinfo  {journal} {Nature}\ }\textbf {\bibinfo {volume} {599}},\ \bibinfo {pages} {582} (\bibinfo {year} {2021})}\BibitemShut {NoStop}%
\bibitem [{\citenamefont {Chen}\ \emph {et~al.}(2022{\natexlab{a}})\citenamefont {Chen}, \citenamefont {He}, \citenamefont {Ruan}, \citenamefont {Hwang}, \citenamefont {Tang}, \citenamefont {Lee}, \citenamefont {Wu}, \citenamefont {Zhu}, \citenamefont {Zhang}, \citenamefont {Ryu}, \citenamefont {Wang}, \citenamefont {Louie}, \citenamefont {Shen}, \citenamefont {Mo}, \citenamefont {Lee},\ and\ \citenamefont {Crommie}}]{Chen2022aa}%
  \BibitemOpen
  \bibfield  {author} {\bibinfo {author} {\bibfnamefont {Y.}~\bibnamefont {Chen}}, \bibinfo {author} {\bibfnamefont {W.-Y.}\ \bibnamefont {He}}, \bibinfo {author} {\bibfnamefont {W.}~\bibnamefont {Ruan}}, \bibinfo {author} {\bibfnamefont {J.}~\bibnamefont {Hwang}}, \bibinfo {author} {\bibfnamefont {S.}~\bibnamefont {Tang}}, \bibinfo {author} {\bibfnamefont {R.~L.}\ \bibnamefont {Lee}}, \bibinfo {author} {\bibfnamefont {M.}~\bibnamefont {Wu}}, \bibinfo {author} {\bibfnamefont {T.}~\bibnamefont {Zhu}}, \bibinfo {author} {\bibfnamefont {C.}~\bibnamefont {Zhang}}, \bibinfo {author} {\bibfnamefont {H.}~\bibnamefont {Ryu}}, \bibinfo {author} {\bibfnamefont {F.}~\bibnamefont {Wang}}, \bibinfo {author} {\bibfnamefont {S.~G.}\ \bibnamefont {Louie}}, \bibinfo {author} {\bibfnamefont {Z.-X.}\ \bibnamefont {Shen}}, \bibinfo {author} {\bibfnamefont {S.-K.}\ \bibnamefont {Mo}}, \bibinfo {author} {\bibfnamefont {P.~A.}\ \bibnamefont {Lee}},\ and\ \bibinfo {author} {\bibfnamefont {M.~F.}\ \bibnamefont {Crommie}},\ }\bibfield
   {title} {\bibinfo {title} {Evidence for a spinon kondo effect in cobalt atoms on single-layer {{1T-TaSe}}$_2$},\ }\href {https://doi.org/10.1038/s41567-022-01751-4} {\bibfield  {journal} {\bibinfo  {journal} {Nature Physics}\ }\textbf {\bibinfo {volume} {18}},\ \bibinfo {pages} {1335} (\bibinfo {year} {2022}{\natexlab{a}})}\BibitemShut {NoStop}%
\bibitem [{\citenamefont {Di~Salvo}\ \emph {et~al.}(1974)\citenamefont {Di~Salvo}, \citenamefont {Maines}, \citenamefont {Waszczak},\ and\ \citenamefont {Schwall}}]{disalvo1974a}%
  \BibitemOpen
  \bibfield  {author} {\bibinfo {author} {\bibfnamefont {F.~J.}\ \bibnamefont {Di~Salvo}}, \bibinfo {author} {\bibfnamefont {R.~G.}\ \bibnamefont {Maines}}, \bibinfo {author} {\bibfnamefont {J.~V.}\ \bibnamefont {Waszczak}},\ and\ \bibinfo {author} {\bibfnamefont {R.~E.}\ \bibnamefont {Schwall}},\ }\bibfield  {title} {\bibinfo {title} {Preparation and properties of {{1T}}-{{TaSe}}$_2$},\ }\href {https://doi.org/10.1016/0038-1098(74)90975-2} {\bibfield  {journal} {\bibinfo  {journal} {Solid State Communications}\ }\textbf {\bibinfo {volume} {14}},\ \bibinfo {pages} {497} (\bibinfo {year} {1974})}\BibitemShut {NoStop}%
\bibitem [{\citenamefont {Wilson}\ \emph {et~al.}(1974)\citenamefont {Wilson}, \citenamefont {Di~Salvo},\ and\ \citenamefont {Mahajan}}]{wilson1974}%
  \BibitemOpen
  \bibfield  {author} {\bibinfo {author} {\bibfnamefont {J.~A.}\ \bibnamefont {Wilson}}, \bibinfo {author} {\bibfnamefont {F.~J.}\ \bibnamefont {Di~Salvo}},\ and\ \bibinfo {author} {\bibfnamefont {S.}~\bibnamefont {Mahajan}},\ }\bibfield  {title} {\bibinfo {title} {Charge-{{Density Waves}} in {{Metallic}}, {{Layered}}, {{Transition-Metal Dichalcogenides}}},\ }\href {https://doi.org/10.1103/PhysRevLett.32.882} {\bibfield  {journal} {\bibinfo  {journal} {Physical Review Letters}\ }\textbf {\bibinfo {volume} {32}},\ \bibinfo {pages} {882} (\bibinfo {year} {1974})}\BibitemShut {NoStop}%
\bibitem [{\citenamefont {Tian}\ \emph {et~al.}(2023)\citenamefont {Tian}, \citenamefont {Huang}, \citenamefont {Jang}, \citenamefont {Guo}, \citenamefont {Yan}, \citenamefont {Gao}, \citenamefont {Yu}, \citenamefont {Hwang}, \citenamefont {Tang}, \citenamefont {Wang}, \citenamefont {Luo}, \citenamefont {Sun}, \citenamefont {Liu}, \citenamefont {Feng}, \citenamefont {Chen}, \citenamefont {Mo}, \citenamefont {Kim}, \citenamefont {Son}, \citenamefont {Shen}, \citenamefont {Ruan},\ and\ \citenamefont {Zhang}}]{tian2023}%
  \BibitemOpen
  \bibfield  {author} {\bibinfo {author} {\bibfnamefont {N.}~\bibnamefont {Tian}}, \bibinfo {author} {\bibfnamefont {Z.}~\bibnamefont {Huang}}, \bibinfo {author} {\bibfnamefont {B.~G.}\ \bibnamefont {Jang}}, \bibinfo {author} {\bibfnamefont {S.}~\bibnamefont {Guo}}, \bibinfo {author} {\bibfnamefont {Y.-J.}\ \bibnamefont {Yan}}, \bibinfo {author} {\bibfnamefont {J.}~\bibnamefont {Gao}}, \bibinfo {author} {\bibfnamefont {Y.}~\bibnamefont {Yu}}, \bibinfo {author} {\bibfnamefont {J.}~\bibnamefont {Hwang}}, \bibinfo {author} {\bibfnamefont {C.}~\bibnamefont {Tang}}, \bibinfo {author} {\bibfnamefont {M.}~\bibnamefont {Wang}}, \bibinfo {author} {\bibfnamefont {X.}~\bibnamefont {Luo}}, \bibinfo {author} {\bibfnamefont {Y.~P.}\ \bibnamefont {Sun}}, \bibinfo {author} {\bibfnamefont {Z.}~\bibnamefont {Liu}}, \bibinfo {author} {\bibfnamefont {D.-L.}\ \bibnamefont {Feng}}, \bibinfo {author} {\bibfnamefont {X.}~\bibnamefont {Chen}}, \bibinfo {author} {\bibfnamefont {S.-K.}\ \bibnamefont {Mo}}, \bibinfo {author}
  {\bibfnamefont {M.}~\bibnamefont {Kim}}, \bibinfo {author} {\bibfnamefont {Y.-W.}\ \bibnamefont {Son}}, \bibinfo {author} {\bibfnamefont {D.}~\bibnamefont {Shen}}, \bibinfo {author} {\bibfnamefont {W.}~\bibnamefont {Ruan}},\ and\ \bibinfo {author} {\bibfnamefont {Y.}~\bibnamefont {Zhang}},\ }\bibfield  {title} {\bibinfo {title} {Dimensionality-driven metal to {{Mott}} insulator transition in two-dimensional {{1T-TaSe}}$_2$},\ }\href {https://doi.org/10.1093/nsr/nwad144} {\bibfield  {journal} {\bibinfo  {journal} {National Science Review}\ ,\ \bibinfo {pages} {nwad144}} (\bibinfo {year} {2023})}\BibitemShut {NoStop}%
\bibitem [{\citenamefont {Perfetti}\ \emph {et~al.}(2003)\citenamefont {Perfetti}, \citenamefont {Georges}, \citenamefont {Florens}, \citenamefont {Biermann}, \citenamefont {Mitrovic}, \citenamefont {Berger}, \citenamefont {Tomm}, \citenamefont {H{\"o}chst},\ and\ \citenamefont {Grioni}}]{perfetti2003}%
  \BibitemOpen
  \bibfield  {author} {\bibinfo {author} {\bibfnamefont {L.}~\bibnamefont {Perfetti}}, \bibinfo {author} {\bibfnamefont {A.}~\bibnamefont {Georges}}, \bibinfo {author} {\bibfnamefont {S.}~\bibnamefont {Florens}}, \bibinfo {author} {\bibfnamefont {S.}~\bibnamefont {Biermann}}, \bibinfo {author} {\bibfnamefont {S.}~\bibnamefont {Mitrovic}}, \bibinfo {author} {\bibfnamefont {H.}~\bibnamefont {Berger}}, \bibinfo {author} {\bibfnamefont {Y.}~\bibnamefont {Tomm}}, \bibinfo {author} {\bibfnamefont {H.}~\bibnamefont {H{\"o}chst}},\ and\ \bibinfo {author} {\bibfnamefont {M.}~\bibnamefont {Grioni}},\ }\bibfield  {title} {\bibinfo {title} {Spectroscopic {{Signatures}} of a {{Bandwidth-Controlled Mott Transition}} at the {{Surface}} of {{1T}}-{{TaSe}}$_2$},\ }\href {https://doi.org/10.1103/PhysRevLett.90.166401} {\bibfield  {journal} {\bibinfo  {journal} {Physical Review Letters}\ }\textbf {\bibinfo {volume} {90}},\ \bibinfo {pages} {166401} (\bibinfo {year} {2003})}\BibitemShut {NoStop}%
\bibitem [{\citenamefont {Sayers}\ \emph {et~al.}(2023)\citenamefont {Sayers}, \citenamefont {Cerullo}, \citenamefont {Zhang}, \citenamefont {Sanders}, \citenamefont {Chapman}, \citenamefont {Wyatt}, \citenamefont {Chatterjee}, \citenamefont {Springate}, \citenamefont {Wolverson}, \citenamefont {Da~Como},\ and\ \citenamefont {Carpene}}]{sayers2023}%
  \BibitemOpen
  \bibfield  {author} {\bibinfo {author} {\bibfnamefont {C.~J.}\ \bibnamefont {Sayers}}, \bibinfo {author} {\bibfnamefont {G.}~\bibnamefont {Cerullo}}, \bibinfo {author} {\bibfnamefont {Y.}~\bibnamefont {Zhang}}, \bibinfo {author} {\bibfnamefont {C.~E.}\ \bibnamefont {Sanders}}, \bibinfo {author} {\bibfnamefont {R.~T.}\ \bibnamefont {Chapman}}, \bibinfo {author} {\bibfnamefont {A.~S.}\ \bibnamefont {Wyatt}}, \bibinfo {author} {\bibfnamefont {G.}~\bibnamefont {Chatterjee}}, \bibinfo {author} {\bibfnamefont {E.}~\bibnamefont {Springate}}, \bibinfo {author} {\bibfnamefont {D.}~\bibnamefont {Wolverson}}, \bibinfo {author} {\bibfnamefont {E.}~\bibnamefont {Da~Como}},\ and\ \bibinfo {author} {\bibfnamefont {E.}~\bibnamefont {Carpene}},\ }\bibfield  {title} {\bibinfo {title} {Exploring the {{Charge Density Wave Phase}} of {{1T}}-{{TaSe}}$_2$: {{Mott}} or {{Charge-Transfer Gap}}?},\ }\href {https://doi.org/10.1103/PhysRevLett.130.156401} {\bibfield  {journal} {\bibinfo  {journal} {Physical Review Letters}\ }\textbf
  {\bibinfo {volume} {130}},\ \bibinfo {pages} {156401} (\bibinfo {year} {2023})}\BibitemShut {NoStop}%
\bibitem [{\citenamefont {Chen}\ \emph {et~al.}(2022{\natexlab{b}})\citenamefont {Chen}, \citenamefont {Ruan}, \citenamefont {Cain}, \citenamefont {Lee}, \citenamefont {Kahn}, \citenamefont {Jia}, \citenamefont {Zettl},\ and\ \citenamefont {Crommie}}]{chen2022}%
  \BibitemOpen
  \bibfield  {author} {\bibinfo {author} {\bibfnamefont {Y.}~\bibnamefont {Chen}}, \bibinfo {author} {\bibfnamefont {W.}~\bibnamefont {Ruan}}, \bibinfo {author} {\bibfnamefont {J.~D.}\ \bibnamefont {Cain}}, \bibinfo {author} {\bibfnamefont {R.~L.}\ \bibnamefont {Lee}}, \bibinfo {author} {\bibfnamefont {S.}~\bibnamefont {Kahn}}, \bibinfo {author} {\bibfnamefont {C.}~\bibnamefont {Jia}}, \bibinfo {author} {\bibfnamefont {A.}~\bibnamefont {Zettl}},\ and\ \bibinfo {author} {\bibfnamefont {M.~F.}\ \bibnamefont {Crommie}},\ }\bibfield  {title} {\bibinfo {title} {Observation of a multitude of correlated states at the surface of bulk {{1T}}-{{TaSe}}$_2$ crystals},\ }\href {https://doi.org/10.1103/PhysRevB.106.075153} {\bibfield  {journal} {\bibinfo  {journal} {Physical Review B}\ }\textbf {\bibinfo {volume} {106}},\ \bibinfo {pages} {075153} (\bibinfo {year} {2022}{\natexlab{b}})}\BibitemShut {NoStop}%
\bibitem [{\citenamefont {Zhang}\ \emph {et~al.}(2022)\citenamefont {Zhang}, \citenamefont {Wu}, \citenamefont {Bu}, \citenamefont {Fei}, \citenamefont {Zheng}, \citenamefont {Gao}, \citenamefont {Luo}, \citenamefont {Liu}, \citenamefont {Sun},\ and\ \citenamefont {Yin}}]{zhang2022}%
  \BibitemOpen
  \bibfield  {author} {\bibinfo {author} {\bibfnamefont {W.}~\bibnamefont {Zhang}}, \bibinfo {author} {\bibfnamefont {Z.}~\bibnamefont {Wu}}, \bibinfo {author} {\bibfnamefont {K.}~\bibnamefont {Bu}}, \bibinfo {author} {\bibfnamefont {Y.}~\bibnamefont {Fei}}, \bibinfo {author} {\bibfnamefont {Y.}~\bibnamefont {Zheng}}, \bibinfo {author} {\bibfnamefont {J.}~\bibnamefont {Gao}}, \bibinfo {author} {\bibfnamefont {X.}~\bibnamefont {Luo}}, \bibinfo {author} {\bibfnamefont {Z.}~\bibnamefont {Liu}}, \bibinfo {author} {\bibfnamefont {Y.-P.}\ \bibnamefont {Sun}},\ and\ \bibinfo {author} {\bibfnamefont {Y.}~\bibnamefont {Yin}},\ }\bibfield  {title} {\bibinfo {title} {Reconciling the bulk metallic and surface insulating state in {{1T}}-{{TaSe}}$_2$},\ }\href {https://doi.org/10.1103/PhysRevB.105.035110} {\bibfield  {journal} {\bibinfo  {journal} {Physical Review B}\ }\textbf {\bibinfo {volume} {105}},\ \bibinfo {pages} {035110} (\bibinfo {year} {2022})}\BibitemShut {NoStop}%
\bibitem [{\citenamefont {Fei}\ \emph {et~al.}(2022)\citenamefont {Fei}, \citenamefont {Wu}, \citenamefont {Zhang},\ and\ \citenamefont {Yin}}]{fei2022}%
  \BibitemOpen
  \bibfield  {author} {\bibinfo {author} {\bibfnamefont {Y.}~\bibnamefont {Fei}}, \bibinfo {author} {\bibfnamefont {Z.}~\bibnamefont {Wu}}, \bibinfo {author} {\bibfnamefont {W.}~\bibnamefont {Zhang}},\ and\ \bibinfo {author} {\bibfnamefont {Y.}~\bibnamefont {Yin}},\ }\bibfield  {title} {\bibinfo {title} {Understanding the {{Mott}} insulating state in {{1T-TaS}}$_2$ and {{1T-TaSe}}$_2$},\ }\href {https://doi.org/10.1007/s43673-022-00049-0} {\bibfield  {journal} {\bibinfo  {journal} {AAPPS Bulletin}\ }\textbf {\bibinfo {volume} {32}},\ \bibinfo {pages} {20} (\bibinfo {year} {2022})}\BibitemShut {NoStop}%
\bibitem [{\citenamefont {Wiegers}\ \emph {et~al.}(2001)\citenamefont {Wiegers}, \citenamefont {de~Boer}, \citenamefont {Meetsma},\ and\ \citenamefont {van Smaalen}}]{wiegers2001}%
  \BibitemOpen
  \bibfield  {author} {\bibinfo {author} {\bibfnamefont {G.~A.}\ \bibnamefont {Wiegers}}, \bibinfo {author} {\bibfnamefont {J.~L.}\ \bibnamefont {de~Boer}}, \bibinfo {author} {\bibfnamefont {A.}~\bibnamefont {Meetsma}},\ and\ \bibinfo {author} {\bibfnamefont {S.}~\bibnamefont {van Smaalen}},\ }\bibfield  {title} {\bibinfo {title} {Domain structure and refinement of the triclinic superstructure of {{1T-TaSe}}$_2$ by single crystal {{X-ray}} diffraction},\ }\href {https://doi.org/10.1524/zkri.216.1.45.18999} {\bibfield  {journal} {\bibinfo  {journal} {Zeitschrift f{\"u}r Kristallographie - Crystalline Materials}\ }\textbf {\bibinfo {volume} {216}},\ \bibinfo {pages} {45} (\bibinfo {year} {2001})}\BibitemShut {NoStop}%
\bibitem [{\citenamefont {Scruby}\ \emph {et~al.}(1975)\citenamefont {Scruby}, \citenamefont {Williams},\ and\ \citenamefont {Parry}}]{scruby1975}%
  \BibitemOpen
  \bibfield  {author} {\bibinfo {author} {\bibfnamefont {C.~B.}\ \bibnamefont {Scruby}}, \bibinfo {author} {\bibfnamefont {P.~M.}\ \bibnamefont {Williams}},\ and\ \bibinfo {author} {\bibfnamefont {G.~S.}\ \bibnamefont {Parry}},\ }\bibfield  {title} {\bibinfo {title} {The role of charge density waves in structural transformations of {{1T-TaS}}$_2$},\ }\bibfield  {journal} {\bibinfo  {journal} {Philosophical Magazine}\ }\href {https://doi.org/10.1080/14786437508228930} {10.1080/14786437508228930} (\bibinfo {year} {1975})\BibitemShut {NoStop}%
\bibitem [{\citenamefont {Wang}\ \emph {et~al.}(2023{\natexlab{a}})\citenamefont {Wang}, \citenamefont {Yu}, \citenamefont {Zhao}, \citenamefont {Li}, \citenamefont {Wang},\ and\ \citenamefont {Lin}}]{wang2023a}%
  \BibitemOpen
  \bibfield  {author} {\bibinfo {author} {\bibfnamefont {G.}~\bibnamefont {Wang}}, \bibinfo {author} {\bibfnamefont {X.}~\bibnamefont {Yu}}, \bibinfo {author} {\bibfnamefont {E.}~\bibnamefont {Zhao}}, \bibinfo {author} {\bibfnamefont {D.}~\bibnamefont {Li}}, \bibinfo {author} {\bibfnamefont {L.}~\bibnamefont {Wang}},\ and\ \bibinfo {author} {\bibfnamefont {J.}~\bibnamefont {Lin}},\ }\bibfield  {title} {\bibinfo {title} {Atomic {{Visualization}} of the {{3D Charge Density Wave Stacking}} in {{1T-TaS}}$_2$ by {{Cryogenic Transmission Electron Microscopy}}},\ }\href {https://doi.org/10.1021/acs.nanolett.3c00556} {\bibfield  {journal} {\bibinfo  {journal} {Nano Letters}\ }\textbf {\bibinfo {volume} {23}},\ \bibinfo {pages} {4318} (\bibinfo {year} {2023}{\natexlab{a}})}\BibitemShut {NoStop}%
\bibitem [{\citenamefont {Ritschel}\ \emph {et~al.}(2018)\citenamefont {Ritschel}, \citenamefont {Berger},\ and\ \citenamefont {Geck}}]{ritschel2018}%
  \BibitemOpen
  \bibfield  {author} {\bibinfo {author} {\bibfnamefont {T.}~\bibnamefont {Ritschel}}, \bibinfo {author} {\bibfnamefont {H.}~\bibnamefont {Berger}},\ and\ \bibinfo {author} {\bibfnamefont {J.}~\bibnamefont {Geck}},\ }\bibfield  {title} {\bibinfo {title} {Stacking-driven gap formation in layered {{1T}}-{{TaS}}$_2$},\ }\href {https://doi.org/10.1103/PhysRevB.98.195134} {\bibfield  {journal} {\bibinfo  {journal} {Physical Review B}\ }\textbf {\bibinfo {volume} {98}},\ \bibinfo {pages} {195134} (\bibinfo {year} {2018})}\BibitemShut {NoStop}%
\bibitem [{\citenamefont {Lee}\ \emph {et~al.}(2019)\citenamefont {Lee}, \citenamefont {Goh},\ and\ \citenamefont {Cho}}]{lee2019}%
  \BibitemOpen
  \bibfield  {author} {\bibinfo {author} {\bibfnamefont {S.-H.}\ \bibnamefont {Lee}}, \bibinfo {author} {\bibfnamefont {J.~S.}\ \bibnamefont {Goh}},\ and\ \bibinfo {author} {\bibfnamefont {D.}~\bibnamefont {Cho}},\ }\bibfield  {title} {\bibinfo {title} {Origin of the {{Insulating Phase}} and {{First-Order Metal-Insulator Transition}} in{{1T}}-{{TaS}}$_2$},\ }\href {https://doi.org/10.1103/PhysRevLett.122.106404} {\bibfield  {journal} {\bibinfo  {journal} {Physical Review Letters}\ }\textbf {\bibinfo {volume} {122}},\ \bibinfo {pages} {106404} (\bibinfo {year} {2019})}\BibitemShut {NoStop}%
\bibitem [{\citenamefont {Butler}\ \emph {et~al.}(2020)\citenamefont {Butler}, \citenamefont {Yoshida}, \citenamefont {Hanaguri},\ and\ \citenamefont {Iwasa}}]{butler2020}%
  \BibitemOpen
  \bibfield  {author} {\bibinfo {author} {\bibfnamefont {C.~J.}\ \bibnamefont {Butler}}, \bibinfo {author} {\bibfnamefont {M.}~\bibnamefont {Yoshida}}, \bibinfo {author} {\bibfnamefont {T.}~\bibnamefont {Hanaguri}},\ and\ \bibinfo {author} {\bibfnamefont {Y.}~\bibnamefont {Iwasa}},\ }\bibfield  {title} {\bibinfo {title} {Mottness versus unit-cell doubling as the driver of the insulating state in {{1T-TaS}}$_2$},\ }\href {https://doi.org/10.1038/s41467-020-16132-9} {\bibfield  {journal} {\bibinfo  {journal} {Nature Communications}\ }\textbf {\bibinfo {volume} {11}},\ \bibinfo {pages} {2477} (\bibinfo {year} {2020})}\BibitemShut {NoStop}%
\bibitem [{\citenamefont {Wu}\ \emph {et~al.}(2022)\citenamefont {Wu}, \citenamefont {Bu}, \citenamefont {Zhang}, \citenamefont {Fei}, \citenamefont {Zheng}, \citenamefont {Gao}, \citenamefont {Luo}, \citenamefont {Liu}, \citenamefont {Sun},\ and\ \citenamefont {Yin}}]{wu2022}%
  \BibitemOpen
  \bibfield  {author} {\bibinfo {author} {\bibfnamefont {Z.}~\bibnamefont {Wu}}, \bibinfo {author} {\bibfnamefont {K.}~\bibnamefont {Bu}}, \bibinfo {author} {\bibfnamefont {W.}~\bibnamefont {Zhang}}, \bibinfo {author} {\bibfnamefont {Y.}~\bibnamefont {Fei}}, \bibinfo {author} {\bibfnamefont {Y.}~\bibnamefont {Zheng}}, \bibinfo {author} {\bibfnamefont {J.}~\bibnamefont {Gao}}, \bibinfo {author} {\bibfnamefont {X.}~\bibnamefont {Luo}}, \bibinfo {author} {\bibfnamefont {Z.}~\bibnamefont {Liu}}, \bibinfo {author} {\bibfnamefont {Y.-P.}\ \bibnamefont {Sun}},\ and\ \bibinfo {author} {\bibfnamefont {Y.}~\bibnamefont {Yin}},\ }\bibfield  {title} {\bibinfo {title} {Effect of stacking order on the electronic state of {{1T}}-{{TaS}}$_2$},\ }\href {https://doi.org/10.1103/PhysRevB.105.035109} {\bibfield  {journal} {\bibinfo  {journal} {Physical Review B}\ }\textbf {\bibinfo {volume} {105}},\ \bibinfo {pages} {035109} (\bibinfo {year} {2022})}\BibitemShut {NoStop}%
\bibitem [{\citenamefont {Ge}\ and\ \citenamefont {Liu}(2010)}]{ge2010}%
  \BibitemOpen
  \bibfield  {author} {\bibinfo {author} {\bibfnamefont {Y.}~\bibnamefont {Ge}}\ and\ \bibinfo {author} {\bibfnamefont {A.~Y.}\ \bibnamefont {Liu}},\ }\bibfield  {title} {\bibinfo {title} {First-principles investigation of the charge-density-wave instability in {{1T}}-{{TaSe}}$_2$},\ }\href {https://doi.org/10.1103/PhysRevB.82.155133} {\bibfield  {journal} {\bibinfo  {journal} {Physical Review B}\ }\textbf {\bibinfo {volume} {82}},\ \bibinfo {pages} {155133} (\bibinfo {year} {2010})}\BibitemShut {NoStop}%
\bibitem [{\citenamefont {Wang}\ \emph {et~al.}(2023{\natexlab{b}})\citenamefont {Wang}, \citenamefont {Zhao}, \citenamefont {Ming},\ and\ \citenamefont {Si}}]{wang2023}%
  \BibitemOpen
  \bibfield  {author} {\bibinfo {author} {\bibfnamefont {W.}~\bibnamefont {Wang}}, \bibinfo {author} {\bibfnamefont {B.}~\bibnamefont {Zhao}}, \bibinfo {author} {\bibfnamefont {X.}~\bibnamefont {Ming}},\ and\ \bibinfo {author} {\bibfnamefont {C.}~\bibnamefont {Si}},\ }\bibfield  {title} {\bibinfo {title} {Multiple {{Quantum States Induced}} in {{1T-TaSe}}$_2$ by {{Controlling}} the {{Stacking Order}} of {{Charge Density Waves}}},\ }\href {https://doi.org/10.1002/adfm.202214583} {\bibfield  {journal} {\bibinfo  {journal} {Advanced Functional Materials}\ }\textbf {\bibinfo {volume} {33}},\ \bibinfo {pages} {2214583} (\bibinfo {year} {2023}{\natexlab{b}})}\BibitemShut {NoStop}%
\bibitem [{\citenamefont {Yang}\ \emph {et~al.}(2022)\citenamefont {Yang}, \citenamefont {He}, \citenamefont {Koo}, \citenamefont {Shen}, \citenamefont {Zhang}, \citenamefont {Liu}, \citenamefont {Liu}, \citenamefont {Chen}, \citenamefont {Liang}, \citenamefont {Huang}, \citenamefont {Wang}, \citenamefont {Gao}, \citenamefont {Luo}, \citenamefont {Yang}, \citenamefont {Liu}, \citenamefont {Sun}, \citenamefont {Yan}, \citenamefont {Yan}, \citenamefont {Chen}, \citenamefont {Xi},\ and\ \citenamefont {Liu}}]{yang2022}%
  \BibitemOpen
  \bibfield  {author} {\bibinfo {author} {\bibfnamefont {H.~F.}\ \bibnamefont {Yang}}, \bibinfo {author} {\bibfnamefont {K.~Y.}\ \bibnamefont {He}}, \bibinfo {author} {\bibfnamefont {J.}~\bibnamefont {Koo}}, \bibinfo {author} {\bibfnamefont {S.~W.}\ \bibnamefont {Shen}}, \bibinfo {author} {\bibfnamefont {S.~H.}\ \bibnamefont {Zhang}}, \bibinfo {author} {\bibfnamefont {G.}~\bibnamefont {Liu}}, \bibinfo {author} {\bibfnamefont {Y.~Z.}\ \bibnamefont {Liu}}, \bibinfo {author} {\bibfnamefont {C.}~\bibnamefont {Chen}}, \bibinfo {author} {\bibfnamefont {A.~J.}\ \bibnamefont {Liang}}, \bibinfo {author} {\bibfnamefont {K.}~\bibnamefont {Huang}}, \bibinfo {author} {\bibfnamefont {M.~X.}\ \bibnamefont {Wang}}, \bibinfo {author} {\bibfnamefont {J.~J.}\ \bibnamefont {Gao}}, \bibinfo {author} {\bibfnamefont {X.}~\bibnamefont {Luo}}, \bibinfo {author} {\bibfnamefont {L.~X.}\ \bibnamefont {Yang}}, \bibinfo {author} {\bibfnamefont {J.~P.}\ \bibnamefont {Liu}}, \bibinfo {author} {\bibfnamefont {Y.~P.}\ \bibnamefont {Sun}},
  \bibinfo {author} {\bibfnamefont {S.~C.}\ \bibnamefont {Yan}}, \bibinfo {author} {\bibfnamefont {B.~H.}\ \bibnamefont {Yan}}, \bibinfo {author} {\bibfnamefont {Y.~L.}\ \bibnamefont {Chen}}, \bibinfo {author} {\bibfnamefont {X.}~\bibnamefont {Xi}},\ and\ \bibinfo {author} {\bibfnamefont {Z.~K.}\ \bibnamefont {Liu}},\ }\bibfield  {title} {\bibinfo {title} {Visualization of {{Chiral Electronic Structure}} and {{Anomalous Optical Response}} in a {{Material}} with {{Chiral Charge Density Waves}}},\ }\href {https://doi.org/10.1103/PhysRevLett.129.156401} {\bibfield  {journal} {\bibinfo  {journal} {Physical Review Letters}\ }\textbf {\bibinfo {volume} {129}},\ \bibinfo {pages} {156401} (\bibinfo {year} {2022})}\BibitemShut {NoStop}%
\bibitem [{\citenamefont {Chen}\ \emph {et~al.}(2020)\citenamefont {Chen}, \citenamefont {Ruan}, \citenamefont {Wu}, \citenamefont {Tang}, \citenamefont {Ryu}, \citenamefont {Tsai}, \citenamefont {Lee}, \citenamefont {Kahn}, \citenamefont {Liou}, \citenamefont {Jia}, \citenamefont {Albertini}, \citenamefont {Xiong}, \citenamefont {Jia}, \citenamefont {Liu}, \citenamefont {Sobota}, \citenamefont {Liu}, \citenamefont {Moore}, \citenamefont {Shen}, \citenamefont {Louie}, \citenamefont {Mo},\ and\ \citenamefont {Crommie}}]{chen2020}%
  \BibitemOpen
  \bibfield  {author} {\bibinfo {author} {\bibfnamefont {Y.}~\bibnamefont {Chen}}, \bibinfo {author} {\bibfnamefont {W.}~\bibnamefont {Ruan}}, \bibinfo {author} {\bibfnamefont {M.}~\bibnamefont {Wu}}, \bibinfo {author} {\bibfnamefont {S.}~\bibnamefont {Tang}}, \bibinfo {author} {\bibfnamefont {H.}~\bibnamefont {Ryu}}, \bibinfo {author} {\bibfnamefont {H.-Z.}\ \bibnamefont {Tsai}}, \bibinfo {author} {\bibfnamefont {R.~L.}\ \bibnamefont {Lee}}, \bibinfo {author} {\bibfnamefont {S.}~\bibnamefont {Kahn}}, \bibinfo {author} {\bibfnamefont {F.}~\bibnamefont {Liou}}, \bibinfo {author} {\bibfnamefont {C.}~\bibnamefont {Jia}}, \bibinfo {author} {\bibfnamefont {O.~R.}\ \bibnamefont {Albertini}}, \bibinfo {author} {\bibfnamefont {H.}~\bibnamefont {Xiong}}, \bibinfo {author} {\bibfnamefont {T.}~\bibnamefont {Jia}}, \bibinfo {author} {\bibfnamefont {Z.}~\bibnamefont {Liu}}, \bibinfo {author} {\bibfnamefont {J.~A.}\ \bibnamefont {Sobota}}, \bibinfo {author} {\bibfnamefont {A.~Y.}\ \bibnamefont {Liu}}, \bibinfo {author}
  {\bibfnamefont {J.~E.}\ \bibnamefont {Moore}}, \bibinfo {author} {\bibfnamefont {Z.-X.}\ \bibnamefont {Shen}}, \bibinfo {author} {\bibfnamefont {S.~G.}\ \bibnamefont {Louie}}, \bibinfo {author} {\bibfnamefont {S.-K.}\ \bibnamefont {Mo}},\ and\ \bibinfo {author} {\bibfnamefont {M.~F.}\ \bibnamefont {Crommie}},\ }\bibfield  {title} {\bibinfo {title} {Strong correlations and orbital texture in single-layer {{1T-TaSe}}$_2$},\ }\href {https://doi.org/10.1038/s41567-019-0744-9} {\bibfield  {journal} {\bibinfo  {journal} {Nature Physics}\ }\textbf {\bibinfo {volume} {16}},\ \bibinfo {pages} {218} (\bibinfo {year} {2020})}\BibitemShut {NoStop}%
\bibitem [{\citenamefont {Nakata}\ \emph {et~al.}(2021)\citenamefont {Nakata}, \citenamefont {Sugawara}, \citenamefont {Chainani}, \citenamefont {Oka}, \citenamefont {Bao}, \citenamefont {Zhou}, \citenamefont {Chuang}, \citenamefont {Cheng}, \citenamefont {Kawakami}, \citenamefont {Saruta}, \citenamefont {Fukumura}, \citenamefont {Zhou}, \citenamefont {Takahashi},\ and\ \citenamefont {Sato}}]{nakata2021}%
  \BibitemOpen
  \bibfield  {author} {\bibinfo {author} {\bibfnamefont {Y.}~\bibnamefont {Nakata}}, \bibinfo {author} {\bibfnamefont {K.}~\bibnamefont {Sugawara}}, \bibinfo {author} {\bibfnamefont {A.}~\bibnamefont {Chainani}}, \bibinfo {author} {\bibfnamefont {H.}~\bibnamefont {Oka}}, \bibinfo {author} {\bibfnamefont {C.}~\bibnamefont {Bao}}, \bibinfo {author} {\bibfnamefont {S.}~\bibnamefont {Zhou}}, \bibinfo {author} {\bibfnamefont {P.-Y.}\ \bibnamefont {Chuang}}, \bibinfo {author} {\bibfnamefont {C.-M.}\ \bibnamefont {Cheng}}, \bibinfo {author} {\bibfnamefont {T.}~\bibnamefont {Kawakami}}, \bibinfo {author} {\bibfnamefont {Y.}~\bibnamefont {Saruta}}, \bibinfo {author} {\bibfnamefont {T.}~\bibnamefont {Fukumura}}, \bibinfo {author} {\bibfnamefont {S.}~\bibnamefont {Zhou}}, \bibinfo {author} {\bibfnamefont {T.}~\bibnamefont {Takahashi}},\ and\ \bibinfo {author} {\bibfnamefont {T.}~\bibnamefont {Sato}},\ }\bibfield  {title} {\bibinfo {title} {Robust charge-density wave strengthened by electron correlations in monolayer
  {{1T}}-{{TaSe}}$_2$ and {{1T}}-{{NbSe}}$_2$},\ }\href {https://doi.org/10.1038/s41467-021-26105-1} {\bibfield  {journal} {\bibinfo  {journal} {Nature Communications}\ }\textbf {\bibinfo {volume} {12}},\ \bibinfo {pages} {5873} (\bibinfo {year} {2021})}\BibitemShut {NoStop}%
\bibitem [{\citenamefont {Ritschel}\ \emph {et~al.}(2015)\citenamefont {Ritschel}, \citenamefont {Trinckauf}, \citenamefont {Koepernik}, \citenamefont {B{\"u}chner}, \citenamefont {v~Zimmermann}, \citenamefont {Berger}, \citenamefont {Joe}, \citenamefont {Abbamonte},\ and\ \citenamefont {Geck}}]{ritschel2015}%
  \BibitemOpen
  \bibfield  {author} {\bibinfo {author} {\bibfnamefont {T.}~\bibnamefont {Ritschel}}, \bibinfo {author} {\bibfnamefont {J.}~\bibnamefont {Trinckauf}}, \bibinfo {author} {\bibfnamefont {K.}~\bibnamefont {Koepernik}}, \bibinfo {author} {\bibfnamefont {B.}~\bibnamefont {B{\"u}chner}}, \bibinfo {author} {\bibfnamefont {M.}~\bibnamefont {v~Zimmermann}}, \bibinfo {author} {\bibfnamefont {H.}~\bibnamefont {Berger}}, \bibinfo {author} {\bibfnamefont {Y.~I.}\ \bibnamefont {Joe}}, \bibinfo {author} {\bibfnamefont {P.}~\bibnamefont {Abbamonte}},\ and\ \bibinfo {author} {\bibfnamefont {J.}~\bibnamefont {Geck}},\ }\bibfield  {title} {\bibinfo {title} {Orbital textures and charge density waves in transition metal dichalcogenides},\ }\href {https://doi.org/10.1038/nphys3267} {\bibfield  {journal} {\bibinfo  {journal} {Nature Physics}\ }\textbf {\bibinfo {volume} {11}},\ \bibinfo {pages} {328} (\bibinfo {year} {2015})}\BibitemShut {NoStop}%
\bibitem [{\citenamefont {Petocchi}\ \emph {et~al.}(2022)\citenamefont {Petocchi}, \citenamefont {Nicholson}, \citenamefont {Salzmann}, \citenamefont {Pasquier}, \citenamefont {Yazyev}, \citenamefont {Monney},\ and\ \citenamefont {Werner}}]{Petocchi2022}%
  \BibitemOpen
  \bibfield  {author} {\bibinfo {author} {\bibfnamefont {F.}~\bibnamefont {Petocchi}}, \bibinfo {author} {\bibfnamefont {C.~W.}\ \bibnamefont {Nicholson}}, \bibinfo {author} {\bibfnamefont {B.}~\bibnamefont {Salzmann}}, \bibinfo {author} {\bibfnamefont {D.}~\bibnamefont {Pasquier}}, \bibinfo {author} {\bibfnamefont {O.~V.}\ \bibnamefont {Yazyev}}, \bibinfo {author} {\bibfnamefont {C.}~\bibnamefont {Monney}},\ and\ \bibinfo {author} {\bibfnamefont {P.}~\bibnamefont {Werner}},\ }\bibfield  {title} {\bibinfo {title} {Mott versus hybridization gap in the low-temperature phase of {{1T}}-{{TaS}}$_2$},\ }\href {https://doi.org/10.1103/PhysRevLett.129.016402} {\bibfield  {journal} {\bibinfo  {journal} {Physical Review Letters}\ }\textbf {\bibinfo {volume} {129}},\ \bibinfo {pages} {016402} (\bibinfo {year} {2022})}\BibitemShut {NoStop}%
\bibitem [{\citenamefont {Lindroos}\ and\ \citenamefont {Bansil}(1996)}]{Lindroos1996}%
  \BibitemOpen
  \bibfield  {author} {\bibinfo {author} {\bibfnamefont {M.}~\bibnamefont {Lindroos}}\ and\ \bibinfo {author} {\bibfnamefont {A.}~\bibnamefont {Bansil}},\ }\bibfield  {title} {\bibinfo {title} {A novel direct method of {{Fermi}} surface determination using constant initial energy angle-scanned photoemission spectroscopy},\ }\href {https://doi.org/10.1103/PhysRevLett.77.2985} {\bibfield  {journal} {\bibinfo  {journal} {Physical Review Letters}\ }\textbf {\bibinfo {volume} {77}},\ \bibinfo {pages} {2985} (\bibinfo {year} {1996})}\BibitemShut {NoStop}%
\bibitem [{\citenamefont {Ekahana}\ \emph {et~al.}(2023)\citenamefont {Ekahana}, \citenamefont {Winata}, \citenamefont {Soh}, \citenamefont {Tamai}, \citenamefont {Milan}, \citenamefont {Aeppli},\ and\ \citenamefont {Shi}}]{ekahana2023}%
  \BibitemOpen
  \bibfield  {author} {\bibinfo {author} {\bibfnamefont {S.~A.}\ \bibnamefont {Ekahana}}, \bibinfo {author} {\bibfnamefont {G.~I.}\ \bibnamefont {Winata}}, \bibinfo {author} {\bibfnamefont {Y.}~\bibnamefont {Soh}}, \bibinfo {author} {\bibfnamefont {A.}~\bibnamefont {Tamai}}, \bibinfo {author} {\bibfnamefont {R.}~\bibnamefont {Milan}}, \bibinfo {author} {\bibfnamefont {G.}~\bibnamefont {Aeppli}},\ and\ \bibinfo {author} {\bibfnamefont {M.}~\bibnamefont {Shi}},\ }\bibfield  {title} {\bibinfo {title} {Transfer learning application of self-supervised learning in {{ARPES}}},\ }\href {https://doi.org/10.1088/2632-2153/aced7d} {\bibfield  {journal} {\bibinfo  {journal} {Machine Learning: Science and Technology}\ }\textbf {\bibinfo {volume} {4}},\ \bibinfo {pages} {035021} (\bibinfo {year} {2023})}\BibitemShut {NoStop}%
\bibitem [{\citenamefont {Jin}\ \emph {et~al.}(2024)\citenamefont {Jin}, \citenamefont {Ren}, \citenamefont {Tan}, \citenamefont {Xie}, \citenamefont {Lu}, \citenamefont {Zhang}, \citenamefont {Ji},\ and\ \citenamefont {Zhang}}]{jin2024}%
  \BibitemOpen
  \bibfield  {author} {\bibinfo {author} {\bibfnamefont {F.}~\bibnamefont {Jin}}, \bibinfo {author} {\bibfnamefont {W.}~\bibnamefont {Ren}}, \bibinfo {author} {\bibfnamefont {M.}~\bibnamefont {Tan}}, \bibinfo {author} {\bibfnamefont {M.}~\bibnamefont {Xie}}, \bibinfo {author} {\bibfnamefont {B.}~\bibnamefont {Lu}}, \bibinfo {author} {\bibfnamefont {Z.}~\bibnamefont {Zhang}}, \bibinfo {author} {\bibfnamefont {J.}~\bibnamefont {Ji}},\ and\ \bibinfo {author} {\bibfnamefont {Q.}~\bibnamefont {Zhang}},\ }\bibfield  {title} {\bibinfo {title} {$\pi$ {{Phase Interlayer Shift}} and {{Stacking Fault}} in the {{Kagome Superconductor}} {{CsV}}$_3${{Sb}}$_5$},\ }\href {https://doi.org/10.1103/PhysRevLett.132.066501} {\bibfield  {journal} {\bibinfo  {journal} {Physical Review Letters}\ }\textbf {\bibinfo {volume} {132}},\ \bibinfo {pages} {066501} (\bibinfo {year} {2024})}\BibitemShut {NoStop}%
\bibitem [{\citenamefont {Watson}\ \emph {et~al.}(2024)\citenamefont {Watson}, \citenamefont {Date}, \citenamefont {Louat},\ and\ \citenamefont {Schr\"oter}}]{watson2024}%
  \BibitemOpen
  \bibfield  {author} {\bibinfo {author} {\bibfnamefont {M.~D.}\ \bibnamefont {Watson}}, \bibinfo {author} {\bibfnamefont {M.}~\bibnamefont {Date}}, \bibinfo {author} {\bibfnamefont {A.}~\bibnamefont {Louat}},\ and\ \bibinfo {author} {\bibfnamefont {N.~B.~M.}\ \bibnamefont {Schr\"oter}},\ }\bibfield  {title} {\bibinfo {title} {Novel electronic structures from anomalous stackings in {{NbS}}$_2$ and {{MoS}}$_2$},\ }\href {https://doi.org/10.1103/PhysRevB.110.L121121} {\bibfield  {journal} {\bibinfo  {journal} {Physical Review B}\ }\textbf {\bibinfo {volume} {110}},\ \bibinfo {pages} {L121121} (\bibinfo {year} {2024})}\BibitemShut {NoStop}%
\end{thebibliography}

\begin{thebibliography}{56}%
\makeatletter
\providecommand \@ifxundefined [1]{%
 \@ifx{#1\undefined}
}%
\providecommand \@ifnum [1]{%
 \ifnum #1\expandafter \@firstoftwo
 \else \expandafter \@secondoftwo
 \fi
}%
\providecommand \@ifx [1]{%
 \ifx #1\expandafter \@firstoftwo
 \else \expandafter \@secondoftwo
 \fi
}%
\providecommand \natexlab [1]{#1}%
\providecommand \enquote  [1]{``#1''}%
\providecommand \bibnamefont  [1]{#1}%
\providecommand \bibfnamefont [1]{#1}%
\providecommand \citenamefont [1]{#1}%
\providecommand \href@noop [0]{\@secondoftwo}%
\providecommand \href [0]{\begingroup \@sanitize@url \@href}%
\providecommand \@href[1]{\@@startlink{#1}\@@href}%
\providecommand \@@href[1]{\endgroup#1\@@endlink}%
\providecommand \@sanitize@url [0]{\catcode `\\12\catcode `\$12\catcode `\&12\catcode `\#12\catcode `\^12\catcode `\_12\catcode `\%12\relax}%
\providecommand \@@startlink[1]{}%
\providecommand \@@endlink[0]{}%
\providecommand \url  [0]{\begingroup\@sanitize@url \@url }%
\providecommand \@url [1]{\endgroup\@href {#1}{\urlprefix }}%
\providecommand \urlprefix  [0]{URL }%
\providecommand \Eprint [0]{\href }%
\providecommand \doibase [0]{https://doi.org/}%
\providecommand \selectlanguage [0]{\@gobble}%
\providecommand \bibinfo  [0]{\@secondoftwo}%
\providecommand \bibfield  [0]{\@secondoftwo}%
\providecommand \translation [1]{[#1]}%
\providecommand \BibitemOpen [0]{}%
\providecommand \bibitemStop [0]{}%
\providecommand \bibitemNoStop [0]{.\EOS\space}%
\providecommand \EOS [0]{\spacefactor3000\relax}%
\providecommand \BibitemShut  [1]{\csname bibitem#1\endcsname}%
\let\auto@bib@innerbib\@empty
\bibitem [{\citenamefont {Di~Salvo}\ \emph {et~al.}(1974)\citenamefont {Di~Salvo}, \citenamefont {Maines}, \citenamefont {Waszczak},\ and\ \citenamefont {Schwall}}]{disalvo1974SM}%
  \BibitemOpen
  \bibfield  {author} {\bibinfo {author} {\bibfnamefont {F.~J.}\ \bibnamefont {Di~Salvo}}, \bibinfo {author} {\bibfnamefont {R.~G.}\ \bibnamefont {Maines}}, \bibinfo {author} {\bibfnamefont {J.~V.}\ \bibnamefont {Waszczak}},\ and\ \bibinfo {author} {\bibfnamefont {R.~E.}\ \bibnamefont {Schwall}},\ }\bibfield  {title} {\bibinfo {title} {Preparation and properties of {{1T}}-{{TaSe}}$_2$},\ }\href {https://doi.org/10.1016/0038-1098(74)90975-2} {\bibfield  {journal} {\bibinfo  {journal} {Solid State Communications}\ }\textbf {\bibinfo {volume} {14}},\ \bibinfo {pages} {497} (\bibinfo {year} {1974})}\BibitemShut {NoStop}%
\bibitem [{\citenamefont {Polley}\ \emph {et~al.}(2024)\citenamefont {Polley}, \citenamefont {Leandersson}, \citenamefont {Adell}, \citenamefont {Osiecki}, \citenamefont {Carbone}, \citenamefont {Ali}, \citenamefont {Fedderwitz},\ and\ \citenamefont {Balasubramanian}}]{Polley2024aa}%
  \BibitemOpen
  \bibfield  {author} {\bibinfo {author} {\bibfnamefont {C.~M.}\ \bibnamefont {Polley}}, \bibinfo {author} {\bibfnamefont {M.}~\bibnamefont {Leandersson}}, \bibinfo {author} {\bibfnamefont {J.}~\bibnamefont {Adell}}, \bibinfo {author} {\bibfnamefont {J.}~\bibnamefont {Osiecki}}, \bibinfo {author} {\bibfnamefont {D.}~\bibnamefont {Carbone}}, \bibinfo {author} {\bibfnamefont {K.}~\bibnamefont {Ali}}, \bibinfo {author} {\bibfnamefont {H.}~\bibnamefont {Fedderwitz}},\ and\ \bibinfo {author} {\bibfnamefont {T.}~\bibnamefont {Balasubramanian}},\ }\bibfield  {title} {\bibinfo {title} {The {{Bloch}} beamline at {{MAX IV}}: Micro-spot {{ARPES}} from a conventional, full-featured beamline},\ }\href {https://doi.org/10.1080/08940886.2024.2391252} {\bibfield  {journal} {\bibinfo  {journal} {Synchrotron Radiation News}\ }\textbf {\bibinfo {volume} {37}},\ \bibinfo {pages} {18} (\bibinfo {year} {2024})}\BibitemShut {NoStop}%
\bibitem [{\citenamefont {Cucchi}\ \emph {et~al.}(2019)\citenamefont {Cucchi}, \citenamefont {Guti{\'e}rrez-Lezama}, \citenamefont {Cappelli}, \citenamefont {McKeown~Walker}, \citenamefont {Bruno}, \citenamefont {Tenasini}, \citenamefont {Wang}, \citenamefont {Ubrig}, \citenamefont {Barreteau}, \citenamefont {Giannini}, \citenamefont {Gibertini}, \citenamefont {Tamai}, \citenamefont {Morpurgo},\ and\ \citenamefont {Baumberger}}]{cucchi2018}%
  \BibitemOpen
  \bibfield  {author} {\bibinfo {author} {\bibfnamefont {I.}~\bibnamefont {Cucchi}}, \bibinfo {author} {\bibfnamefont {I.}~\bibnamefont {Guti{\'e}rrez-Lezama}}, \bibinfo {author} {\bibfnamefont {E.}~\bibnamefont {Cappelli}}, \bibinfo {author} {\bibfnamefont {S.}~\bibnamefont {McKeown~Walker}}, \bibinfo {author} {\bibfnamefont {F.~Y.}\ \bibnamefont {Bruno}}, \bibinfo {author} {\bibfnamefont {G.}~\bibnamefont {Tenasini}}, \bibinfo {author} {\bibfnamefont {L.}~\bibnamefont {Wang}}, \bibinfo {author} {\bibfnamefont {N.}~\bibnamefont {Ubrig}}, \bibinfo {author} {\bibfnamefont {C.}~\bibnamefont {Barreteau}}, \bibinfo {author} {\bibfnamefont {E.}~\bibnamefont {Giannini}}, \bibinfo {author} {\bibfnamefont {M.}~\bibnamefont {Gibertini}}, \bibinfo {author} {\bibfnamefont {A.}~\bibnamefont {Tamai}}, \bibinfo {author} {\bibfnamefont {A.~F.}\ \bibnamefont {Morpurgo}},\ and\ \bibinfo {author} {\bibfnamefont {F.}~\bibnamefont {Baumberger}},\ }\bibfield  {title} {\bibinfo {title} {Microfocus laser--angle-resolved
  photoemission on encapsulated mono-, bi-, and few-layer {{1T}}'-{{WTe}}$_2$},\ }\href {https://doi.org/10.1021/acs.nanolett.8b04534} {\bibfield  {journal} {\bibinfo  {journal} {Nano Letters}\ }\textbf {\bibinfo {volume} {19}},\ \bibinfo {pages} {554} (\bibinfo {year} {2019})},\ \Eprint {https://arxiv.org/abs/https://doi.org/10.1021/acs.nanolett.8b04534} {https://doi.org/10.1021/acs.nanolett.8b04534} \BibitemShut {NoStop}%
\bibitem [{\citenamefont {Giannozzi}\ \emph {et~al.}(2009)\citenamefont {Giannozzi}, \citenamefont {Baroni}, \citenamefont {Bonini}, \citenamefont {Calandra}, \citenamefont {Car}, \citenamefont {Cavazzoni}, \citenamefont {Ceresoli}, \citenamefont {Chiarotti}, \citenamefont {Cococcioni}, \citenamefont {Dabo}, \citenamefont {Corso}, \citenamefont {de~Gironcoli}, \citenamefont {Fabris}, \citenamefont {Fratesi}, \citenamefont {Gebauer}, \citenamefont {Gerstmann}, \citenamefont {Gougoussis}, \citenamefont {Kokalj}, \citenamefont {Lazzeri}, \citenamefont {Martin-Samos}, \citenamefont {Marzari}, \citenamefont {Mauri}, \citenamefont {Mazzarello}, \citenamefont {Paolini}, \citenamefont {Pasquarello}, \citenamefont {Paulatto}, \citenamefont {Sbraccia}, \citenamefont {Scandolo}, \citenamefont {Sclauzero}, \citenamefont {Seitsonen}, \citenamefont {Smogunov}, \citenamefont {Umari},\ and\ \citenamefont {Wentzcovitch}}]{QEref}%
  \BibitemOpen
  \bibfield  {author} {\bibinfo {author} {\bibfnamefont {P.}~\bibnamefont {Giannozzi}}, \bibinfo {author} {\bibfnamefont {S.}~\bibnamefont {Baroni}}, \bibinfo {author} {\bibfnamefont {N.}~\bibnamefont {Bonini}}, \bibinfo {author} {\bibfnamefont {M.}~\bibnamefont {Calandra}}, \bibinfo {author} {\bibfnamefont {R.}~\bibnamefont {Car}}, \bibinfo {author} {\bibfnamefont {C.}~\bibnamefont {Cavazzoni}}, \bibinfo {author} {\bibfnamefont {D.}~\bibnamefont {Ceresoli}}, \bibinfo {author} {\bibfnamefont {G.~L.}\ \bibnamefont {Chiarotti}}, \bibinfo {author} {\bibfnamefont {M.}~\bibnamefont {Cococcioni}}, \bibinfo {author} {\bibfnamefont {I.}~\bibnamefont {Dabo}}, \bibinfo {author} {\bibfnamefont {A.~D.}\ \bibnamefont {Corso}}, \bibinfo {author} {\bibfnamefont {S.}~\bibnamefont {de~Gironcoli}}, \bibinfo {author} {\bibfnamefont {S.}~\bibnamefont {Fabris}}, \bibinfo {author} {\bibfnamefont {G.}~\bibnamefont {Fratesi}}, \bibinfo {author} {\bibfnamefont {R.}~\bibnamefont {Gebauer}}, \bibinfo {author} {\bibfnamefont
  {U.}~\bibnamefont {Gerstmann}}, \bibinfo {author} {\bibfnamefont {C.}~\bibnamefont {Gougoussis}}, \bibinfo {author} {\bibfnamefont {A.}~\bibnamefont {Kokalj}}, \bibinfo {author} {\bibfnamefont {M.}~\bibnamefont {Lazzeri}}, \bibinfo {author} {\bibfnamefont {L.}~\bibnamefont {Martin-Samos}}, \bibinfo {author} {\bibfnamefont {N.}~\bibnamefont {Marzari}}, \bibinfo {author} {\bibfnamefont {F.}~\bibnamefont {Mauri}}, \bibinfo {author} {\bibfnamefont {R.}~\bibnamefont {Mazzarello}}, \bibinfo {author} {\bibfnamefont {S.}~\bibnamefont {Paolini}}, \bibinfo {author} {\bibfnamefont {A.}~\bibnamefont {Pasquarello}}, \bibinfo {author} {\bibfnamefont {L.}~\bibnamefont {Paulatto}}, \bibinfo {author} {\bibfnamefont {C.}~\bibnamefont {Sbraccia}}, \bibinfo {author} {\bibfnamefont {S.}~\bibnamefont {Scandolo}}, \bibinfo {author} {\bibfnamefont {G.}~\bibnamefont {Sclauzero}}, \bibinfo {author} {\bibfnamefont {A.~P.}\ \bibnamefont {Seitsonen}}, \bibinfo {author} {\bibfnamefont {A.}~\bibnamefont {Smogunov}}, \bibinfo {author}
  {\bibfnamefont {P.}~\bibnamefont {Umari}},\ and\ \bibinfo {author} {\bibfnamefont {R.~M.}\ \bibnamefont {Wentzcovitch}},\ }\bibfield  {title} {\bibinfo {title} {{{QUANTUM ESPRESSO}}: a modular and open-source software project for quantum simulations of materials},\ }\href {https://doi.org/10.1088/0953-8984/21/39/395502} {\bibfield  {journal} {\bibinfo  {journal} {Journal of Physics: Condensed Matter}\ }\textbf {\bibinfo {volume} {21}},\ \bibinfo {pages} {395502} (\bibinfo {year} {2009})}\BibitemShut {NoStop}%
\bibitem [{\citenamefont {Perdew}\ \emph {et~al.}(1996)\citenamefont {Perdew}, \citenamefont {Burke},\ and\ \citenamefont {Ernzerhof}}]{GGAref}%
  \BibitemOpen
  \bibfield  {author} {\bibinfo {author} {\bibfnamefont {J.~P.}\ \bibnamefont {Perdew}}, \bibinfo {author} {\bibfnamefont {K.}~\bibnamefont {Burke}},\ and\ \bibinfo {author} {\bibfnamefont {M.}~\bibnamefont {Ernzerhof}},\ }\bibfield  {title} {\bibinfo {title} {Generalized gradient approximation made simple},\ }\href {https://doi.org/10.1103/PhysRevLett.77.3865} {\bibfield  {journal} {\bibinfo  {journal} {Physical Review Letters}\ }\textbf {\bibinfo {volume} {77}},\ \bibinfo {pages} {3865} (\bibinfo {year} {1996})}\BibitemShut {NoStop}%
\bibitem [{\citenamefont {Zhang}\ \emph {et~al.}(2020)\citenamefont {Zhang}, \citenamefont {Si}, \citenamefont {Lian}, \citenamefont {Zhou},\ and\ \citenamefont {Sun}}]{KangZhang2020SM}%
  \BibitemOpen
  \bibfield  {author} {\bibinfo {author} {\bibfnamefont {K.}~\bibnamefont {Zhang}}, \bibinfo {author} {\bibfnamefont {C.}~\bibnamefont {Si}}, \bibinfo {author} {\bibfnamefont {C.-S.}\ \bibnamefont {Lian}}, \bibinfo {author} {\bibfnamefont {J.}~\bibnamefont {Zhou}},\ and\ \bibinfo {author} {\bibfnamefont {Z.}~\bibnamefont {Sun}},\ }\bibfield  {title} {\bibinfo {title} {{Mottness collapse in monolayer 1T-TaSe$_2$ with persisting charge density wave order}},\ }\href {https://doi.org/10.1039/d0tc01719a} {\bibfield  {journal} {\bibinfo  {journal} {Journal of Materials Chemistry C}\ }\textbf {\bibinfo {volume} {8}},\ \bibinfo {pages} {9742} (\bibinfo {year} {2020})}\BibitemShut {NoStop}%
\bibitem [{\citenamefont {Mostofi}\ \emph {et~al.}(2014)\citenamefont {Mostofi}, \citenamefont {Yates}, \citenamefont {Pizzi}, \citenamefont {Lee}, \citenamefont {Souza}, \citenamefont {Vanderbilt},\ and\ \citenamefont {Marzari}}]{W90ref}%
  \BibitemOpen
  \bibfield  {author} {\bibinfo {author} {\bibfnamefont {A.~A.}\ \bibnamefont {Mostofi}}, \bibinfo {author} {\bibfnamefont {J.~R.}\ \bibnamefont {Yates}}, \bibinfo {author} {\bibfnamefont {G.}~\bibnamefont {Pizzi}}, \bibinfo {author} {\bibfnamefont {Y.-S.}\ \bibnamefont {Lee}}, \bibinfo {author} {\bibfnamefont {I.}~\bibnamefont {Souza}}, \bibinfo {author} {\bibfnamefont {D.}~\bibnamefont {Vanderbilt}},\ and\ \bibinfo {author} {\bibfnamefont {N.}~\bibnamefont {Marzari}},\ }\bibfield  {title} {\bibinfo {title} {An updated version of {{wannier90}}: A tool for obtaining maximally-localised {{Wannier}} functions},\ }\href {https://doi.org/https://doi.org/10.1016/j.cpc.2014.05.003} {\bibfield  {journal} {\bibinfo  {journal} {Computer Physics Communications}\ }\textbf {\bibinfo {volume} {185}},\ \bibinfo {pages} {2309} (\bibinfo {year} {2014})}\BibitemShut {NoStop}%
\bibitem [{\citenamefont {Nakamura}\ \emph {et~al.}(2021)\citenamefont {Nakamura}, \citenamefont {Yoshimoto}, \citenamefont {Nomura}, \citenamefont {Tadano}, \citenamefont {Kawamura}, \citenamefont {Kosugi}, \citenamefont {Yoshimi}, \citenamefont {Misawa},\ and\ \citenamefont {Motoyama}}]{RespackRef}%
  \BibitemOpen
  \bibfield  {author} {\bibinfo {author} {\bibfnamefont {K.}~\bibnamefont {Nakamura}}, \bibinfo {author} {\bibfnamefont {Y.}~\bibnamefont {Yoshimoto}}, \bibinfo {author} {\bibfnamefont {Y.}~\bibnamefont {Nomura}}, \bibinfo {author} {\bibfnamefont {T.}~\bibnamefont {Tadano}}, \bibinfo {author} {\bibfnamefont {M.}~\bibnamefont {Kawamura}}, \bibinfo {author} {\bibfnamefont {T.}~\bibnamefont {Kosugi}}, \bibinfo {author} {\bibfnamefont {K.}~\bibnamefont {Yoshimi}}, \bibinfo {author} {\bibfnamefont {T.}~\bibnamefont {Misawa}},\ and\ \bibinfo {author} {\bibfnamefont {Y.}~\bibnamefont {Motoyama}},\ }\bibfield  {title} {\bibinfo {title} {{{RESPACK}}: An ab initio tool for derivation of effective low-energy model of material},\ }\href {https://doi.org/https://doi.org/10.1016/j.cpc.2020.107781} {\bibfield  {journal} {\bibinfo  {journal} {Computer Physics Communications}\ }\textbf {\bibinfo {volume} {261}},\ \bibinfo {pages} {107781} (\bibinfo {year} {2021})}\BibitemShut {NoStop}%
\bibitem [{\citenamefont {Tian}\ \emph {et~al.}(2023)\citenamefont {Tian}, \citenamefont {Huang}, \citenamefont {Jang}, \citenamefont {Guo}, \citenamefont {Yan}, \citenamefont {Gao}, \citenamefont {Yu}, \citenamefont {Hwang}, \citenamefont {Tang}, \citenamefont {Wang}, \citenamefont {Luo}, \citenamefont {Sun}, \citenamefont {Liu}, \citenamefont {Feng}, \citenamefont {Chen}, \citenamefont {Mo}, \citenamefont {Kim}, \citenamefont {Son}, \citenamefont {Shen}, \citenamefont {Ruan},\ and\ \citenamefont {Zhang}}]{tian2023SM}%
  \BibitemOpen
  \bibfield  {author} {\bibinfo {author} {\bibfnamefont {N.}~\bibnamefont {Tian}}, \bibinfo {author} {\bibfnamefont {Z.}~\bibnamefont {Huang}}, \bibinfo {author} {\bibfnamefont {B.~G.}\ \bibnamefont {Jang}}, \bibinfo {author} {\bibfnamefont {S.}~\bibnamefont {Guo}}, \bibinfo {author} {\bibfnamefont {Y.-J.}\ \bibnamefont {Yan}}, \bibinfo {author} {\bibfnamefont {J.}~\bibnamefont {Gao}}, \bibinfo {author} {\bibfnamefont {Y.}~\bibnamefont {Yu}}, \bibinfo {author} {\bibfnamefont {J.}~\bibnamefont {Hwang}}, \bibinfo {author} {\bibfnamefont {C.}~\bibnamefont {Tang}}, \bibinfo {author} {\bibfnamefont {M.}~\bibnamefont {Wang}}, \bibinfo {author} {\bibfnamefont {X.}~\bibnamefont {Luo}}, \bibinfo {author} {\bibfnamefont {Y.~P.}\ \bibnamefont {Sun}}, \bibinfo {author} {\bibfnamefont {Z.}~\bibnamefont {Liu}}, \bibinfo {author} {\bibfnamefont {D.-L.}\ \bibnamefont {Feng}}, \bibinfo {author} {\bibfnamefont {X.}~\bibnamefont {Chen}}, \bibinfo {author} {\bibfnamefont {S.-K.}\ \bibnamefont {Mo}}, \bibinfo {author}
  {\bibfnamefont {M.}~\bibnamefont {Kim}}, \bibinfo {author} {\bibfnamefont {Y.-W.}\ \bibnamefont {Son}}, \bibinfo {author} {\bibfnamefont {D.}~\bibnamefont {Shen}}, \bibinfo {author} {\bibfnamefont {W.}~\bibnamefont {Ruan}},\ and\ \bibinfo {author} {\bibfnamefont {Y.}~\bibnamefont {Zhang}},\ }\bibfield  {title} {\bibinfo {title} {Dimensionality-driven metal to {{Mott}} insulator transition in two-dimensional {{1T-TaSe}}$_2$},\ }\href {https://doi.org/10.1093/nsr/nwad144} {\bibfield  {journal} {\bibinfo  {journal} {National Science Review}\ ,\ \bibinfo {pages} {nwad144}} (\bibinfo {year} {2023})}\BibitemShut {NoStop}%
\bibitem [{\citenamefont {Lin}\ \emph {et~al.}(2020)\citenamefont {Lin}, \citenamefont {Huang}, \citenamefont {Zhao}, \citenamefont {Qiao}, \citenamefont {Liu}, \citenamefont {Wu}, \citenamefont {Chen},\ and\ \citenamefont {Ji}}]{STM_ml}%
  \BibitemOpen
  \bibfield  {author} {\bibinfo {author} {\bibfnamefont {H.}~\bibnamefont {Lin}}, \bibinfo {author} {\bibfnamefont {W.}~\bibnamefont {Huang}}, \bibinfo {author} {\bibfnamefont {K.}~\bibnamefont {Zhao}}, \bibinfo {author} {\bibfnamefont {S.}~\bibnamefont {Qiao}}, \bibinfo {author} {\bibfnamefont {Z.}~\bibnamefont {Liu}}, \bibinfo {author} {\bibfnamefont {J.}~\bibnamefont {Wu}}, \bibinfo {author} {\bibfnamefont {X.}~\bibnamefont {Chen}},\ and\ \bibinfo {author} {\bibfnamefont {S.-H.}\ \bibnamefont {Ji}},\ }\bibfield  {title} {\bibinfo {title} {Scanning tunneling spectroscopic study of monolayer {{1T}}-{{TaS}}$_2$ and {{1T}}-{{TaSe}}$_2$},\ }\href {https://doi.org/10.1007/s12274-019-2584-4} {\bibfield  {journal} {\bibinfo  {journal} {Nano Research}\ }\textbf {\bibinfo {volume} {13}},\ \bibinfo {pages} {133} (\bibinfo {year} {2020})}\BibitemShut {NoStop}%
\bibitem [{\citenamefont {Petocchi}\ \emph {et~al.}(2022)\citenamefont {Petocchi}, \citenamefont {Nicholson}, \citenamefont {Salzmann}, \citenamefont {Pasquier}, \citenamefont {Yazyev}, \citenamefont {Monney},\ and\ \citenamefont {Werner}}]{Petocchi2022SM}%
  \BibitemOpen
  \bibfield  {author} {\bibinfo {author} {\bibfnamefont {F.}~\bibnamefont {Petocchi}}, \bibinfo {author} {\bibfnamefont {C.~W.}\ \bibnamefont {Nicholson}}, \bibinfo {author} {\bibfnamefont {B.}~\bibnamefont {Salzmann}}, \bibinfo {author} {\bibfnamefont {D.}~\bibnamefont {Pasquier}}, \bibinfo {author} {\bibfnamefont {O.~V.}\ \bibnamefont {Yazyev}}, \bibinfo {author} {\bibfnamefont {C.}~\bibnamefont {Monney}},\ and\ \bibinfo {author} {\bibfnamefont {P.}~\bibnamefont {Werner}},\ }\bibfield  {title} {\bibinfo {title} {Mott versus hybridization gap in the low-temperature phase of {{1T}}-{{TaS}}$_2$},\ }\href {https://doi.org/10.1103/PhysRevLett.129.016402} {\bibfield  {journal} {\bibinfo  {journal} {Physical Review Letters}\ }\textbf {\bibinfo {volume} {129}},\ \bibinfo {pages} {016402} (\bibinfo {year} {2022})}\BibitemShut {NoStop}%
\bibitem [{\citenamefont {Perfetti}\ \emph {et~al.}(2003)\citenamefont {Perfetti}, \citenamefont {Georges}, \citenamefont {Florens}, \citenamefont {Biermann}, \citenamefont {Mitrovic}, \citenamefont {Berger}, \citenamefont {Tomm}, \citenamefont {H{\"o}chst},\ and\ \citenamefont {Grioni}}]{perfetti2003SM}%
  \BibitemOpen
  \bibfield  {author} {\bibinfo {author} {\bibfnamefont {L.}~\bibnamefont {Perfetti}}, \bibinfo {author} {\bibfnamefont {A.}~\bibnamefont {Georges}}, \bibinfo {author} {\bibfnamefont {S.}~\bibnamefont {Florens}}, \bibinfo {author} {\bibfnamefont {S.}~\bibnamefont {Biermann}}, \bibinfo {author} {\bibfnamefont {S.}~\bibnamefont {Mitrovic}}, \bibinfo {author} {\bibfnamefont {H.}~\bibnamefont {Berger}}, \bibinfo {author} {\bibfnamefont {Y.}~\bibnamefont {Tomm}}, \bibinfo {author} {\bibfnamefont {H.}~\bibnamefont {H{\"o}chst}},\ and\ \bibinfo {author} {\bibfnamefont {M.}~\bibnamefont {Grioni}},\ }\bibfield  {title} {\bibinfo {title} {Spectroscopic {{Signatures}} of a {{Bandwidth-Controlled Mott Transition}} at the {{Surface}} of {{1T}}-{{TaSe}}$_2$},\ }\href {https://doi.org/10.1103/PhysRevLett.90.166401} {\bibfield  {journal} {\bibinfo  {journal} {Physical Review Letters}\ }\textbf {\bibinfo {volume} {90}},\ \bibinfo {pages} {166401} (\bibinfo {year} {2003})}\BibitemShut {NoStop}%
\bibitem [{\citenamefont {Chen}\ \emph {et~al.}(2020)\citenamefont {Chen}, \citenamefont {Ruan}, \citenamefont {Wu}, \citenamefont {Tang}, \citenamefont {Ryu}, \citenamefont {Tsai}, \citenamefont {Lee}, \citenamefont {Kahn}, \citenamefont {Liou}, \citenamefont {Jia}, \citenamefont {Albertini}, \citenamefont {Xiong}, \citenamefont {Jia}, \citenamefont {Liu}, \citenamefont {Sobota}, \citenamefont {Liu}, \citenamefont {Moore}, \citenamefont {Shen}, \citenamefont {Louie}, \citenamefont {Mo},\ and\ \citenamefont {Crommie}}]{chen2020SM}%
  \BibitemOpen
\bibfield  {author} {\bibinfo {author} {\bibfnamefont {Y.}~\bibnamefont {Chen}}, \bibinfo {author} {\bibfnamefont {W.}~\bibnamefont {Ruan}}, \bibinfo {author} {\bibfnamefont {M.}~\bibnamefont {Wu}}, \bibinfo {author} {\bibfnamefont {S.}~\bibnamefont {Tang}}, \bibinfo {author} {\bibfnamefont {H.}~\bibnamefont {Ryu}}, \bibinfo {author} {\bibfnamefont {H.-Z.}\ \bibnamefont {Tsai}}, \bibinfo {author} {\bibfnamefont {R.~L.}\ \bibnamefont {Lee}}, \bibinfo {author} {\bibfnamefont {S.}~\bibnamefont {Kahn}}, \bibinfo {author} {\bibfnamefont {F.}~\bibnamefont {Liou}}, \bibinfo {author} {\bibfnamefont {C.}~\bibnamefont {Jia}}, \bibinfo {author} {\bibfnamefont {O.~R.}\ \bibnamefont {Albertini}}, \bibinfo {author} {\bibfnamefont {H.}~\bibnamefont {Xiong}}, \bibinfo {author} {\bibfnamefont {T.}~\bibnamefont {Jia}}, \bibinfo {author} {\bibfnamefont {Z.}~\bibnamefont {Liu}}, \bibinfo {author} {\bibfnamefont {J.~A.}\ \bibnamefont {Sobota}}, \bibinfo {author} {\bibfnamefont {A.~Y.}\ \bibnamefont {Liu}}, \bibinfo {author}
  {\bibfnamefont {J.~E.}\ \bibnamefont {Moore}}, \bibinfo {author} {\bibfnamefont {Z.-X.}\ \bibnamefont {Shen}}, \bibinfo {author} {\bibfnamefont {S.~G.}\ \bibnamefont {Louie}}, \bibinfo {author} {\bibfnamefont {S.-K.}\ \bibnamefont {Mo}},\ and\ \bibinfo {author} {\bibfnamefont {M.~F.}\ \bibnamefont {Crommie}},\ }\bibfield  {title} {\bibinfo {title} {Strong correlations and orbital texture in single-layer {{1T-TaSe}}$_2$},\ }\href {https://doi.org/10.1038/s41567-019-0744-9} {\bibfield  {journal} {\bibinfo  {journal} {Nature Physics}\ }\textbf {\bibinfo {volume} {16}},\ \bibinfo {pages} {218} (\bibinfo {year} {2020})}\BibitemShut {NoStop}%
  \bibitem [{\citenamefont {Nakata}\ \emph {et~al.}(2021)\citenamefont {Nakata}, \citenamefont {Sugawara}, \citenamefont {Chainani}, \citenamefont {Oka}, \citenamefont {Bao}, \citenamefont {Zhou}, \citenamefont {Chuang}, \citenamefont {Cheng}, \citenamefont {Kawakami}, \citenamefont {Saruta}, \citenamefont {Fukumura}, \citenamefont {Zhou}, \citenamefont {Takahashi},\ and\ \citenamefont {Sato}}]{nakata2021SM}%
  \BibitemOpen
  \bibfield  {author} {\bibinfo {author} {\bibfnamefont {Y.}~\bibnamefont {Nakata}}, \bibinfo {author} {\bibfnamefont {K.}~\bibnamefont {Sugawara}}, \bibinfo {author} {\bibfnamefont {A.}~\bibnamefont {Chainani}}, \bibinfo {author} {\bibfnamefont {H.}~\bibnamefont {Oka}}, \bibinfo {author} {\bibfnamefont {C.}~\bibnamefont {Bao}}, \bibinfo {author} {\bibfnamefont {S.}~\bibnamefont {Zhou}}, \bibinfo {author} {\bibfnamefont {P.-Y.}\ \bibnamefont {Chuang}}, \bibinfo {author} {\bibfnamefont {C.-M.}\ \bibnamefont {Cheng}}, \bibinfo {author} {\bibfnamefont {T.}~\bibnamefont {Kawakami}}, \bibinfo {author} {\bibfnamefont {Y.}~\bibnamefont {Saruta}}, \bibinfo {author} {\bibfnamefont {T.}~\bibnamefont {Fukumura}}, \bibinfo {author} {\bibfnamefont {S.}~\bibnamefont {Zhou}}, \bibinfo {author} {\bibfnamefont {T.}~\bibnamefont {Takahashi}},\ and\ \bibinfo {author} {\bibfnamefont {T.}~\bibnamefont {Sato}},\ }\bibfield  {title} {\bibinfo {title} {Robust charge-density wave strengthened by electron correlations in monolayer
  {{1T}}-{{TaSe}}$_2$ and {{1T}}-{{NbSe}}$_2$},\ }\href {https://doi.org/10.1038/s41467-021-26105-1} {\bibfield  {journal} {\bibinfo  {journal} {Nature Communications}\ }\textbf {\bibinfo {volume} {12}},\ \bibinfo {pages} {5873} (\bibinfo {year} {2021})}\BibitemShut {NoStop}%
\bibitem [{\citenamefont {Fecher}\ \emph {et~al.}(2022)\citenamefont {Fecher}, \citenamefont {K{\"u}bler},\ and\ \citenamefont {Felser}}]{fecher2022}%
  \BibitemOpen
  \bibfield  {author} {\bibinfo {author} {\bibfnamefont {G.~H.}\ \bibnamefont {Fecher}}, \bibinfo {author} {\bibfnamefont {J.}~\bibnamefont {K{\"u}bler}},\ and\ \bibinfo {author} {\bibfnamefont {C.}~\bibnamefont {Felser}},\ }\bibfield  {title} {\bibinfo {title} {Chirality in the {{Solid State}}: {{Chiral Crystal Structures}} in {{Chiral}} and {{Achiral Space Groups}}},\ }\href {https://doi.org/10.3390/ma15175812} {\bibfield  {journal} {\bibinfo  {journal} {Materials}\ }\textbf {\bibinfo {volume} {15}},\ \bibinfo {pages} {5812} (\bibinfo {year} {2022})}\BibitemShut {NoStop}%
\bibitem [{\citenamefont {Louat}\ \emph {et~al.}(2024)\citenamefont {Louat}, \citenamefont {Watson}, \citenamefont {Kim}, \citenamefont {Ni}, \citenamefont {Cava},\ and\ \citenamefont {Cacho}}]{Louat2024aa}%
  \BibitemOpen
  \bibfield  {author} {\bibinfo {author} {\bibfnamefont {A.}~\bibnamefont {Louat}}, \bibinfo {author} {\bibfnamefont {M.~D.}\ \bibnamefont {Watson}}, \bibinfo {author} {\bibfnamefont {T.~K.}\ \bibnamefont {Kim}}, \bibinfo {author} {\bibfnamefont {D.}~\bibnamefont {Ni}}, \bibinfo {author} {\bibfnamefont {R.~J.}\ \bibnamefont {Cava}},\ and\ \bibinfo {author} {\bibfnamefont {C.}~\bibnamefont {Cacho}},\ }\bibfield  {title} {\bibinfo {title} {The pseudochiral {{Fermi}} surface of $\alpha$-{{RuI}}$_3$},\ }\href {https://doi.org/10.1038/s42005-024-01533-9} {\bibfield  {journal} {\bibinfo  {journal} {Communications Physics}\ }\textbf {\bibinfo {volume} {7}},\ \bibinfo {pages} {43} (\bibinfo {year} {2024})}\BibitemShut {NoStop}%
\bibitem [{\citenamefont {Margot}\ \emph {et~al.}(2023)\citenamefont {Margot}, \citenamefont {Lisi}, \citenamefont {Cucchi}, \citenamefont {Cappelli}, \citenamefont {Hunter}, \citenamefont {{Guti{\'e}rrez-Lezama}}, \citenamefont {Ma}, \citenamefont {{von Rohr}}, \citenamefont {Berthod}, \citenamefont {Petocchi}, \citenamefont {Ponc{\'e}}, \citenamefont {Marzari}, \citenamefont {Gibertini}, \citenamefont {Tamai}, \citenamefont {Morpurgo},\ and\ \citenamefont {Baumberger}}]{margot2023}%
  \BibitemOpen
  \bibfield  {author} {\bibinfo {author} {\bibfnamefont {F.}~\bibnamefont {Margot}}, \bibinfo {author} {\bibfnamefont {S.}~\bibnamefont {Lisi}}, \bibinfo {author} {\bibfnamefont {I.}~\bibnamefont {Cucchi}}, \bibinfo {author} {\bibfnamefont {E.}~\bibnamefont {Cappelli}}, \bibinfo {author} {\bibfnamefont {A.}~\bibnamefont {Hunter}}, \bibinfo {author} {\bibfnamefont {I.}~\bibnamefont {{Guti{\'e}rrez-Lezama}}}, \bibinfo {author} {\bibfnamefont {K.}~\bibnamefont {Ma}}, \bibinfo {author} {\bibfnamefont {F.}~\bibnamefont {{von Rohr}}}, \bibinfo {author} {\bibfnamefont {C.}~\bibnamefont {Berthod}}, \bibinfo {author} {\bibfnamefont {F.}~\bibnamefont {Petocchi}}, \bibinfo {author} {\bibfnamefont {S.}~\bibnamefont {Ponc{\'e}}}, \bibinfo {author} {\bibfnamefont {N.}~\bibnamefont {Marzari}}, \bibinfo {author} {\bibfnamefont {M.}~\bibnamefont {Gibertini}}, \bibinfo {author} {\bibfnamefont {A.}~\bibnamefont {Tamai}}, \bibinfo {author} {\bibfnamefont {A.~F.}\ \bibnamefont {Morpurgo}},\ and\ \bibinfo {author} {\bibfnamefont
  {F.}~\bibnamefont {Baumberger}},\ }\bibfield  {title} {\bibinfo {title} {Electronic {{Structure}} of {{Few-Layer Black Phosphorus}} from {$\mu$}-{{ARPES}}},\ }\href {https://doi.org/10.1021/acs.nanolett.3c01226} {\bibfield  {journal} {\bibinfo  {journal} {Nano Letters}\ }\textbf {\bibinfo {volume} {23}},\ \bibinfo {pages} {6433} (\bibinfo {year} {2023})}\BibitemShut {NoStop}%
\bibitem [{\citenamefont {Milun}\ \emph {et~al.}(2002)\citenamefont {Milun}, \citenamefont {Pervan},\ and\ \citenamefont {Woodruff}}]{milun2002}%
  \BibitemOpen
  \bibfield  {author} {\bibinfo {author} {\bibfnamefont {M.}~\bibnamefont {Milun}}, \bibinfo {author} {\bibfnamefont {P.}~\bibnamefont {Pervan}},\ and\ \bibinfo {author} {\bibfnamefont {D.~P.}\ \bibnamefont {Woodruff}},\ }\bibfield  {title} {\bibinfo {title} {Quantum well structures in thin metal films: Simple model physics in reality?},\ }\href {https://doi.org/10.1088/0034-4885/65/2/201} {\bibfield  {journal} {\bibinfo  {journal} {Reports on Progress in Physics}\ }\textbf {\bibinfo {volume} {65}},\ \bibinfo {pages} {99} (\bibinfo {year} {2002})}\BibitemShut {NoStop}%
\bibitem [{\citenamefont {Kawakami}\ \emph {et~al.}(1999)\citenamefont {Kawakami}, \citenamefont {Rotenberg}, \citenamefont {Choi}, \citenamefont {{Escorcia-Aparicio}}, \citenamefont {Bowen}, \citenamefont {Wolfe}, \citenamefont {Arenholz}, \citenamefont {Zhang}, \citenamefont {Smith},\ and\ \citenamefont {Qiu}}]{kawakami1999}%
  \BibitemOpen
  \bibfield  {author} {\bibinfo {author} {\bibfnamefont {R.~K.}\ \bibnamefont {Kawakami}}, \bibinfo {author} {\bibfnamefont {E.}~\bibnamefont {Rotenberg}}, \bibinfo {author} {\bibfnamefont {H.~J.}\ \bibnamefont {Choi}}, \bibinfo {author} {\bibfnamefont {E.~J.}\ \bibnamefont {{Escorcia-Aparicio}}}, \bibinfo {author} {\bibfnamefont {M.~O.}\ \bibnamefont {Bowen}}, \bibinfo {author} {\bibfnamefont {J.~H.}\ \bibnamefont {Wolfe}}, \bibinfo {author} {\bibfnamefont {E.}~\bibnamefont {Arenholz}}, \bibinfo {author} {\bibfnamefont {Z.~D.}\ \bibnamefont {Zhang}}, \bibinfo {author} {\bibfnamefont {N.~V.}\ \bibnamefont {Smith}},\ and\ \bibinfo {author} {\bibfnamefont {Z.~Q.}\ \bibnamefont {Qiu}},\ }\bibfield  {title} {\bibinfo {title} {Quantum-well states in copper thin films},\ }\href {https://doi.org/10.1038/18178} {\bibfield  {journal} {\bibinfo  {journal} {Nature}\ }\textbf {\bibinfo {volume} {398}},\ \bibinfo {pages} {132} (\bibinfo {year} {1999})}\BibitemShut {NoStop}%
\end{thebibliography}
%

\end{document}